\documentclass[fleqn,usenatbib]{mnras}

\usepackage{newtxtext,newtxmath}
\usepackage{graphicx} 
  \graphicspath{ {./images/} }
\usepackage{amsmath}
\usepackage[detect-all]{siunitx}
\usepackage[T1]{fontenc}

\DeclareRobustCommand{\VAN}[3]{#2}
\let\VANthebibliography\thebibliography
\def\thebibliography{\DeclareRobustCommand{\VAN}[3]{##3}\VANthebibliography}

\usepackage{graphicx}	
\usepackage{amsmath}	

\title[Tracing cosmic structure with post-EoR HI]{Tracing cosmic structure with neutral hydrogen after the Epoch of Reionization}

\author[J. Incley and L. Wolz]{
Jamie Incley$^{1}$\thanks{E-mail: jamie.incley@manchester.ac.uk} and Laura Wolz$^{1}$
\\
$^{1}$Jodrell Bank Centre for Astrophysics, Department of Physics and Astronomy, The University of Manchester, Manchester M13 9PL, UK\\
}

\date{Accepted XXX. Received YYY; in original form ZZZ}

\pubyear{\the\year{}}

\begin{document}
\label{firstpage}
\pagerange{\pageref{firstpage}--\pageref{lastpage}}
\maketitle

\begin{abstract}
We present a study of the transition of Neutral Hydrogen (HI) gas from the end of the Epoch of Reionization (EoR) to late-time large-scale structure. We examine the signature of the transition as traced through the redshifted 21-cm line with SKA-Low at $3 < z < 7$. To do so, we use the semi-numerical simulation \textsc{21cmFAST} to model the HI during the EoR and add a HI-halo based post-processing model of the late-time HI. This approach gives a robust estimate of the amplitude of the HI temperature field and predicts the observable power spectrum during the transition period. We find that our simulation pipeline reproduces the expected power spectrum trends from existing observations and theory, in addition to replicating current observational constraints on $\Omega_{\text{HI}}$. Our simulations predict a drop in power of four orders of magnitude between $4 < z < 7$. Assuming an inhomogeneous recombination model, we find a flattening of the power due to lingering neutral islands masking the late-time HI signal for $5 < z< 6.5$. 
Using SKA-Low deep survey parameters, we find HI power spectrum detectability at scales $k \leq 1$ $h$ Mpc$^{-1}$ for redshifts $3< z < 7$, even when using the horizon limit to mitigate foregrounds. Our results suggest a sufficient SNR of the HI power spectrum tracing the underlying halos $z < 5$, which can be used for late-time cosmology. 
Our results suggest that the resulting $\Omega_{\rm HI}$ constraints can trace different reionization scenarios such as a decreased escape fraction. This study implies that deep SKA-Low observations for $3< z< 7$ will be an important probe to constrain reionization parameters as well as cosmological models.

\end{abstract}

\begin{keywords}
large-scale structure of universe -- cosmology: theory -- software: simulations
\end{keywords}



\section{Introduction}

The gas in the universe has undergone an array of phase transitions in its lifetime, starting as a hot opaque cosmic `soup' of photons, protons, and electrons in its early life. Once it had expanded and cooled sufficiently, recombination occurred. Protons and electrons combined to form hydrogen atoms, leaving behind a neutral transparent medium through which the cosmic microwave background was able to free-stream. At \emph{Cosmic Dawn}, the first stars started to form, and the neutral universe was gradually `reionized' by the ionizing photons produced from these stars during an era appropriately termed the \emph{Epoch of Reionization} (EoR). This leaves us in the period of the universe we are currently living in: late-time structure formation and accelerating expansion, driven by a mysterious component of our universe called `dark energy'. If we wish to understand the nature of this component, we must track the behaviour of our universe across cosmic time.

The 21-cm signal from \emph{neutral hydrogen} (HI) is a rich source of information about our universe. It is the only probe of the cosmological dark ages (e.g. \citealt{2006PhR...433..181F, 2007PhRvD..76h3005L, 2012RPPh...75h6901P}) and produced by the most abundant baryon in our universe (approximately 75 per cent of baryonic mass in our universe is hydrogen, e.g. \citealt{1996MNRAS.282..263E, 2010ApJ...724..341H, 2012RPPh...75h6901P}); and it is a good tracer of the overall large-scale structure. However, the spin-flip transition that produces the 21-cm line is forbidden, with a half-life of order $10^7$ years. As such, the signal is weak and hidden behind much stronger galactic and extragalactic foregrounds (e.g. \citealt{2002ApJ...564..576D, 2008MNRAS.390.1496G, 2010ApJ...724..526D, 2012ApJ...757..101T, 2012ApJ...745..176V, 2014arXiv1408.4695C}). To extract statistically-significant information from the data, accurate modelling of the signal is crucial.

The EoR takes place across redshifts $5< z < 15$ (e.g. \citealt{2000ApJ...534..507C, 2001MNRAS.320..153B, 2004ApJ...610....9G}). It is characterized by significant ionizing UV photon production by stars in newly-formed structures such as galaxies and galaxy clusters. 
Prior to the conclusion of the EoR, HI is distributed continuously across the universe throughout the \emph{intergalactic medium} (IGM), and so the primary contributor to the spatial variation of the 21-cm signal during the EoR is the formation of ionized bubbles around dark matter \emph{haloes}, which have a greatly reduced emission in the HI line compared to the neutral background IGM. As such, the 21-cm signal during the EoR primarily tells us about the topology of these ionized bubbles and the underlying astrophysics that dictate their growth. 

After the EoR has concluded, only small pockets of the universe are left neutral. A significant proportion of these pockets are dense regions within galaxies, wherein recombination and ionization rates are comparable \citep{2000ApJ...530....1M}, shielding the bulk of the hydrogen contained within from the ionizing photons. 

Since only neutral hydrogen can undergo the 21-cm spin-flip transition, these pockets contribute the vast majority of the HI signal post-reionization. \cite{2018ApJ...866..135V} used outputs from the hydrodynamical \textsc{IllustrisTNG} simulation \citep{2018MNRAS.475..624N} to deduce that by $z=5$, 88 per cent of HI in the universe is found inside haloes (and 80 per cent inside galaxies within the haloes), and by $z=0$, these values are 99 and 97 per cent respectively. 

As a result, the post-EoR 21-cm signal traces the dark matter field more closely than the EoR signal, with only a minor modulation due to small-scale fluctuations in the background UV ionization field \citep{2009MNRAS.397.1926W}. 


\emph{HI intensity mapping} is a technique used for statistically significant observation of the HI signal at redshifts beyond our local Universe (e.g. \citealt{2001JApA...22...21B, 2004MNRAS.355.1339B, chang2008baryon, wyithe2008baryonic}). 
It removes the need to resolve individual galaxies by integrating the entire redshifted HI emission from multiple galaxies in each pixel on the sky. This allows for much deeper surveys to be carried out, beyond the local universe and out to high redshift. This allows us to map structure evolution in the universe post-EoR and therefore determine the cosmology in this era, since cosmological structure lies on the largest scales and can be mapped by a low-resolution survey.
This technique has already been used by numerous telescopes, including the \emph{Green Bank Telescope} (e.g. \citealt{2013MNRAS.434L..46S, 2013ApJ...763L..20M, 2022MNRAS.510.3495W}), \emph{CHIME} (Canadian Hydrogen Intensity Mapping Experiment) (e.g. \citealt{2014SPIE.9145E..22B, 2022ApJS..261...29C}), the \emph{Parkes Telescope} (e.g. \citealt{2018MNRAS.476.3382A}), and \emph{MeerKAT} (e.g. \citealt{2023MNRAS.518.6262C, 2023arXiv230111943P, 2025MNRAS.541..476M}). Additionally, \emph{HIRAX} (Hydrogen Intensity and Real-time Analysis eXperiment) \citep{2016SPIE.9906E..5XN} is a future instrument utilising HI intensity mapping, which is currently under construction in South Africa. 

In preparation for data analysis of upcoming intensity mapping surveys, considerable effort has been put into building theoretical models of the anticipated HI field as observed by various instruments. A commonly-taken approach applies a halo model of HI (see \autoref{halo-hi-theory} for theoretical details) to simulate the 21-cm field. This method is used by e.g. \cite{2009arXiv0908.2854W}, \cite{2019MNRAS.485.4060P}, \cite{2019MNRAS.484.1007W}, \cite{2021MNRAS.502.5259C}, and \cite{2025JCAP...04..003H}. 

Observing the evolution of the 21-cm field across redshift allows us to map the expansion history of the universe, which, in the era of precision cosmology, we need to measure to less than $1$ per cent uncertainty out to $z=6$ in order to deduce the cosmological dark energy density to a precision of $10$ per cent \citep{2018arXiv181009572C}.
HI intensity mapping has historically been used for $z < 1$, but in this work, we focus on the upcoming surveys that will be carried out by next-generation telescopes.

We are entering an era of information influx, with the \emph{SKA Observatory} \footnote{For detailed telescope specifications and science goals, see \url{https://www.skao.int/en/science-users/122/relevant-documents}.} set to begin full operation before the decade ends. This will be the largest radio telescope ever built, opening a door to far more stringent constraints on key cosmological parameters and enabling access to earlier epochs of the universe. The SKA will be comprised of telescopes arrays, the mid-frequency array SKA-Mid which will comprise MeerKAT in South Africa, and the low-frequency array SKA-Low set in Australia.

In particular, SKA-Low will be able to image frequencies down to $z\approx 3$ and up to $z\approx 27$, allowing it to observe not just the entire EoR, but also the post-EoR era $3 < z < 6$. 
This is an important transitionary period in our universe’s history, wherein reionization concludes, leaving behind a completely ionized intergalactic medium. HI remains only in small islands located towards the centre of haloes, where the medium is sufficiently dense to self-shield from ionizing photons. 
Consequently, during this time, the HI brightness temperature field goes from being a measure of the astrophysics of reionization to a tracer of the cosmic web. Therefore, the HI power spectrum in the post-reionization epoch presents itself as not only a rich source of information on cosmic structure at high redshift, but also an additional constraint on reionization end, due to its drastic change in behaviour during this period.

Current observational constraints on the HI transition period are limited owing to their inherent non-locality. However, we do have some tentative constraints on reionization end, namely the presence of a \emph{Gunn-Peterson trough} \citep{1965ApJ...142.1633G} in quasar spectra. This is a drop in flux at the long-wavelength end of the Lyman-$\alpha$ forest. The available data suggests a fairly late end to the EoR, concluding in the redshift range $5 < z < 6$ \citep{2015PASA...32...45B, 2020MNRAS.491.1736K, 2021ApJ...923...87C}. Constraints on the midpoint of reionization (where the mean neutral fraction $x_{\text{HI}}\approx 0.5$), from the Cosmic Microwave Background optical depth (see e.g. \citeauthor{2020A&A...641A...6P} \citeyear{2020A&A...641A...6P}) are in line with this, sitting around $7< z < 8$. 

\cite{2024MNRAS.533.2364G} simulate the HI transition period numerically, using an N-body code \textsc{PKDGRAV3} \citep{2017ComAC...4....2P}, a Friends-of-Friends halo finder (see \autoref{Halo-find}), and the radiative transfer code \textsc{pyC\textsuperscript{2}Ray} \citep{2006NewA...11..374M, 2024A&C....4800861H}. They anticipate a `knee-like' feature in the dimensionless 21-cm power spectrum to arise during the HI transition period as a result of neutral IGM islands persisting to low redshifts via recombination. On top of this, they predict a flattening of the redshift evolution of the 21-cm power towards the end of the transition, as the contribution to the total power from halo-based HI dominates over the IGM HI contribution. 

In this work, we investigate the use of the 21-cm power spectrum as a probe of the HI transition from IGM to haloes with the upcoming SKA-Low telescope. We aim to constrain how and when the power spectrum evolves due to this transition, and how this may vary with changes in the astrophysics of reionization.
To do this, we semi-numerically simulate the HI field during this transitionary period using the reionization code \textsc{21cmFAST} and a post-processing pipeline, and deduce the projected \emph{signal-to-noise ratio} (SNR) of the HI power spectrum across the range of observable scales and redshifts of interest. By taking a semi-numerical approach, we reduce the computational expense of the simulation, making it more suited for use in data analysis.
We also investigate whether the 21-cm power spectrum over this HI transition period may give us additional astrophysical information, such as constraints on the ionizing photon escape fraction and the star formation rate in galaxies, and how we can maximise the statistical significance of survey results through varying the redshift binning of the observations: balancing the range of $k_{\parallel}$ measured with the degree of thermal noise in each bin.

This paper is structured as follows. In \autoref{background}, we discuss the theory behind the modelling undertaken in this work. In \autoref{hi-im}, we describe the observational set-up and noise modelling. In \autoref{21cmfast-theory}, we detail the semi-numerical simulation \textsc{21cmFAST}, whose work we build on in this paper. In \autoref{Halo-find}, we describe our methods for finding haloes and assigning HI around them. Finally, in \autoref{forecast}, we discuss our results and make predictions of the observability of the HI signal by SKA-Low, for differing reionization astrophysics and foreground avoidance regimes. 

The cosmology assumed throughout is [h, $\Omega_{\text{m}}$, $\Omega_{\text{b}}$, $\Omega_{\Lambda}$, $f_{\text{H,c}}$] = [$0.6766$, $0.3096$, $0.04897$, $0.6904$, $0.75$], based upon the \emph{Planck}18 results \citep{2020A&A...641A...6P}.

\section{Theoretical background}\label{background}
\subsection{Power spectrum}\label{post-eor}
The post-EoR HI signal is primarily dependent upon the matter distribution of the universe. 
We can describe the two-point correlation statistics of a field using the \emph{power spectrum}: the HI power spectrum $P_{\text{HI}}(k, z)$ therefore describes a biased version of the underlying matter power spectrum $P_{\text{m}}(k,z)$.
\begin{equation}\label{bias-eq}
    P_{\text{HI}}(k,z)=b_{\text{HI}}^2(k,z)P_{\text{m}}(k,z)
\end{equation}
Application of a simple \emph{bias model} is a computationally-efficient method of generating the HI field from a given matter field. 
These models assume that the HI power spectrum traces the underlying matter power spectrum, using a function of proportionality called the \emph{HI bias}, $b_{\text{HI}}(k,z)$.
This bias is, in general, both scale- and redshift-dependent. However, at the largest scales, the HI bias is approximately constant, and is referred to as the \emph{linear bias}.
The HI bias becomes increasingly non-linear as the scales we observe decrease in size: linear theory is unable to fully explain the HI-halo density relation for scales $k> 0.3$ $h$ Mpc$^{-1}$ at any redshift \citep{2018ApJ...866..135V}. This makes the application of a bias model complicated for small-scale modelling of the HI, requiring in-depth simulations to deduce how the bias varies at these scales. 

One interesting avenue to look into is the use of the 21-cm power spectrum as a probe of reionization end. \cite{2016MNRAS.462..804G} use the \textsc{Tiamat} suite of N-body simulations \citep{2016MNRAS.459.3025P}, the \textsc{MERAXES} semi-numerical galaxy formation model \citep{2016MNRAS.462..250M}, and the reionization code \textsc{21cmFAST} \citep{2011MNRAS.411..955M, 2020JOSS....5.2582M, 2025arXiv250417254D} to investigate the power spectrum as a diagnostic of its progress, and find that power at small (large) scales is suppressed (amplified) as reionization progresses. In particular, it finds that the power at scales $k \approx \numrange{0.1}{1}$ Mpc$^{-1}$ reaches a maximum at the midpoint of reionization, and the slope of the power flattens beyond this point. If we can observe the changing shape of the power spectrum, this will give us key information about how the neutral fraction varies with time, and allow us to more precisely constrain when and how the EoR occurred.

\subsection{Halo models of HI}\label{halo-hi-theory}
We use a HI-halo model to describe the distribution of HI throughout the universe at $z < 7$. HI-halo models assume that the remaining HI after reionization is contained exclusively within dark matter haloes, with the total HI mass entirely determined by the mass of the halo in which it resides.
By combining a halo field with a deterministic HI-halo relation and HI profile, we can model the late-time 21-cm field.

Although there is a dependence of the HI mass within a halo on the local environment, this effect is expected to average out over the sample sizes and resolution we observe at using intensity mapping (see \autoref{hi-im}). As such, a HI-halo mass relation is a useful method of modelling the late-time HI field for our use case.

\cite{2017MNRAS.464.4008P} describe a HI-halo mass relation that includes lower and upper virial velocity cutoffs. 
We expect there to be a reduction in the HI mass fraction of low-mass haloes (with low virial velocities) as they will be insufficiently dense to shield the hydrogen from ionizing photons; and similarly a reduction at high masses due to the halo being sufficiently dense for hydrogen atoms to combine into molecular hydrogen.
For halo mass $M_{\text{h}}$, the corresponding HI mass $M_{\text{HI}}$ is given by \autoref{HI-halo-pad-a}:
\begin{equation}\label{HI-halo-pad-a}\begin{split}
    M_{\text{HI}}(M_{\text{h}}) = \alpha f_{\text{H,c}}M_{\text{h}}\left(\frac{M_{\text{h}}}{10^{11}h^{-1}\text{M}_{\odot}}\right)^{\beta} \exp{\left[-\left(\frac{v_{\text{c0}}}{v_{\text{c}}(M_{\text{h}})}\right)^3\right]} \\ \times \exp{\left[-\left(\frac{v_{\text{c}}(M_{\text{h}})}{v_{\text{c1}}}\right)^3\right]}
\end{split}\end{equation}
where $f_{\text{H,c}}$ is the cosmic hydrogen fraction, and $v_{\text{c}}$ is the virial velocity of the halo, defined as the circular velocity of the halo at its virial radius $r_{\text{v}}$.

The best-fit parameters are shown in the central column of \autoref{Padmanabhan-table}. $c_{\text{HI}}$ and $\gamma$ are used later in \autoref{conc}.

These parameters have been tuned to observational data from $0\leq z \leq 2.3$, including the HI mass function at $z\approx0$ obtained from HIPASS \citep{2004MNRAS.350.1195M}, and the DLA column density and incidence.
\begin{table}
 \caption{Best-fitting parameters used in the OM1 \citep{2017MNRAS.464.4008P} and OM2 \citep{2017MNRAS.469.2323P} HI-halo mass relations.}
 \label{Padmanabhan-table}
 \begin{tabular}{ |m{0.25\linewidth}|m{0.25\linewidth}|m{0.25\linewidth}| } 
 \hline
 \textbf{Parameter} & \textbf{OM1 value} & \textbf{OM2 value} \\ 
 \hline \hline
 c$_{\text{HI}}$ & $113.80$ & $28.65$ \\ 
 \hline
 $\alpha$ & $0.17$ & $0.09$ \\ 
 \hline
 $\log v_{\text{c0}}$ (km s$^{-1}$) & $1.57$ & $1.56$ \\ 
 \hline
 $\log v_{\text{c1}}$ (km s$^{-1}$) & $4.39$ & N/A \\ 
 \hline
 $\beta$ & $-0.55$ & $-0.58$ \\ 
 \hline
 $\gamma$ & $0.22$ & $1.45$ \\ 
 \hline
\end{tabular}
\end{table}
We will hereafter refer to this model as OM1 (Observation Model 1).

This model is evolved further in \cite{2017MNRAS.469.2323P}: removing the high $v_{\text{c}}$ exponential cutoff term (as the best fit value of the parameter $v_{\text{c1}}$ was found to be too high to affect the outputs), and fitting to a wider range of observational data up to $z\approx5$, using this to update the best-fitting parameter values in \autoref{Padmanabhan-table}. We will hereafter refer to this model as OM2 (Observation Model 2).

\cite{2020MNRAS.493.5434S} describe an alternative halo model, with parameters that evolve with redshift:
\begin{equation}
\begin{split}
    M_{\text{HI}}(M_{\text{h}})=M_{\text{h}} \left\{a_1\left(\frac{M_{\text{h}}}{10^{10}}\right)^{\beta}\exp{\left[-\left(\frac{M_{\text{h}}}{M_{\text{break}}}\right)^{\alpha}\right]}+a_2\right\} \\ \times\exp{\left[-\left(\frac{M_{\text{min}}}{M_{\text{h}}}\right)^{\gamma}\right]}
    \end{split}
\end{equation}
where $a_1$, $M_{\text{break}}$, $a_2$, and $M_{\text{min}}$ are free parameters that may be fit to data. They fit these to simulation data from the \textsc{GAlaxy and Evolution Assembly} (GAEA) semi-numerical model \citep{2016MNRAS.461.1760H, 2017MNRAS.466L..88D, 2017MNRAS.464.3812F}, which was run on the merger trees output from the \emph{Millennium} simulations \citep{2005Natur.435..629S, 2009MNRAS.398.1150B}. These free parameters evolve with redshift and are defined at integer redshift between $0 \leq z \leq 5$ (see \cite{2020MNRAS.493.5434S} for the values of these parameters). We will hereafter refer to this model as SM (Simulation Model).
We compare the outputs from these three analytic models in \autoref{Halo-find}.

It has been found that a rapid evolution of haloes containing a significant HI population must occur after $z\approx 2$ in order to match low-z observations, which is potentially unphysical \citep{2017MNRAS.464.4008P}; and \cite{2014JCAP...09..050V} find that halo models of the neutral hydrogen tend to under-predict the abundance of DLAs. However, for our purposes of modelling the HI field as seen through the 21-cm line at $z \geq 3$, a halo model is suitable.

The density profile of neutral hydrogen inside the host dark matter haloes varies widely between haloes, as the HI is sensitive to processes such as AGN feedback and tidal stripping which occur within the halo environment \citep{2018ApJ...866..135V}. However, the mean density profile is universal, and may be modelled using a modified \emph{Navarro-Frenk-White} (NFW) profile \citep{1996ApJ...462..563N, 1997ApJ...490..493N, 2017MNRAS.464.4008P}:
\begin{equation}\label{halo-hi-prof}
    \rho_{\text{HI}}(r)=\frac{\rho_0r_{\text{s}}^3}{(r+0.75r_{\text{s}})(r+r_{\text{s}})^2}
\end{equation}
where $r_{\text{s}} = \frac{r_{\text{v}}(M)}{c(M,z)}$, and $\rho_0$ is a normalisation parameter, obtained through solving \autoref{norm} for a given total HI mass and virial radius.
\begin{equation}\label{norm}
    \int^{r_{\text{v}}(M)}_04\pi r^2\rho_{\text{HI}}(r)dr=M_{\text{HI}}(M)
\end{equation}
$c(M_{\text{h}},z)$ is the HI concentration parameter, defined as:
\begin{equation}\label{conc}
    c(M_{\text{h}},z)=c_{\text{HI}}\left(\frac{M_{\text{h}}}{10^{11}\text{M}_{\odot}}\right)^{-0.109}\frac{4}{(1+z)^{\gamma}}
\end{equation}
where $c_{\text{HI}}$ is a free parameter, and $r_{\text{v}}$ is the virial radius of the halo.
We used this modified profile due to the effects of baryonic heating on the HI, which the NFW profile, describing the dark matter profile, does not account for.

\section{Observational set-up}\label{hi-im}
In this section, we review the instrument used in our forecasts, and the instrumental noise contribution to the observed power spectrum.
The three surveys used to calculate the SKA-Low instrumental noise in this paper are referred to as \emph{Deep-AA4} (AA4 stage, 100 deg$^2$ survey area, 5000 h observing time), \emph{Deep-AA*} (AA* stage, 100 deg$^2$ survey area, 5000 h observing time), and \emph{Shallow-AA*} (AA* stage, 100 deg$^2$ survey area, 1000 h observing time).
\subsection{SKA Observatory}
The SKA Observatory is a next-generation radio telescope currently under construction, with sites in South Africa (\emph{SKA-Mid}) and Australia (\emph{SKA-Low}). 
SKA-Low is a low-frequency array, able to observe in the frequency range $\numrange{50}{350}$ MHz.
It will be comprised of 307 (512) stations for the AA* (AA4) configuration\footnote{Via the SKA staged delivery memo, retrieved from \url{https://www.skao.int/en/science-users/ska-tools/494/ska-staged-delivery-array-assemblies-and-subarrays}.}, each of which is in turn composed of 256 antennas. 
The signals received by each antenna may be combined with a controllable delay, which allows a beam to be formed and its pointing be varied across the sky, without having to physically move any of the antennas.
The stations are arranged into a dense core surrounded by log-spaced stations out to a diameter of approximately $80$ km. This core enables high sensitivity to large low-brightness regions on the sky\footnote{Via the SKA Design Baseline Description memo, retrieved from \url{https://www.skao.int/en/science-users/118/ska-telescope-specifications}.}.

Pairs of stations, separated by a distance $D$, form interferometric baselines with each other, with a corresponding $uv$-coordinate distance $|\textbf{u}|=u=\frac {D} {\lambda_{21}(z)}$. Each of these baselines measures the HI visibility on the sky for a given perpendicular wavenumber $k_{\perp} = \frac{2\pi u}{\chi(z)}$, with $\chi(z)$ the comoving cosmological distance to the observed redshift $z$. This may be combined with the parallel wavenumber $k_{\parallel} = \frac{2\pi}{\Delta \chi}$, where $\Delta \chi$ is the line-of-sight cosmological distance corresponding to the separation between frequency channels, to form the overall 1D wavenumber $k=\sqrt{k_{\perp}^2+k_{\parallel}^2}$.
In order to remove contamination by discrete radio sources, it is projected that baselines out to 65 km will need to be processed \citep{2020PASA...37....7S}, although the cosmological signals of interest (i.e. those corresponding to large-scale structure formation) are probed via the shortest baselines.

We compare the baseline number density\footnote{This is calculated through the \emph{ska-ost-array-config} package, retrieved from \url{https://gitlab.com/ska-telescope/ost/ska-ost-array-config}. This package is used to determine the number of station pairs that correspond to a given $u$.} $n(u)$ for the AA4 and AA\textsuperscript{*} stages of SKA-Low in \autoref{fig:therm-ps}.

The minimum observable wavenumber $k_{\text{min}}$ is set by the \emph{field-of-view} (FoV) of the telescope. Using the definition from \cite{2015ApJ...803...21B}:
\begin{equation}
    \text{FoV} \approx \frac{\lambda^2(z)}{A_e}
\end{equation}
where $A_e$ is the effective area of a single array element and $\lambda(z) = \lambda_{21}(1+z)$ is the observed wavelength at redshift $z$, we find that $k_{\text{min}}=0.026$ $h$ Mpc$^{-1}$ at $z=5$.

The maximum observable wavenumber is set by the maximum baseline of the interferometer, $k_{\text{max}}=\frac{2\pi D_{\text{max}}}{\lambda(z)\chi(z)}$, where $\chi(z)$ is the comoving cosmological distance to redshift $z$. For SKA-Low, this is $81$ $h$ Mpc$^{-1}$ for $z=5$.

\subsection{Noise power spectrum}
We use the radiometer equation to calculate the thermal noise per baseline \citep{2015ApJ...803...21B}:
\begin{equation}\label{therm-per-bl-eq}
    \sigma_{\text{T}}^2(z, u)=\frac{\lambda_{21}^4(z)T_{\text{sys}}^2(z)}{A_{\text{e}}^2\Delta\nu t_{\text{p}}n(u)2\pi u \mathrm{d}u}\frac{1}{N_{\text{beam}}N_{\text{pol}}}
\end{equation}
where $\lambda_{21}(z)$ is the wavelength of the observed HI signal at redshift $z$, $\Delta\nu$ is the frequency bin width, $t_{\text{p}}$ is the total observation time, $N_{\text{beam}}$ is the number of beams of the instrument, $N_{\text{pol}}$ is the number of polarisations of the instrument, and $T_{\text{sys}}$ is the system temperature of the survey \citep{2019arXiv191212699B}:
\begin{equation}\label{eq-T_sys}
    T_{\text{sys}} = \left[T_{\text{sky}}(z)+T_{\text{spl}}+T_{\text{rcv}}(z)\right]\frac{x}{e^x-1}
\end{equation}
for $x = \frac{h\nu_{21}}{k_{\text{B}}(T_{\text{rcv}}(z)+T_{\text{sky}}(z)+T_{\text{spl}})}$. The constituent temperatures are defined in \autoref{temp-table}: the equation for $T_{\text{sky}}(z)$ is from \cite{2019arXiv191212699B}; and the equations for $T_{\text{gal}}(z)$ and $T_{\text{rcv}}(z)$ are from \cite{2020PASA...37....7S}. We normalise the half-plane integral of $n(u)$ to the total number of baseline pairs, as in \autoref{half-plane}.
\begin{equation}\label{half-plane}
    N_{\text{tot}} = \frac{N_{\text{dish}}(N_{\text{dish}}-1)}{2} = \int n(u)2 \pi u \text{d}u
\end{equation}
The thermal noise power spectrum is given as \citep{2015ApJ...803...21B}:
\begin{equation}\label{therm-ps-eq}
    P_{\text{noise}}(k)=\frac{\sigma_{\text{T}}^2(u) V_{\text{pix}} 2\pi u \mathrm{d}u}{\Omega_{\text{FoV}}}
\end{equation}
where $\Omega_{\text{FoV}}$ is the FoV solid angle, and $V_{\text{pix}}$ is the comoving volume observed during a single pointing of the telescope.
The FoV of SKA-Low for a single pointing is approximately $5$ deg$^2$ at $z\approx 5$, and so this limit sets how deep our survey may be for a given observation time.
\begin{table}
 \caption{Temperatures used in \autoref{eq-T_sys}. Here, for simplicity, we neglect the contribution from spillover and atmospheric temperature.}
 \label{temp-table}
 \begin{tabular}{ |m{0.2\linewidth}|m{0.2\linewidth}|m{0.4\linewidth}| } 
 \hline
 \textbf{Parameter} & \textbf{Type} & \textbf{Value, K} \\
 \hline\hline
 $T_{\text{sky}}(z)$ & Sky & $T_{\text{CMB}}+T_{\text{gal}}(z)+T_{\text{atm}}$\\ 
 \hline
 $T_{\text{gal}}(z)$ & Galaxy & $25 \left(\frac{408\text{MHz}}{\nu_{21}(z)}\right)^{2.75}$\\
 \hline
 $T_{\text{rcv}}(z)$ & Receiver & $0.1T_{\text{gal}}(z)+40$\\ 
 \hline
 $T_{\text{CMB}}$ & CMB & $2.73$\\ 
 \hline
 $T_{\text{spl}}$ & Spillover & $0$\\ 
 \hline
 $T_{\text{atm}}$ & Atmospheric & $0$\\ 
 \hline
\end{tabular}
\end{table}
\begin{figure}
    \centering
    \includegraphics[width=\linewidth]{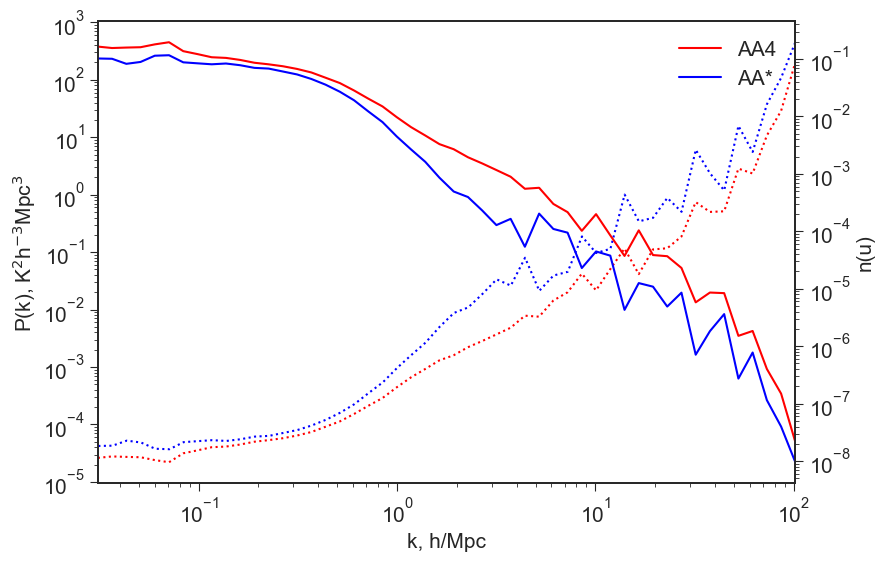}
    \caption{The thermal noise power spectrum at $z=4.5$ (solid line, left axis) and the baseline number density (dashed line, right axis) for the AA4 (red) and AA* (blue) stages of SKA-Low. The thermal noise power spectra were calculated assuming a deep survey (100 deg$^2$ survey area, 5000 h observing time).}
    \label{fig:therm-ps}
\end{figure}
We show the thermal noise power spectrum (alongside the corresponding baseline number density) for AA* vs AA4 for a mean redshift of $z=4.5$ in \autoref{fig:therm-ps}. As expected, the significance of thermal noise increases with decreasing scale, due to a lack of baselines observing at the smallest scales, requiring deeper observations to get a significant detection at higher wavenumbers.

\section{21cmFAST}\label{21cmfast-theory}
In this section, we describe the reionization simulation \textsc{21cmFAST} and the settings that we use for our purposes of modelling the transition in the distribution of HI in the universe that takes place at the end of the EoR.

A full N-body or hydrodynamic simulation of the universe on the Mpc-Gpc scales required for intensity mapping requires significant computational expense. As such, analytic methods are frequently used within simulations used in data analysis and forecasting, often being combined with numerical methods in what are called \emph{semi-numerical} simulations.
We build from the semi-numerical reionization code \textsc{21cmFAST} to fully model the HI transition from the EoR to late-time large-scale structure.

\textsc{21cmFAST}\footnote{\url{https://github.com/21cmfast/21cmFAST}} \citep{2011MNRAS.411..955M, 2020JOSS....5.2582M, 2025arXiv250417254D} utilises Lagrangian perturbation theory and the excursion set formalism to simulate the 21-cm signal throughout the EoR. As such, it does not need to explicitly resolve source haloes\footnote{As of \textsc{21cmFAST}v4 \citep{2025arXiv250417254D}, stochastic sampling of conditional halo mass functions is utilised in addition to the excursion set formalism to model the halo field. However, we started this project before this discrete halo finder was publicly available, and upon release, we found our halo finder to produce similar outputs to the built-in functionality. As a result, we do not use the native halo finder in this work.}, and instead can work directly with the overdensity field. This allows for a large dynamic range to be computed, with only a small loss in accuracy.

\textsc{21cmFAST} employs the \emph{excursion set formalism} \citep{1991ApJ...379..440B, 2004ApJ...613....1F} to describe the formation of ionized regions. It is based upon the principle of collapsed regions of space forming ionizing sources and subsequently producing ionizing photons. In its simplest form, it requires only two input parameters: the ionizing efficiency of a collapsed object and the minimum permissible halo mass. The ionizing efficiency is oftentimes represented through an ionization parameter $\zeta$, via the relation $M_{\text{ion}} = \zeta M_{\text{coll}}$, where $M_{\text{coll}}$ is the mass within a region that is assumed to have collapsed into dark matter haloes, and $M_{\text{ion}}$ is the corresponding mass of matter that has been ionized within that region.

\textsc{21cmFAST} supports the use of \emph{2\textsuperscript{nd}-order Lagrangian Perturbation Theory} (2LPT) for evolution of the density field, which increases the memory requirements 6-fold. 2LPT is recommended for the simulation of structure at low redshifts, when the matter distribution has become strongly non-linear. 

\textsc{21cmFAST} outputs zero brightness temperature after reionization has concluded and all cells are fully ionized.
We develop a post-processing pipeline to add the brightness temperature contribution from the late-time HI found in haloes.
The settings used to run \textsc{21cmFAST} are shown in \autoref{21cmFAST-table}, in addition to the fiducial astrophysical parameters (discussed in \autoref{impact}).

\begin{table*}
 \caption{\textsc{21cmFAST} settings (top half of table), alongside the fiducial values of the astrophysical parameters varied in \autoref{impact} (bottom half of table).}
 \label{21cmFAST-table}
 \begin{tabular}{ |m{0.25\textwidth}|m{0.55\textwidth}|m{0.1\textwidth}| } 
 \hline
 \textbf{Parameter} & \textbf{Definition} & \textbf{Value} \\
 \hline\hline
  USE\_HALO\_FIELD & Use a halo field to model discrete galaxy populations & False\\
 \hline
 USE\_RELATIVE\_VELOCITIES & Include DM-baryon relative velocities & True\\
 \hline
 PERTURB\_ALGORITHM & The perturbation regime used to evolve the overdensity field & "2LPT"\\
 \hline
 INHOMO\_RECO & Model inhomogeneous recombination & True\\
 \hline
 USE\_TS\_FLUCT & Calculate spin temperature fluctuations & False\\
 \hline
 \hline
 NU\_X\_THRESH & X-ray energy threshold for self-absorption by host & $500$\\ 
 \hline
 X\_RAY\_SPEC\_INDEX & X-ray luminosity spectral index & $1.0$\\
 \hline
 t\_STAR & Characteristic timescale for star formation in galaxies, as a fraction of the Hubble time & $0.5$\\ 
 \hline
 F\_STAR10 & Fraction of gas in stars for halos of mass $10^{10}\text{ M}_{\odot}$ in $\log_{10}$ units & $-1.3$\\ 
 \hline
 ALPHA\_ESC & Index of escape fraction as function of halo mass & $-0.5$\\ 
 \hline
 F\_ESC10 & Escape fraction for halos of mass $10^{10}\text{ M}_{\odot}$ in $\log_{10}$ units & $-1.0$\\ 
 \hline
\end{tabular}
\end{table*}
We do not perform spin temperature fluctuations, since the spin temperature across the IGM can be assumed saturated by the end stages of reionization.
2LPT is used to evolve the overdensities to the non-linear regimes of late times and smaller scales.
We use inhomogeneous recombination \citep{2014MNRAS.440.1662S} since the reionization process becomes absorption-dominated during its final stages and thus the effects of recombination become significant \citep{2025PASA...42..107C}.

It is important to ensure the actual size of the cells is small enough to enable sufficient resolution and halo modelling. If the cells are too large, multiple haloes will likely lie within a single cell, and the HI-halo mass relation becomes invalid. To avoid this occurring, a cell size of $V_{\text{cell}}=(0.16 \text{ }h^{-1} \text{Mpc})^3$ is used when running the static simulation boxes discussed in \autoref{Halo-find} and \autoref{power-section}, corresponding to a mean resolution of order $10^{9}$$ h^{-1} \text{M}_{\odot}$.
This is in line with \cite{2017MNRAS.465..111K}, who find that in halo-based models at redshifts $z > 0.5$, a halo mass resolution of at least $\approx 10^{10}\text{ }h^{-1}\text{M}_{\odot}$ must be used in simulations in order for the resultant HI distribution to be well-converged.

\section{Halo finding and HI assignment}\label{Halo-find}
In this section, we describe the dark matter haloes within the gridded density field output by \textsc{21cmFAST}, and subsequently distribute the neutral hydrogen about these haloes. We discuss the evolution of the halo mass function and $\Omega_{\text{HI}}$ as a function of redshift. The volume of the simulation box used in this section is $V_{\text{box}}=(40\text{ }h ^{-1}\text{Mpc})^3$, comprised of $250^3$ cells.

Halo finders are a ubiquitous tool in cosmological simulations, owing to the importance of dark matter haloes in dictating overall structure formation. However, many halo finders are designed to act on discrete particle distributions generated by N-body simulations. One commonly-used example is the \textsc{Amiga Halo Finder} (AHF) \citep{2009ApJS..182..608K}. This utilises an \emph{adaptive mesh refinement} scheme to find density peaks, wherein the resolution of the mesh that the particles are `painted' on to increases in higher-density regions of the simulation. This enables a high dynamic range (which is a requirement to locate subhaloes within a larger halo structure) while minimizing computational resources. Once an initial list of particles for each halo has been generated, gravitationally unbound particles are removed iteratively until the entire structure is bound.
AHF is an example of a \emph{Spherical Overdensity} (SO) halo finder: the other main category of halo finder, \emph{Friends-of-Friends}, will be discussed later in this section\footnote{For a more in-depth review of different halo finders, see e.g. \cite{2011MNRAS.415.2293K} and \cite{2013MNRAS.435.1618K}.}.

Our halo finder operates directly on a gridded overdensity field. We utilise the \emph{watershed} image segmentation algorithm to identify individual overdense regions, corresponding to each diffuse halo. Watershed is applied to the negative of the overdensity field, such that the densest regions correspond to the deepest `troughs'. 
This algorithm has previously been used in other cosmological structure finders: not only in halo finding (see e.g. \textsc{PHEW} in \citealt{2015ComAC...2....5B, 2025MNRAS.537..321G}), but also more commonly in cosmic void finding (see e.g. \textsc{ZOBOV} in \citealt{2008MNRAS.386.2101N} or the \textsc{Watershed Void Finder} in \citealt{2007MNRAS.380..551P}).
It follows the steepest gradient of a field to find regions associated with each other on a 3D image. These regions are grouped around local maxima of the image. By applying this algorithm to an overdensity field, we may locate density maxima and deduce whether these correspond to haloes, based on the total mass contained within.
Within the watershed function itself, we vary the connectivity parameter. This dictates which order of `nearest neighbours' are classed as part of the same object, to a maximum effective value of 3.

To remove extrema that do not correspond to haloes, we apply an overdensity cap to the field before application of the watershed algorithm. This cap is required to vary with cell size to avoid neglecting low-mass haloes. Through comparison to the theoretical \emph{halo mass function} (HMF), we find that the optimal overdensity cap for a cell size of $V_{\text{cell}}=(0.16\text{ }h^{-1} \text{Mpc})^3$ is $8$. For scaling to different cell sizes, we use an inverse linear scaling between cell length and overdensity cap.

The total halo mass corresponding to each region is obtained by summing the masses from the cells within that region which are sufficiently overdense, and finally the halo centre is set as the cell containing the density maximum of that region.

The mass of each cell is calculated using \autoref{cell-mass}. We assume the mean density in the universe is equal to the critical overdensity (taking the universe to be flat), and use a constant critical density (the value at current time, $z=0$), as the box is comoving:
\begin{equation}\label{cell-mass}
    M_{\text{cell}}=(\delta+1)\times \rho_{\text{crit}}(z=0) \times \frac{\Omega_{\text{m}}}{\Omega_{\text{m}} + \Omega_{\Lambda}} \times l_{\text{cell}}^3
\end{equation}
where $\delta$ is the overdensity of the cell, $\rho_{\text{crit}}(z=0)$ is the critical density of the universe evaluated at $z = 0$, $\Omega_{\text{m}}$ is the dimensionless matter density parameter at $z = 0$, $\Omega_{\Lambda}$ is the dimensionless dark energy density parameter at $z = 0$, and $l_{\text{cell}}$ is the physical length of a cell.

\begin{figure}
    \centering
    \includegraphics[width=\linewidth]{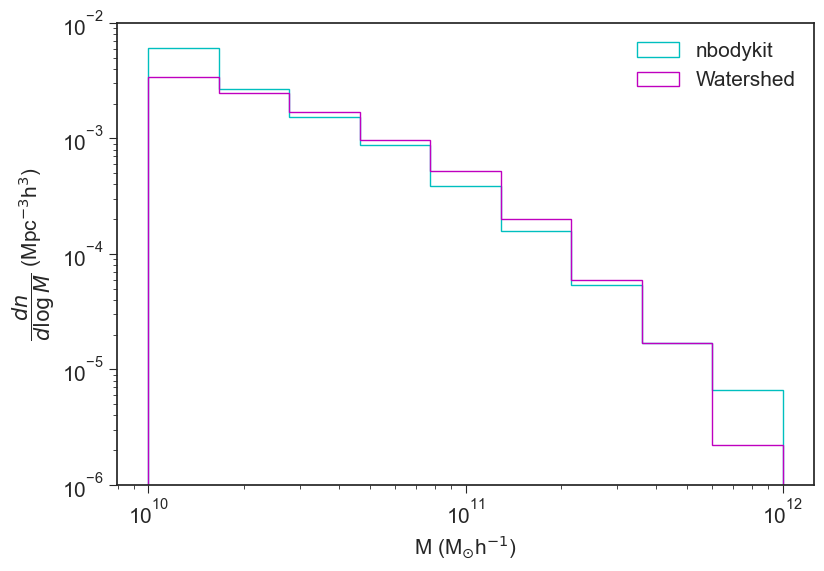}
    \caption{Consistency check between the HMFs of our watershed halo finder and of the FoF halo finder. Apart from divergences at mass extremes, the two halo finders overall show good agreement.}
    \label{fig:hmf-comp}
\end{figure}

The HMF produced by this halo finder is shown in \autoref{fig:hmf-comp}, alongside the HMF produced by a \emph{Friends-of-Friends} (FoF) halo finder from the package \textsc{nbodykit}\footnote{\url{https://github.com/bccp/nbodykit}} \citep{2018AJ....156..160H}. 

The FoF halo finder is an example of a \emph{particle collector}: it links and collects particles near each other. Two particles within each other’s linking length (usually set to be a fifth of the mean interparticle separation) are considered to be `friends'. If two particles aren’t within each others’ linking length, but have a mutual `friend', they are considered `friends-of-friends' and are thus part of the same object. Since it is based on particle separation, it can only be applied to N-body simulations and not gridded fields. However, the initial density field from \textsc{nbodykit} can be transferred to a grid, for use with the watershed halo finder. 

For this comparison, we use an overdensity cutoff of 8 and a connectivity of 1 for watershed; and a minimum particle number per halo of 6 and a linking length $b=0.2$ for \textsc{nbodykit}. 

\begin{figure}
    \centering
    \includegraphics[width=\linewidth]{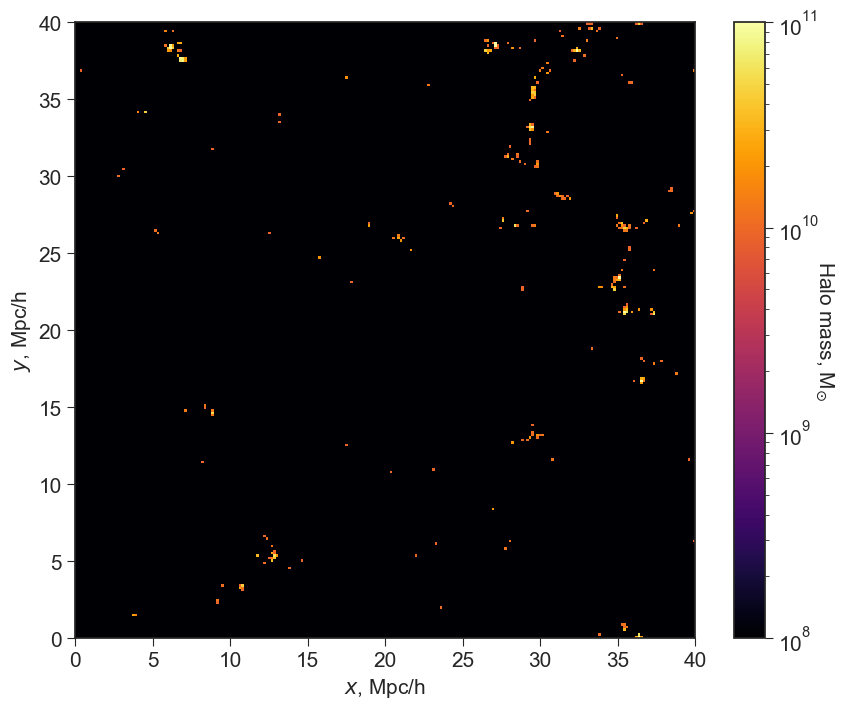}
    \includegraphics[width=\linewidth]{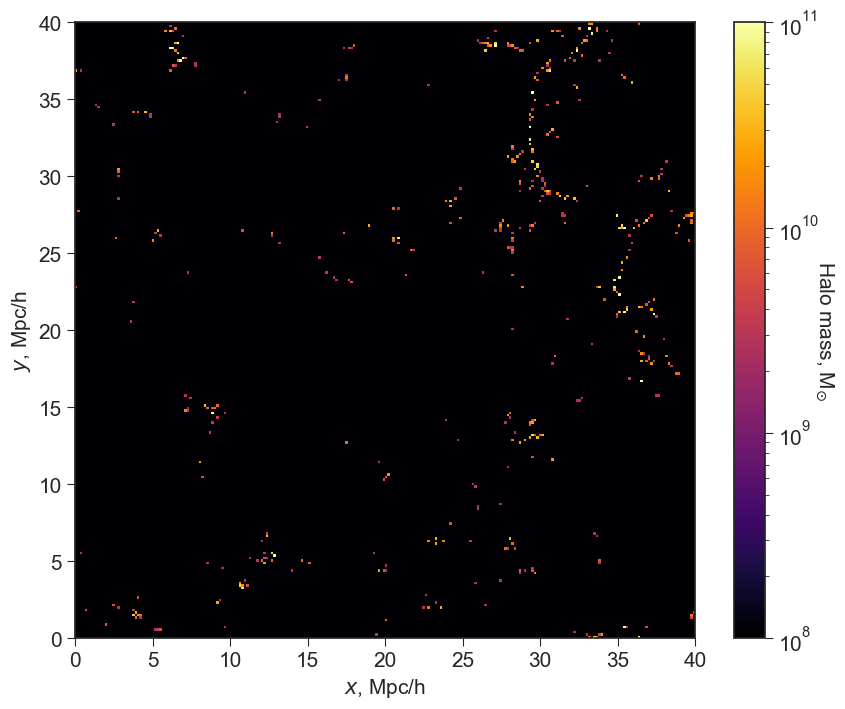}
    \caption{Slices (0.8 $h^{-1}$Mpc thick) of the halo field found using each algorithm. \textbf{Top:} Halo field from \emph{nbodykit}'s FoF halo finder. \textbf{Bottom:} Halo field from the watershed halo finder, acting on the same density field. Due to slices being shown, and slight differences in halo centre position, some haloes located by one halo finder appear `missing' in the other, but are present in adjacent cells. Additionally, subhaloes are resolved by the watershed halo finder, but not by FoF.}
    \label{fig:nb-halo-comp}
\end{figure}

In \autoref{fig:hmf-comp} we show that the watershed-based halo finder broadly reproduces the HMF produced by the FoF halo finder, although they differ at mass extrema, where resolution or volume limits become significant.
\begin{figure}
    \centering
    \includegraphics[width=\linewidth]{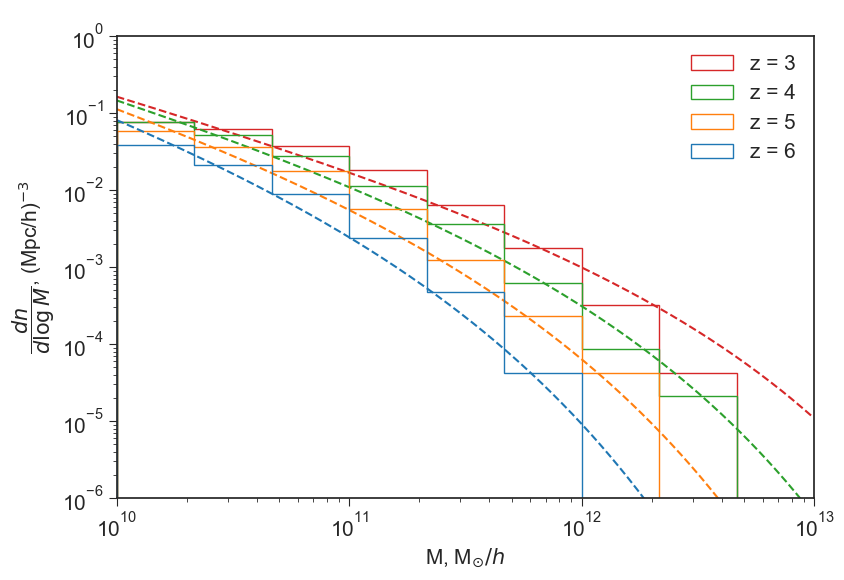}
    \caption{Evolution of the HMF across redshift. The dashed lines represent the theoretical Watson HMF \citep{2013MNRAS.433.1230W} generated using nbodykit \citep{2013A&C.....3...23M}, and the solid lines are the binned HMFs computed from the simulation.}
    \label{fig:hmf-evo}
\end{figure}
We also review how the HMF of our simulation evolves with redshift, and compare it to the predictions of the SO-derived HMF from \cite{2013MNRAS.433.1230W}, via the \textsc{hmf}\footnote{\url{https://github.com/halomod/hmf}} Python package \citep{2013A&C.....3...23M}. We choose to compare to this HMF as it accounts for small deviations from universality (i.e. due to redshift and cosmology) and has been fit to large N-body simulation data from $0\leq z\leq 30$. In \autoref{fig:hmf-evo}, the HMF from our code evolves a little slower with redshift than the theoretical. However, it follows the same overall trend, and within the redshift range our code focuses on, it does not diverge significantly from the theoretical. We also see a divergence from the theoretical at the highest halo masses, but this is likely due to the scarcity of these haloes combined with the small box size we used for testing our halo finder. The number of haloes of this mass likely to be resolved within our box is O$(1)$, and so their abundance is noticeably affected by cosmic variance, since we are sampling over such a small portion of the universe.

Through this, we find that the optimal parameters for matching our halo finder outputs to the theoretical are a connectivity parameter of 1 and an overdensity cap of 8 (as discussed previously). These values will be used throughout the remainder of this work.

Overall, we find that our watershed halo finder is an adequate algorithm to identify halo centres and assign mass on a gridded density field to a reasonable degree of accuracy considering our use case.

Once the halo masses and centres are obtained, we apply the HI-halo mass relations (see \autoref{halo-hi-theory}) to the halo field, to determine the total HI mass associated with each halo. 

The next step is to distribute this total HI mass around each halo centre by generating the radial density profile of the HI (taken as the modified Navarro-Frenk-White profile from \autoref{halo-hi-prof}) onto a grid, and then convolving this profile with the positions of the halo centres. This profile generation and convolution is applied to the haloes falling in each of the 25 linearly-spaced halo mass bins, following \cite{2019MNRAS.484.1007W}, using the central mass within each bin to build the radial profile for efficiency.

\begin{figure}
    \centering
    \includegraphics[width=\linewidth]{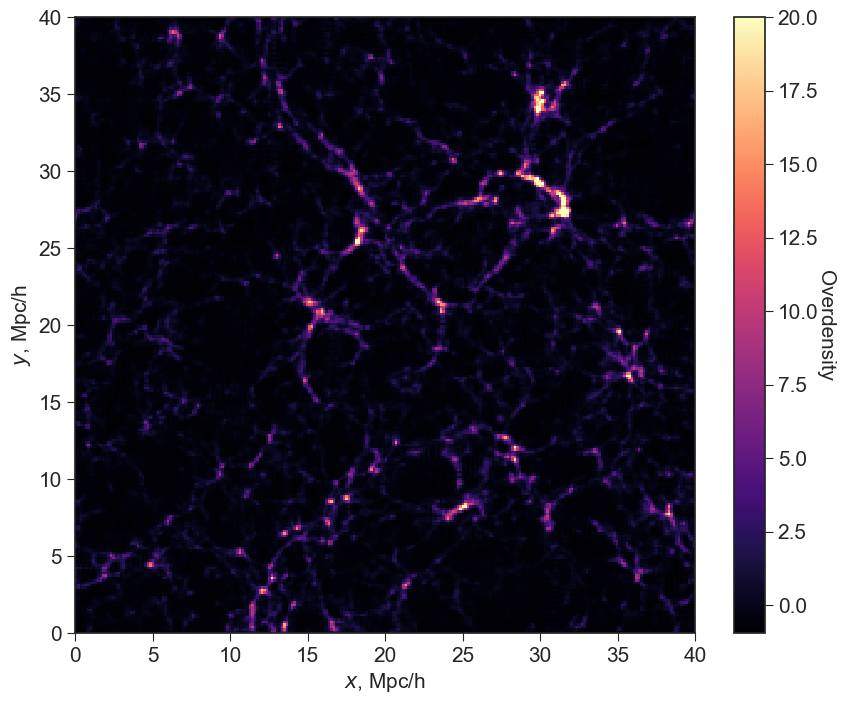}
    \includegraphics[width=\linewidth]{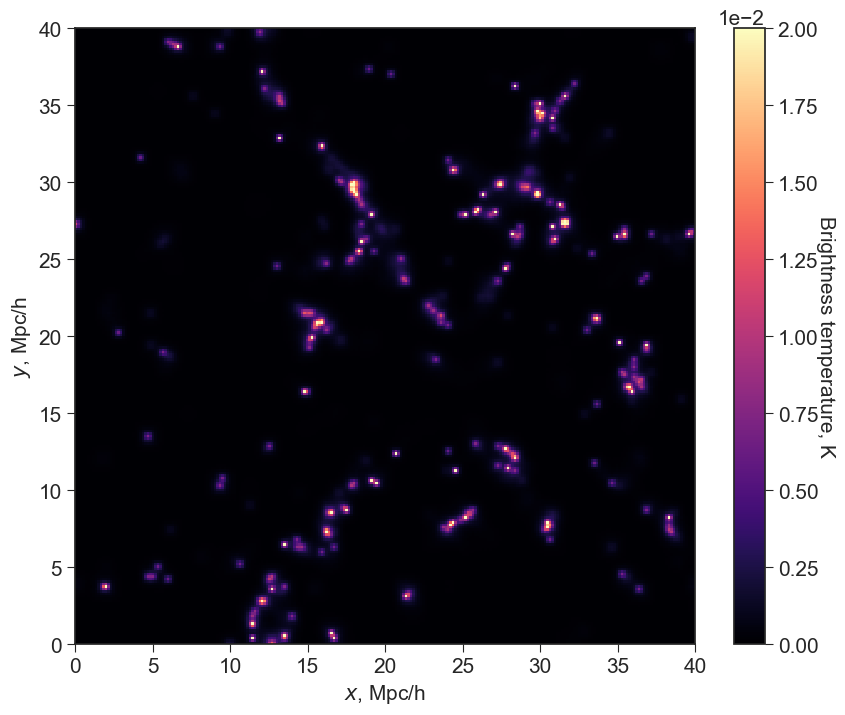}
    \caption{\textbf{Top:} Slice of the overdensity field output by \textsc{21cmFAST} at $z = 4$. \textbf{Bottom:} Corresponding brightness temperature field at this position and redshift, using OM2. $z=4$ was chosen to highlight the contribution to the signal from halo-based HI.}
    \label{fig:bt-field}
\end{figure}

The final step is to determine the 21-cm brightness temperature, which corresponds directly to the observed signal. We use the form given by \cite{2017MNRAS.470.3220W}:
\begin{equation}
    T_{\text{HI}}(x, z) = \frac{3h_{\text{p}}c^3A_{10}}{32\pi m_{\text{h}}k_{\text{B}}\nu_{21}^2}\frac{(1+z)^2}{H(z)}\rho_{\text{HI}}(x)
\end{equation}
where $h_p$ is the Planck constant, and take $\rho_{\text{HI}}(x) = \frac{M_{\text{HI}}(x)}{V_{\text{cell}}}$. The brightness temperature field is shown in \autoref{fig:bt-field}, and the corresponding power spectrum in \autoref{fig:pad-b-ps} (discussed in \autoref{power-section}).

\begin{figure}
    \centering
    \includegraphics[width=\linewidth]{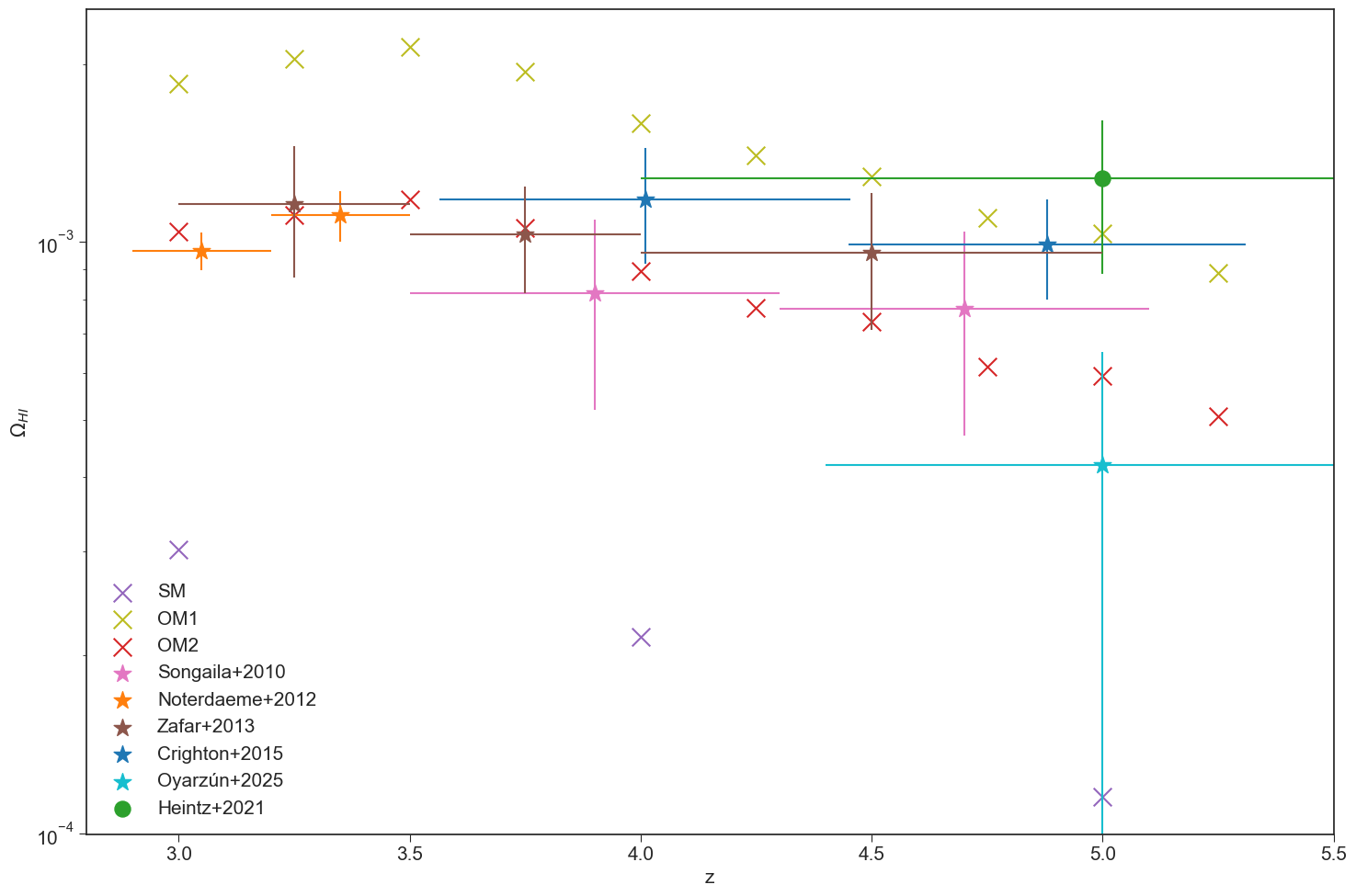}
    \caption{Evolution of halo-based $\Omega_{\text{HI}}$ across redshift for analytic models OM1, OM2, and SM, compared to observational data from DLA column density (stars) and from [CII] emission (circle). We only plot the $\Omega_{\text{HI}}$ for $z=[3, 4, 5]$ for SM, as its parameters are redshift-dependent and defined only for integer redshifts $z\leq5$.}
    \label{fig:omega-hi-comp}
\end{figure}

We also calculate and compare how the dimensionless HI density parameter $\Omega_{\text{HI}}$ evolves across redshift for all three HI-halo models, in \autoref{fig:omega-hi-comp}. We plot current observational constraints alongside our model values.

In \autoref{fig:omega-hi-comp} we show that the $\Omega_{\text{HI}}$ output by SM is smaller than the other two models and the observational constraints. This is likely due to the model being tuned to data at $z\approx0$, and it is known to predict a decreased amount of HI at high redshifts with respect to observational data \citep{2020MNRAS.493.5434S}. 
OM1, on the other hand, outputs a slightly higher $\Omega_{\text{HI}}$ than the observational constraints: again, likely as a result of being fit to low-redshift data.
We will focus on the outputs of OM2 in the remainder of this work.

\section{Forecasting and discussion}\label{forecast}
In this section, we present the outcomes of our HI transition simulation, and consider them in the context of SKA-Low instrumental noise. We present results for redshifts $3 < z < 7$.
The redshifts listed in this section are characteristic of the specific realisations of the initial density field used, but the overall behaviour is applicable regardless of the actual timing of reionization.

\subsection{HI power spectra}\label{power-section}
In this section, we present our results on the HI bias across redshift and scales, and predictions on the detectability of the signal, using a theoretical forecast of the HI power and instrumental noise.

\begin{figure}
    \centering
    \includegraphics[width=\linewidth]{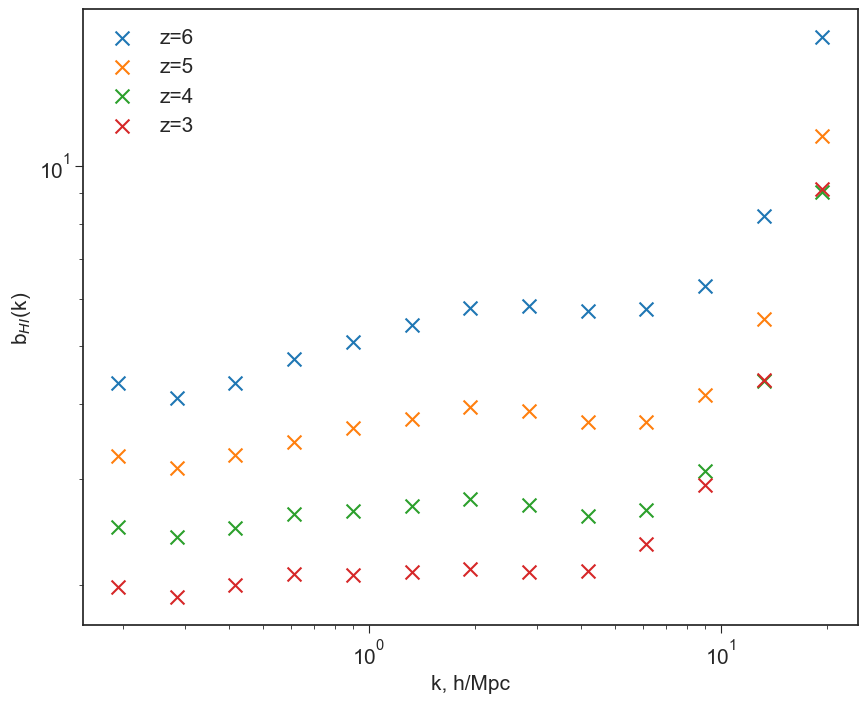}
    \caption{Halo-based HI bias across $3 < z < 6$, produced by the small box run of volume $V =(40 \text{ }h^{-1} \text{Mpc})^3$ using analytic model OM2. The bias across all scales becomes flatter and smaller with decreasing redshift.}
    \label{fig:pad-b-ps}
\end{figure}
In \autoref{fig:pad-b-ps}, we plot the HI bias (derived solely from the HI found in haloes) as a function of scale, from the small-box ($V =(40 \text{ }h^{-1} \text{Mpc})^3$) run. We only evaluate the halo-based HI as the box is too small to contain the large photon mean free paths during the late stages of reionization, and thus does not fully capture the reionization process. At late times $z\approx 4$, the halo-based HI bias remains approximately constant at large scales but becomes increasingly non-linear at smaller scales, agreeing with the theory in \cite{2015aska.confE..19S} and simulation results from \cite{2018ApJ...866..135V}. 

\begin{figure}
    \centering
    \includegraphics[width=\linewidth]{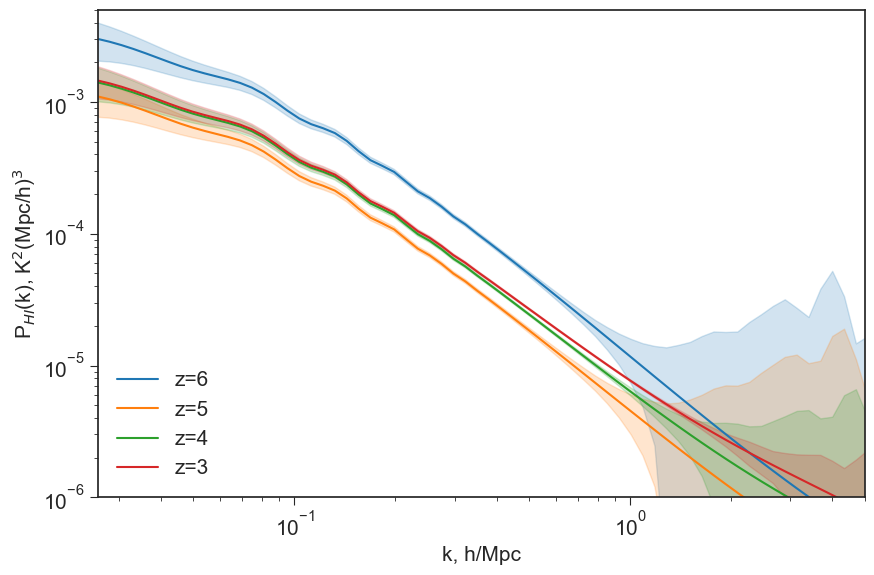}
    \caption{Evolution of the theoretical power spectrum (solid lines) and the anticipated thermal noise and cosmic variance (error bands) across the redshift range observed by the planned SKA-Low deep survey, for analytic model OM2. Thermal noise is calculated assuming a redshift bin width of $\Delta z = 0.2$ and using the AA* stage of SKA-Low, and cosmic variance is calculated assuming Gaussian noise and a wavenumber bin width of $\Delta k = 0.01 \text{ }h\text{Mpc}^{-1}$.}
    \label{fig:theoretical-PS}
\end{figure}

We also calculate the HI SNR, using the thermal noise power $P_{\text{N}}(k)$ for the AA* stage of SKA-Low from \autoref{therm-ps-eq}, assuming a wavenumber binning of $\Delta k = 0.01 \text{ }h\text{Mpc}^{-1}$, and a Gaussian sample variance as in \autoref{zhaoting}, with $N_{\text{modes}}=\frac{V_{\text{surv}}}{(2\pi)^3}2\pi k^2 \Delta k$ \citep{2021MNRAS.502.5259C}. This is shown as error bands in \autoref{fig:theoretical-PS}.

\begin{equation}\label{zhaoting}
    \sigma_{\text{N}} = \frac{1}{\sqrt{N_{\text{modes}}}} \left(P_{\text{HI}}(k)+P_{\text{N}}(k)\right)
\end{equation}

To produce our forecasts across a wider range of scales and remove extraneous features arising from the characteristics of a specific simulation box, we combine the mean brightness temperature of the small box runs ($V_{\text{box}}=(40\text{ }h^{-1}\text{ Mpc})^3$) to account for the late-time HI contribution which presents at smaller scales; and the mean brightness temperature of a large box run ($V_{\text{box}}=(150\text{ }h^{-1} \text{Mpc})^3$) to account for the EoR contribution which presents at larger scales. We then multiply this mean brightness temperature with the linear bias $b_{\text{HI}}$ (redshift-dependent, derived from the bias at small $k$ from \autoref{fig:pad-b-ps}), and finally multiply this by the non-linear \emph{halofit} power spectrum from the package \textsc{CAMB}\footnote{\url{https://github.com/cmbant/camb}} \citep{Lewis_2000}. This aligns with our halo-based model of HI which we have used throughout.

We present forecasts using this description for redshifts $z=[6, 5, 4, 3]$ in \autoref{fig:theoretical-PS}. Within this redshift range is where reionization is expected to completely conclude, and also demonstrates the range of behaviours we expect to see across our entire survey: rapidly decreasing power across all scales with decreasing redshift up until reionization has completely concluded; and steadily increasing power across all scales as redshift continues to decrease beyond this critical time. This also covers the full range of redshifts observed in the planned SKA-Low deep survey.

\autoref{fig:theoretical-PS} shows a survey area of $100$ deg$^2$ and collecting time $5000$ h (matching the planned SKA1-Low deep survey at redshifts $3 \leq z \leq 6$ from \citeauthor{2020PASA...37....7S} \citeyear{2020PASA...37....7S}) for the AA* stage, and with a redshift bin width of $\Delta z = 0.2$, the SNR remains above $1$ for large scales $k\lesssim 1$ $h$ Mpc$^{-1}$, for all redshifts which we are interested in.

When reionization has completely concluded and the neutral IGM has been fully ionized, the HI signal becomes indistinguishable from noise at smaller scales, although this improves as redshift continues to decrease beyond the endpoint of reionization (since more haloes grow in mass to the point that they are able to host HI). From \autoref{fig:theoretical-PS}, at $z=5$, the thermal noise dominates for scales $k\gtrsim 1$ $h$ Mpc$^{-1}$, but this limit increases to $k\gtrsim 4$ $h$ Mpc$^{-1}$ for $z=3$.

It is important to note the limitations of these theoretical forecasts: since they are based upon a non-linear halo power spectrum, they do not capture the shape of the IGM contribution to the power spectrum. We use them here to give a fast conservative limit on the smallest accessible scale as a function of redshift, independent of the initial density realisation of a given simulation box.

\subsection{Observing the HI transition}\label{lightcone-section}
In this section, we investigate the prospects of detecting the transition from the continuously distributed HI throughout the IGM during the EoR to the late-time HI in haloes tracing large-scale structure.

To do this, we use \textsc{21cmFAST} to generate a rectilinear lightcone $4 \leq z \leq 35$, and then apply the post-processing pipeline to the resulting density field for $z\leq7.5$, as described in \autoref{Halo-find}. Due to the additional computational expense of generating a lightcone vs. a static box, and the importance of resolving the ionized bubbles from the EoR during the HI transition, we increase the length of the spatial dimensions to $102$ $h^{-1}$Mpc, and decrease the number of cells along each spatial dimension to 160. To account for the lower resolution of the lightcone, the overdensity cap when finding haloes is reduced to $2$.
Although our algorithm supports the generation of the HI field down to $z=3$, we focus here on the redshift range $4 < z < 7$, as this is where reionization is currently anticipated to reach its end stages (see \autoref{post-eor}).
We multiply the halo-based HI density in the lightcone by a correction factor of $5$, to account for unresolved haloes and thus unresolved HI. The value of this correction factor was derived by comparison of the $\Omega_{\text{HI}}$ parameter between the higher-resolution box and the uncorrected lightcone. We have verified that this correction factor is scale-independent on the relevant scales to our study, i.e. $k<1$ $h^{-1}$Mpc.

We bin the lightcone density field along the line-of-sight after applying our halo finder, to account for the redshift evolution across the field. We verify that using the binned redshift values made no significant difference compared to using the exact redshift values by analysing the radial HI profiles across different redshifts for haloes of a given mass. Even for the lowest resolved halo masses in our simulation ($\approx 10^9\text{ } h^{-1}\text{M}_{\odot}$) which are the most susceptible to redshift evolution, the difference in HI profiles is negligible across the redshift bins used ($\Delta z=0.07$).

\begin{figure*}
    \centering
    \includegraphics[width=\textwidth]{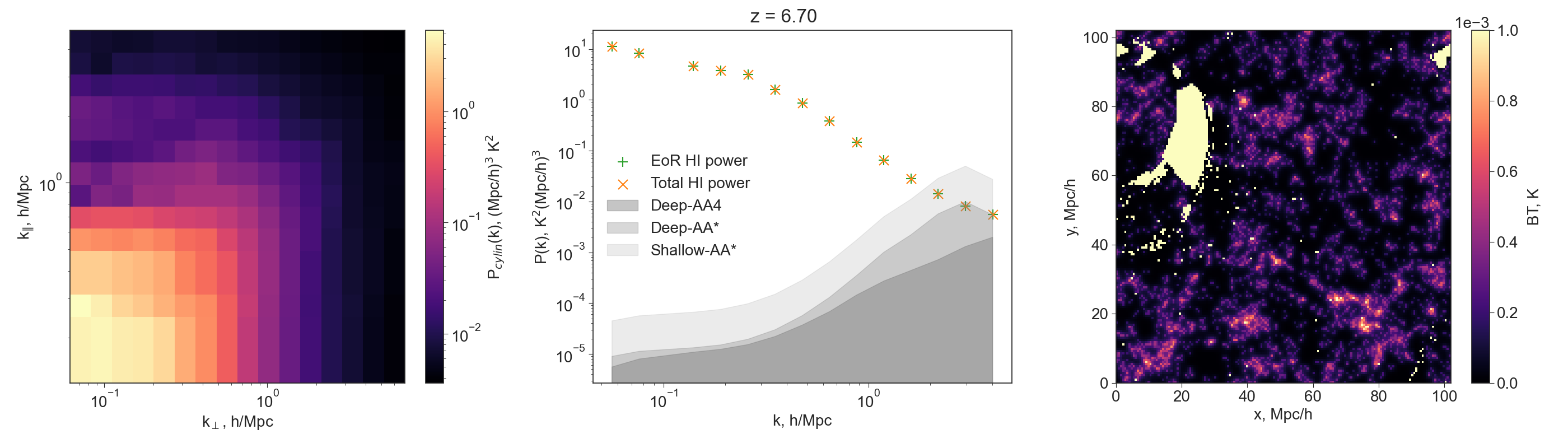}
    \includegraphics[width=\textwidth]{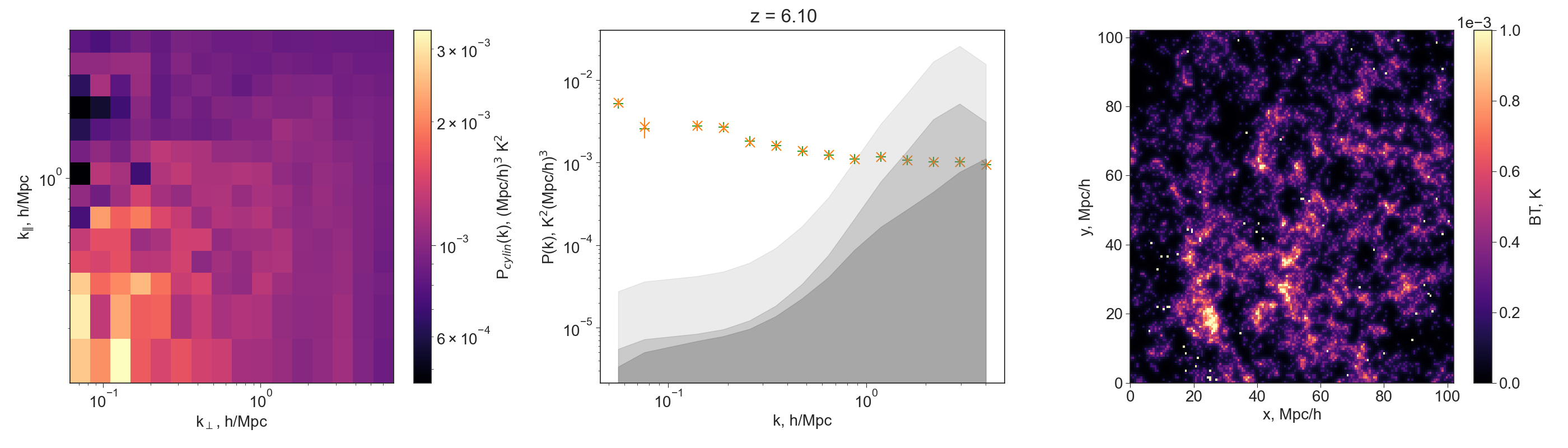}
    \includegraphics[width=\textwidth]{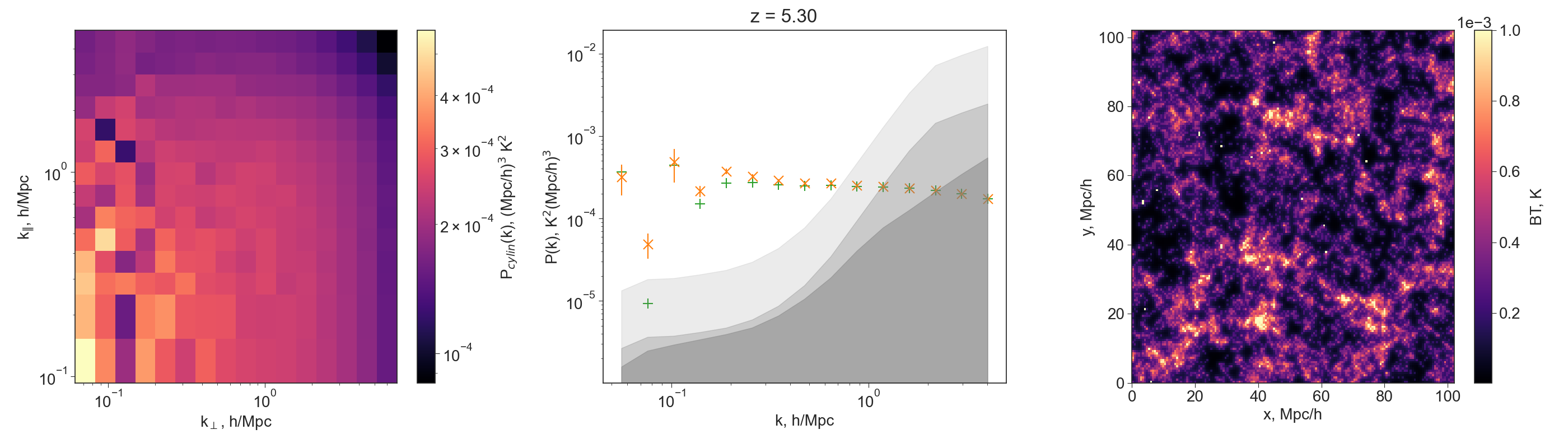}
    \includegraphics[width=\textwidth]{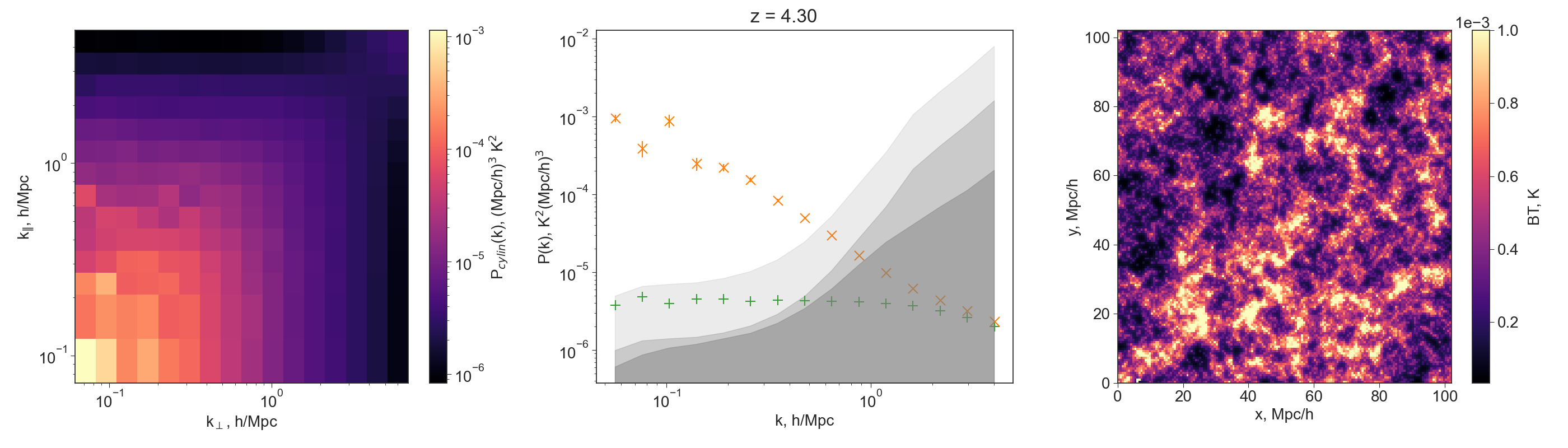}
    \caption{Cylindrically-averaged power spectrum (\textbf{left}), spherically-averaged power spectrum (\textbf{middle}), and brightness temperature field (\textbf{right}) for central redshifts $z=6.7$ (\textbf{top}), $z=6.1$ (\textbf{middle-top}), $z=5.3$ (\textbf{middle-bottom}), and $z=4.3$ (\textbf{bottom}), evaluated on the lightcone split into redshift bins of width $z=0.2$. The error bars show the standard deviation of the power within each wavenumber bin. The shaded regions correspond to the instrumental noise contribution from Deep-AA4 (dark grey), Deep-AA* (middle grey), and Shallow-AA* (light grey). We also plot the EoR-only contribution to the HI power. We see the final stages of reionization at $z=6.7$ with large neutral regions still present, followed by significant fragmented neutral regions during the transition period at $z=6.1$ and $z=5.3$, ending with the final stages of the transition at $z=4.3$ where there are almost no fragments remaining.}
    \label{fig:2d-ps-0.1-1}
\end{figure*}

We estimate the cylindrically-averaged power spectra from the processed lightcones, by binning the 3D power spectrum into line-of-sight ($k_{\parallel}$) and planar ($k_{\perp}$) modes. In \autoref{fig:2d-ps-0.1-1}, we show the cylindrically-averaged power spectra across the point of transition from the EoR to late-time large-scale structure, alongside the corresponding brightness temperature field and spherically-averaged power spectrum for redshift bins of size $\Delta z = 0.2$. The central redshifts are chosen to show the pre- ($z\approx6.7$), mid- ($z\approx6.1$), end- ($z\approx 5.3$), and post-transition ($z\approx 4.3$) stages, and the bin width is chosen as it gives us a good signal-to-noise ratio and is a realistic value for observations.

Qualitatively, we expect a large drop in HI power at all scales after the last regions of neutral IGM are completely reionized. 
When studying the power spectrum evolution from redshift 7 towards 4, we see not only this reduction in power, but also a change in its distribution.
During the HI transition starting around $z\approx 6$, the power flattens across all scales due to small regions of neutral IGM, which generally occupy no more than a single cell, and reside around the edges of the lowest-density regions (middle two panels of \autoref{fig:2d-ps-0.1-1}). By the end of the EoR, starting around $z\approx5$, we find that the power from the HI residing in haloes starts to dominate the largest scales. Nevertheless, a small constant contribution from the neutral fragments remains, even by $z\approx4$ (seen in the non-zero EoR contribution to the power in the bottom panel of \autoref{fig:2d-ps-0.1-1}), albeit few and with little HI left in these regions. 
These neutral IGM regions and their contribution to the evolution of the 21-cm power match the `neutral islands' seen in numerical simulations of the HI transition period \citep{2024MNRAS.533.2364G}.

\begin{figure}
    \centering
    \includegraphics[width=\linewidth]{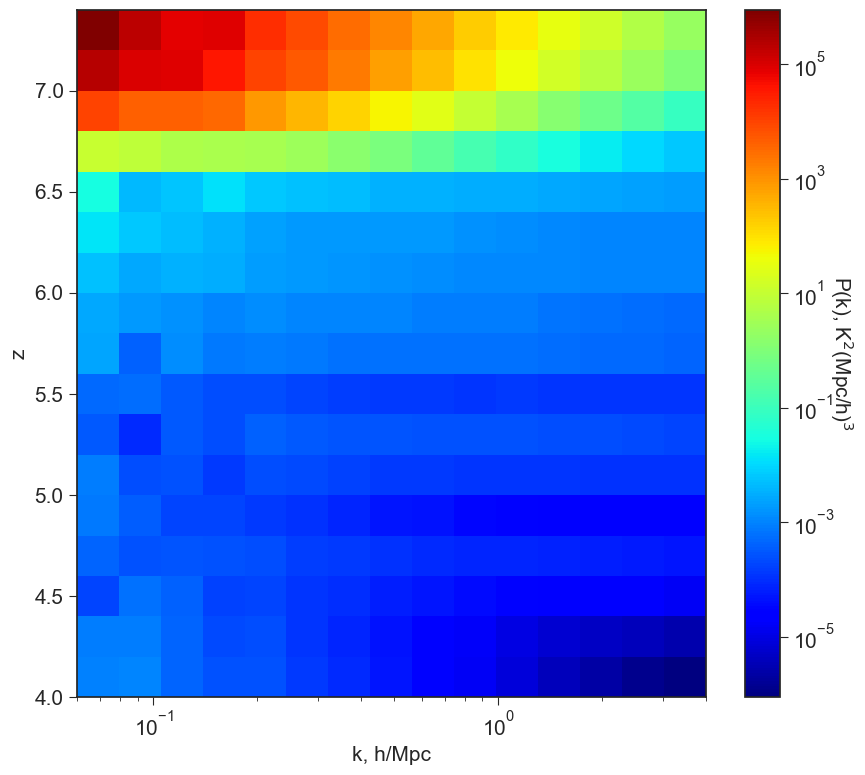}
    \caption{Average power across scales and redshifts within the lightcone output by our simulation, for a redshift bin width of $\Delta z = 0.2$. Power is concentrated at the largest scales pre- and post-transition, but flattens during the transition period.}
    \label{fig:mega-power-plot-0.3}
\end{figure}

In \autoref{fig:mega-power-plot-0.3}, we show a schematic demonstrating the mean power within a given wavenumber and redshift bin.
We can see that power is concentrated at the largest scales pre- and post-reionization end, but flattens during the HI transition period due to the fragmented IGM dominating the power.

\begin{figure}
    \centering
    \includegraphics[width=\linewidth]{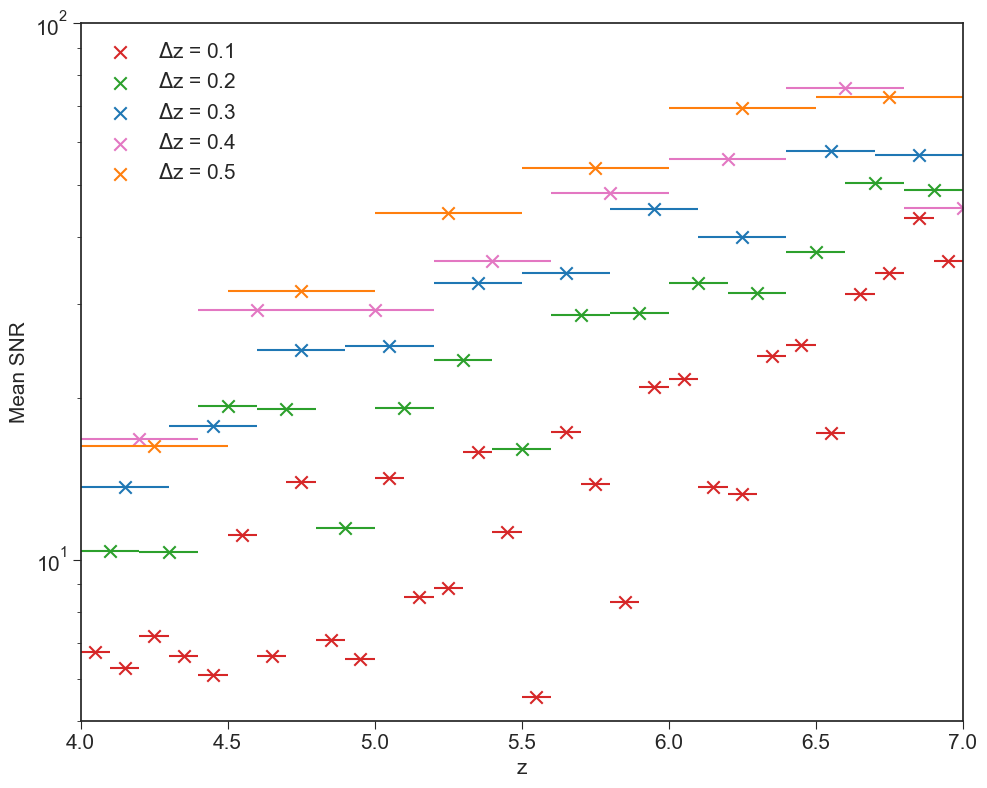}
    \caption{Mean SNR across all scales as a function of redshift from $4 \leq z \leq 7$, using Deep-AA* and assuming a wavenumber binning of $\Delta k = 0.01 \text{ }h\text{Mpc}^{-1}$, shown for five different redshift bin widths in the range $\Delta z = \numrange{0.1}{0.5}$. The mean SNR remains above $10$ at all redshifts for all the bin widths apart from the smallest, $\Delta z = 0.1$. The horizontal error bars represent the uncertainty on the redshift measurement arising from the choice of bin width.}
    \label{fig:mean-snr}
\end{figure}

\autoref{fig:mean-snr} shows that when we increase the width of the redshift bins, we make more line of sight scales accessible and also decrease the thermal noise on the perpendicular power. The mean SNR across all observable scales is above $10$ at all redshifts for all but the $\Delta z = 0.1$ redshift bin.
However, the larger the width, the more the universe (and therefore the cosmology) evolves over the observation period. Additionally, a wider bin width makes it harder to pinpoint the exact redshift we are observing. 

\begin{figure*}
    \centering
    \includegraphics[width=\textwidth]{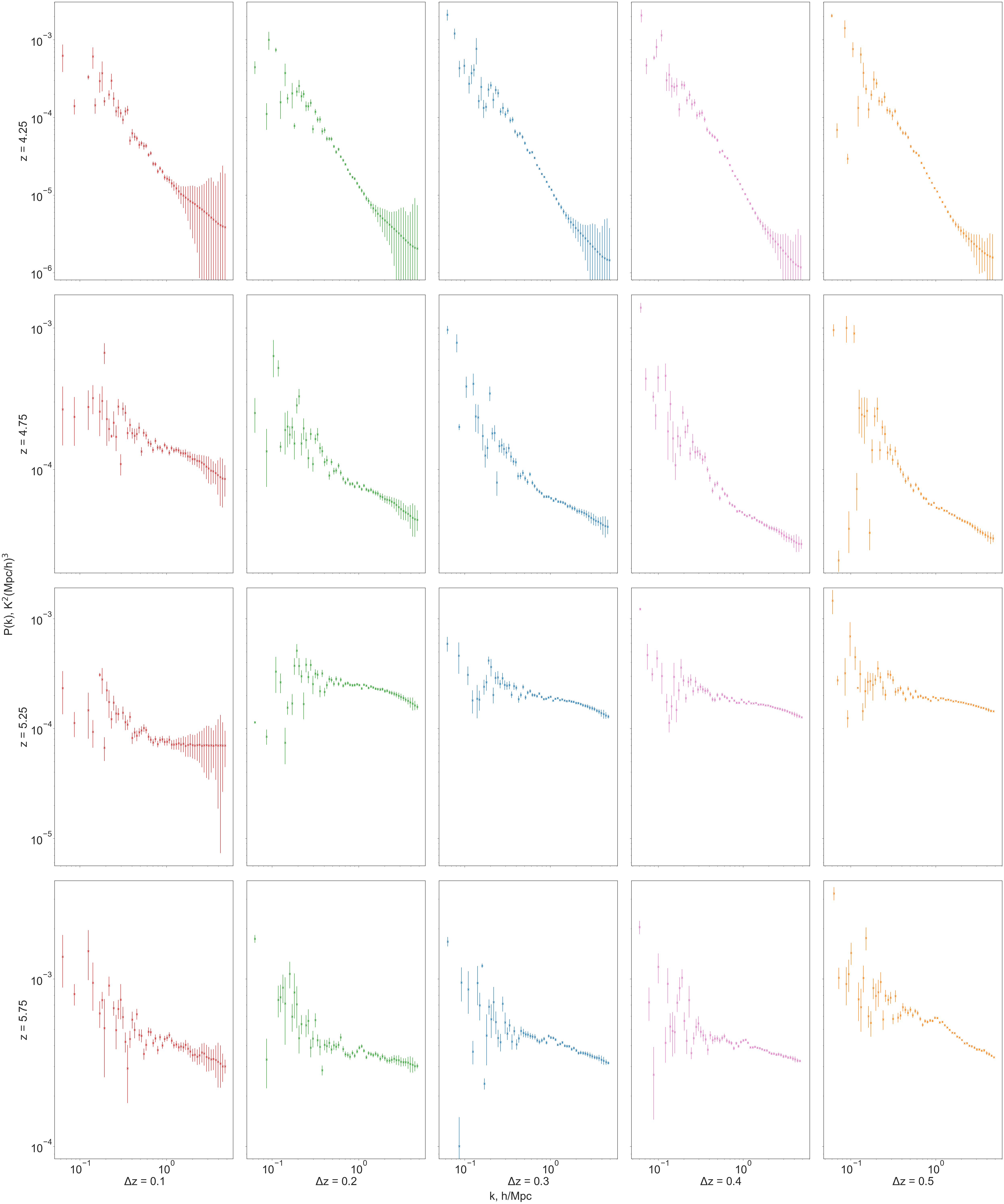}
    \caption{HI power spectra across redshift and varying redshift bins, with the instrumental noise and cosmic variance shown by the error bars, assuming a wavenumber bin width of $\Delta k = 0.01 \text{ }h\text{Mpc}^{-1}$. Instrumental noise is calculated using Deep-AA*. Although larger bin widths display a better SNR across all scales, significant cosmological evolution starts to take place with widening bin size.}
    \label{fig:matrix-ps}
\end{figure*}

Considering \autoref{fig:2d-ps-0.1-1}, \autoref{fig:mean-snr}, and \autoref{fig:matrix-ps}, our results suggest that the optimum redshift bin width for observing the HI transition is $\Delta z \approx 0.2$, for an SKA-Low deep survey. This is the smallest bin width where the detected signal will still be statistically significant (assuming perfect foreground removal) up to scales of $\approx 0.5$ $h^{-1}$Mpc, even for a shallow survey. It also does not have the significant cosmological evolution in a single bin seen with the larger bin widths considered here. 

\subsection{Impact of astrophysics}\label{impact}
Finally, we investigate if and how the astrophysics of reionization can affect the transition HI signal. To this end, we vary the less-constrained parameters in \textsc{21cmFAST}, which we list and provide fiducial values for in \autoref{21cmFAST-table}.

When varying the X-ray parameters NU\_X\_THRESH and X\_RAY\_SPEC\_INDEX, we saw no change to the HI field or power across $4\leq z\leq 7$, even at the upper and lower allowed limits of these parameters. This is as expected: X-ray heating impacts the kinetic (and thus spin) temperature of the HI gas, but even at the start of our redshift range of interest, the spin temperature has long since saturated. As a result, changes to X-ray heating will not impact the late-time HI signal, and so we use the fiducial values given in \autoref{21cmFAST-table} for NU\_X\_THRESH and X\_RAY\_SPEC\_INDEX from here on out.

We focus here on delayed reionization (i.e. decreased star formation rate and decreased escape fraction), as a late reionization scenario is currently favoured by observational data. 
For the decreased escape fraction runs, ALPHA\_ESC is set to $-0.5$, and F\_ESC10 is set to $-1.046$.
For the decreased star formation runs, t\_STAR is set to $0.5$, and F\_STAR10 is set to $-1.35$.

\begin{figure*}
    \centering
    \includegraphics[width=\textwidth]{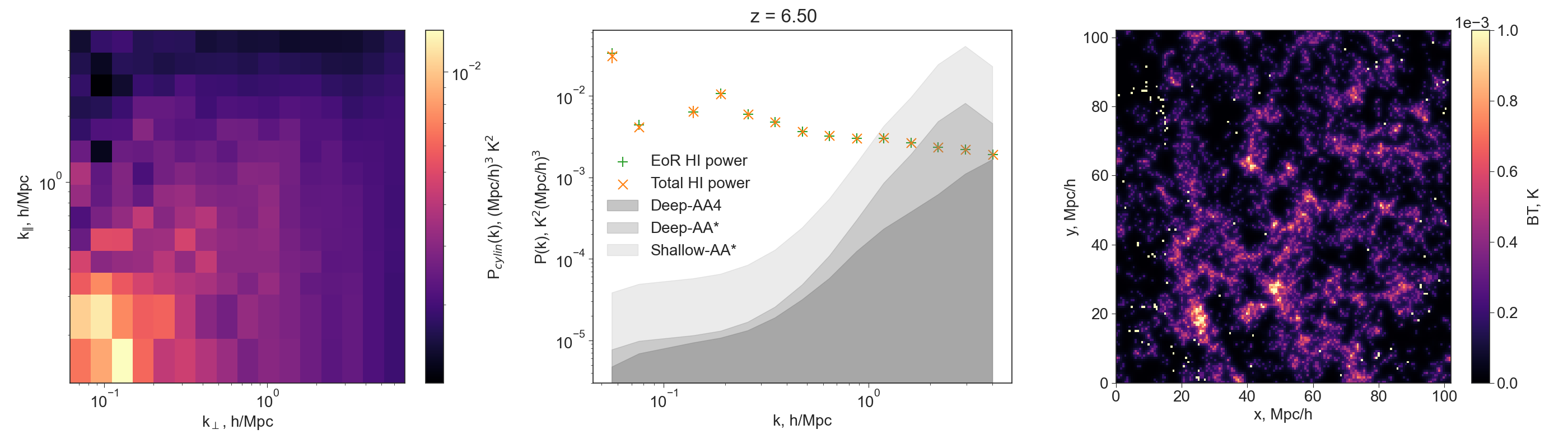}
    \includegraphics[width=\textwidth]{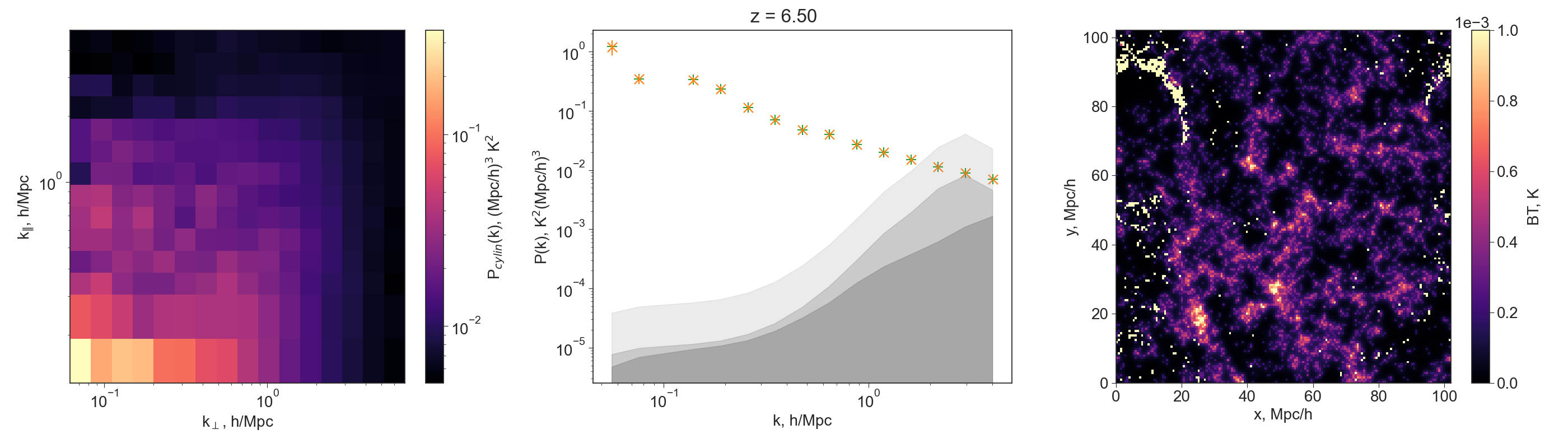}
    \includegraphics[width=\textwidth]{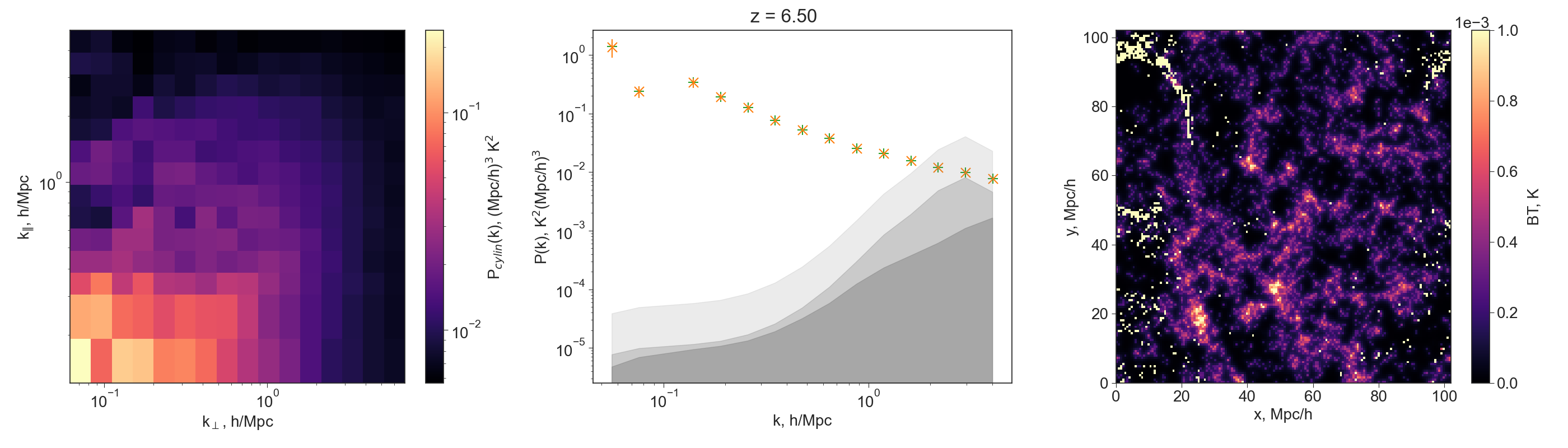}
    \caption{Cylindrically-averaged power spectrum (\textbf{left}), spherically-averaged power spectrum (\textbf{middle}), and brightness temperature field (\textbf{right}) for the fiducial run (\textbf{top}), decreased escape fraction (\textbf{middle}), and decreased star formation rate (\textbf{bottom}). The central redshift is $z=6.5$, and the statistics were evaluated on the lightcone split into redshift bins of width $\Delta z=0.2$. The error bars show the standard deviation of the power within each wavenumber bin. The grey regions represent instrumental noise. The reduced rate of reionization can be seen in the bottom two plots to change how the reionization process progresses, becoming more patchy and increasing the power across all scales.}
    \label{fig:astro-more}
\end{figure*}

\begin{figure*}
    \centering
    \includegraphics[width=\textwidth]{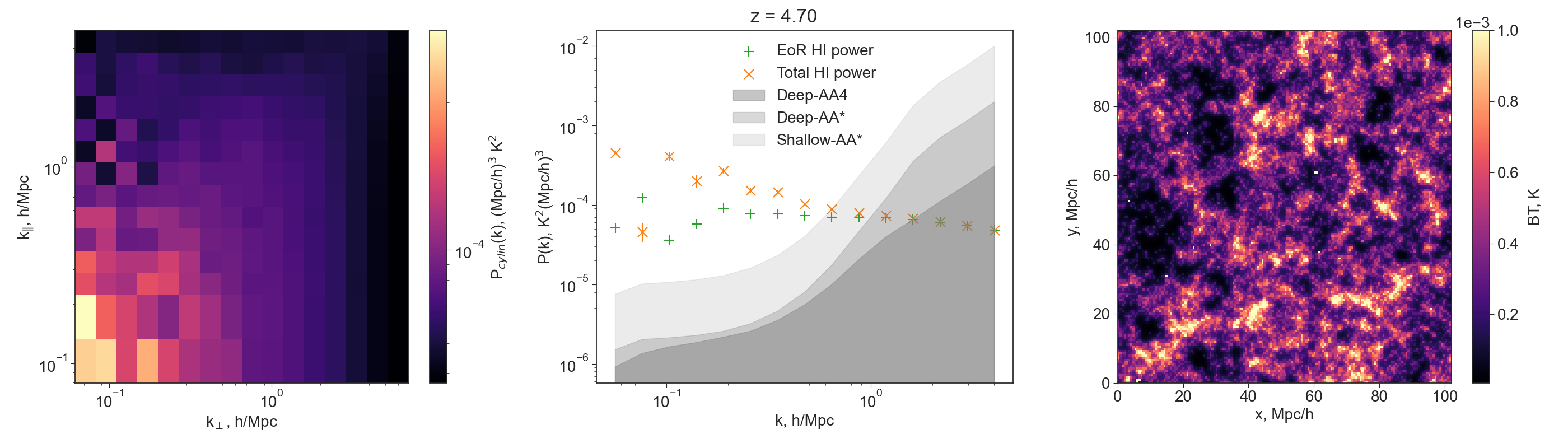}
    \includegraphics[width=\textwidth]{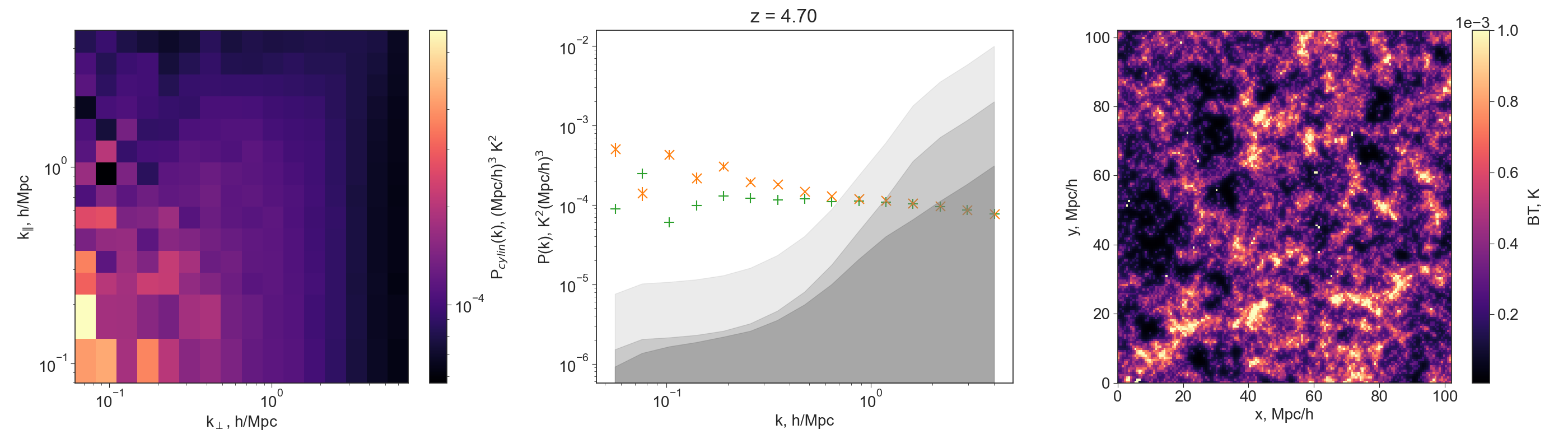}
    \includegraphics[width=\textwidth]{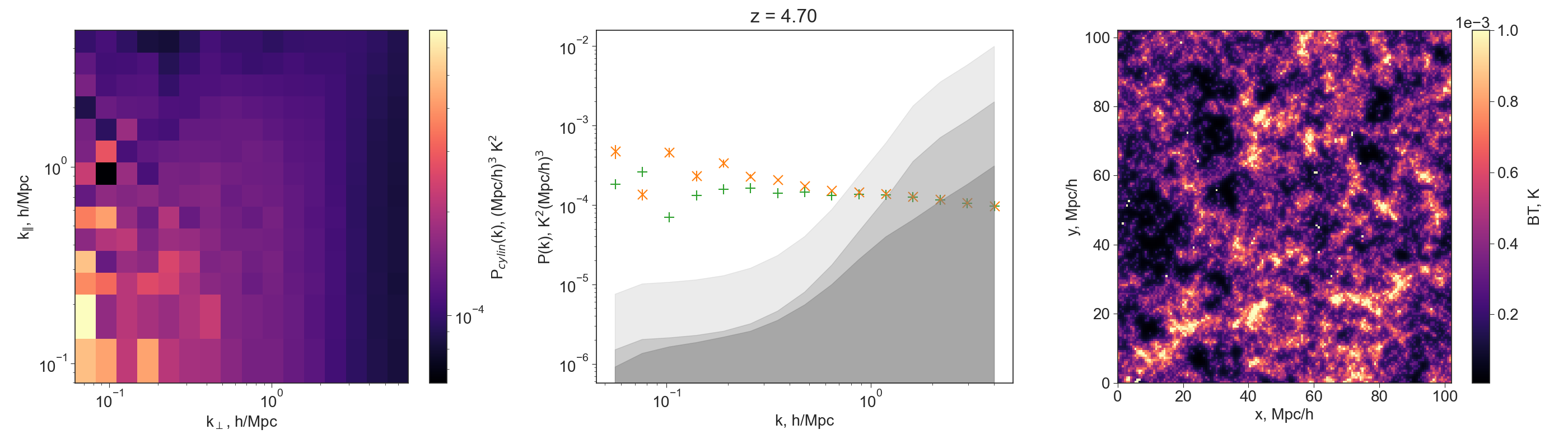}
    \caption{Cylindrically-averaged power spectrum (\textbf{left}), spherically-averaged power spectrum (\textbf{middle}), and brightness temperature field (\textbf{right}) for the fiducial run (\textbf{top}), decreased escape fraction (\textbf{middle}), and decreased star formation rate (\textbf{bottom}). The central redshift is $z=4.7$, and the statistics were evaluated on the lightcone split into redshift bins of width $\Delta z=0.2$. The error bars show the standard deviation of the power within each wavenumber bin. The grey regions represent instrumental noise.}
    \label{fig:astro-less}
\end{figure*}

This is demonstrated in \autoref{fig:astro-more} and \autoref{fig:astro-less}, which show the effects of reduced ionizing photon production in the middle of and after reionization respectively.
From \autoref{fig:astro-more} and \autoref{fig:astro-less}, reducing ionizing photon production not only delays ionization (as anticipated), but results in a far more patchy and fragmented reionization scenario. The drop in power post-EoR occurs much later and more gradually, and the evolution of the shape of the spherically-averaged power spectrum displays an additional stage wherein power at intermediate scales drops (relative to the fiducial's shape at the same degree of reionization). In terms of the distribution of power along the line-of-sight vs. planar directions, we see a flatter power spectrum along the line-of-sight, but with the power concentrated at the largest planar scales, for the reduced ionization runs. 

However, there appears to be a degeneracy in the effects of ionizing photon escape fraction and star formation rate. The change that these parameters both effect on the HI power spectrum is very similar, and therefore separate constraints on their values will be difficult to extract from HI power measurements. Nevertheless, we may still be able to place constraints on their product.

\begin{figure}
    \centering
    \includegraphics[width=\linewidth]{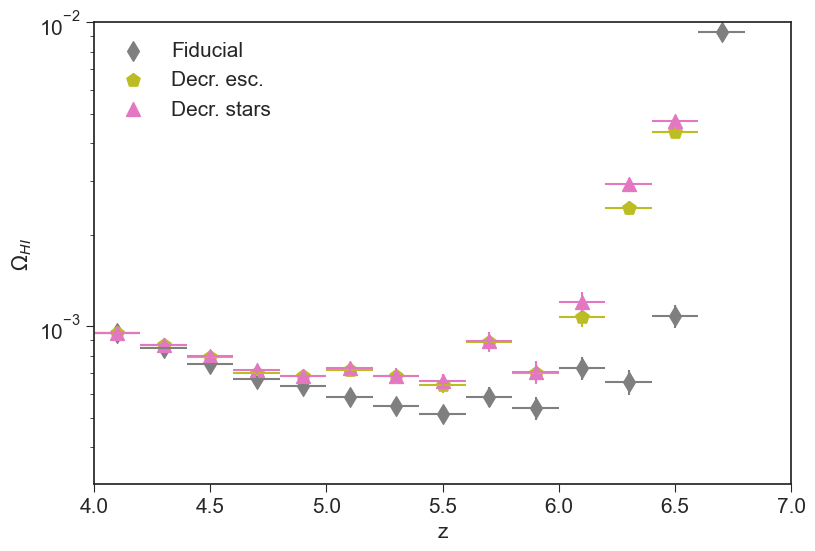}
    \caption{Evolution of the $\Omega_{\text{HI}}$ parameter with redshift for three different simulation parameter sets: fiducial astrophysics, a decreased escape fraction, and a decreased star formation rate.}
    \label{fig:omega_HI_astro}
\end{figure}

Finally, from \autoref{fig:omega_HI_astro}, we find that the choice of reionization astrophysics significantly affects the $\Omega_{\text{HI}}$ obtained during the HI transition period, changing not just the timing at which it decreases but also the rate. The minimum at $z\approx 5.5$ in the fiducial run is still visible in the reduced ionization runs, but is delayed to $z\approx 5.0$ and is less pronounced, as the IGM HI contribution for the decreased parameter sets is comparable to the halo-based HI contribution for longer.

\subsection{Foreground avoidance}
In this section, we explore the impact of foreground removal on the transition detectability.

In order to mitigate the influence of astrophysical foregrounds on our observed spherically-averaged power spectrum, two approaches are predominantly taken: \emph{subtraction}, where the foregrounds are modelled and then subtracted from the data (e.g \citealt{2013ApJ...763L..20M, 2013PASA...30...31B, 2014MNRAS.441.3271W}); and \emph{avoidance}, where the spherically-averaged power spectrum is determined by averaging only from scales that are least affected by foreground contamination (e.g. \citealt{2014PhRvD..90b3018L, 2020PASP..132f2001L, 2025A&A...693A.276M}). 
Here, we focus on foreground avoidance.

We apply two different foreground avoidance regimes. The first (hereafter referred to as R1) removes only those modes beneath the horizon limit \citep{2014PhRvD..90b3018L}:
\begin{equation}
    k_{\parallel}=\frac{H(z)D_{\text{c}}(z)\theta_{\text{B}}}{c(1+z)}k_{\perp}
\end{equation}
where $D_{\text{c}}(z)$ is the comoving cosmological distance to redshift $z$, and $\theta_{\text{B}}$ is the primary beam size in radians.

\begin{figure}
    \centering
    \includegraphics[width=\linewidth]{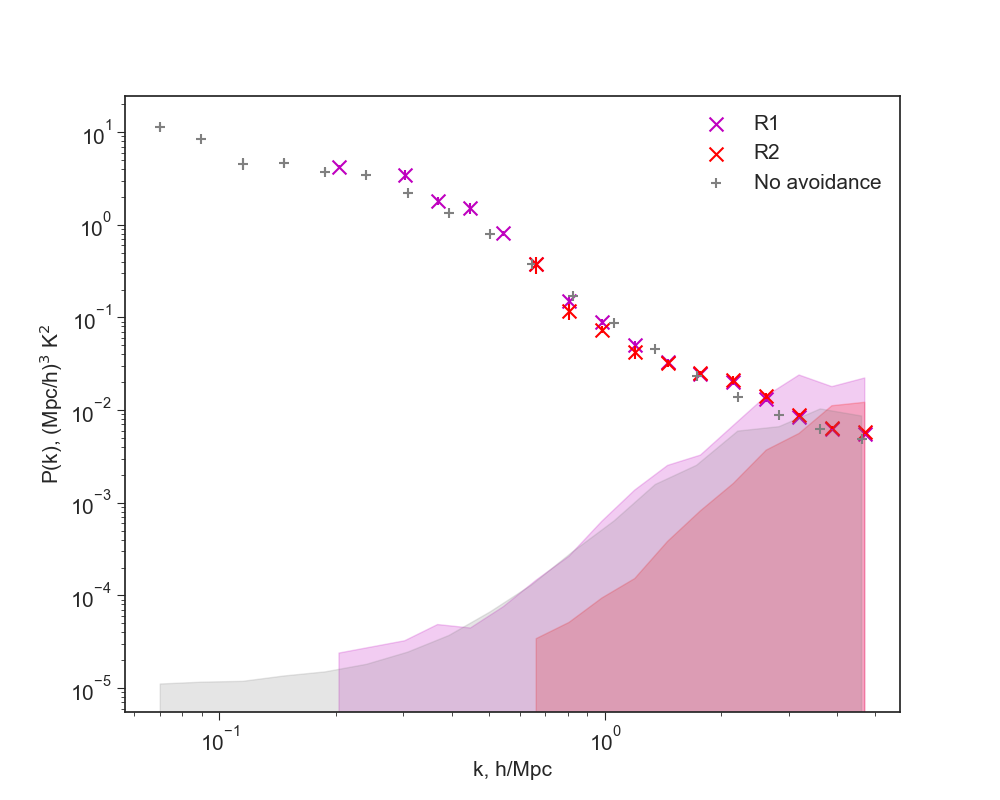}
    \includegraphics[width=\linewidth]{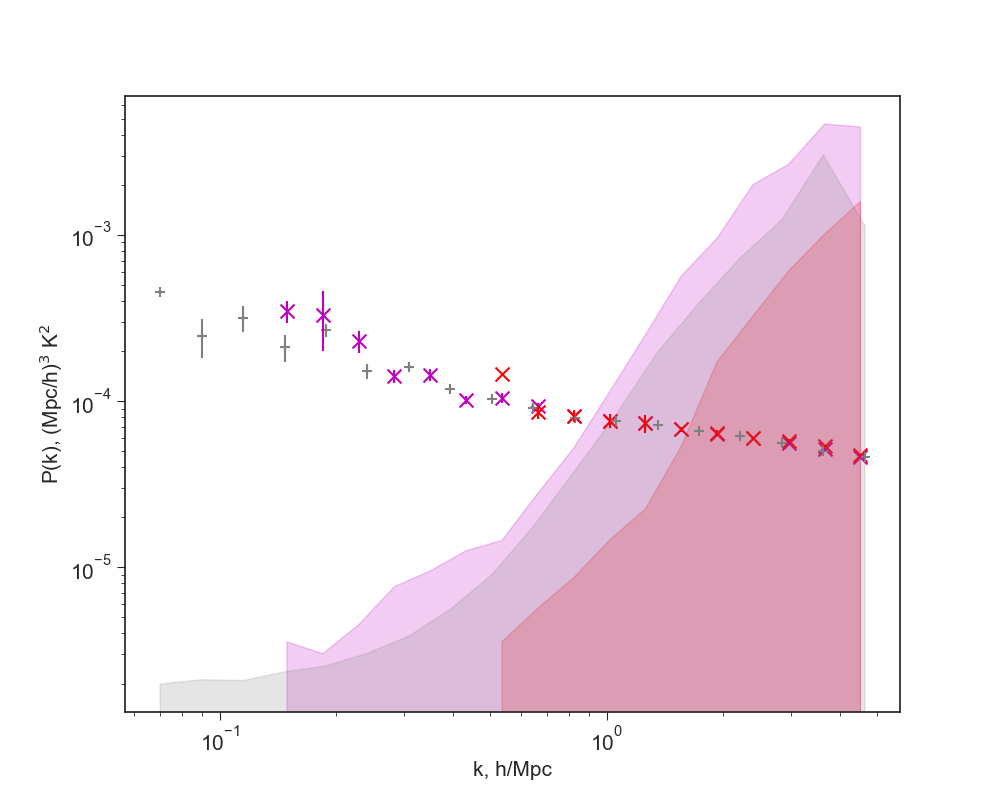}
    \caption{Comparison of the spherically-averaged power spectrum obtained for different foreground avoidance limits, in addition to no avoidance regime. These are shown at central redshifts $z=6.7$ (\textbf{top}) and $z=4.7$ (\textbf{bottom}), for the fiducial lightcone with redshift bin widths of $\Delta z = 0.2$. The shaded regions represent the instrumental noise in the two avoidance regimes, for survey Deep-AA*.}
    \label{fig:fg-removal}
\end{figure}

The other regime is a conservative limit, following \cite{2025MNRAS.541..476M}: $k_{\parallel}=0.5k_{\perp}+0.5$ (hereafter referred to as R2) which mitigates for additional contaminations. We show the results of these avoidance regimes to the spherically-averaged power spectrum of the fiducial lightcone in \autoref{fig:fg-removal}.
Not only does foreground avoidance limit the accessible range of scales, but impacts the shape of the power spectrum we recover across the HI transition period. The smallest scales are most affected, with the foreground-removed power spectra overestimating small-scale power early in the HI transition period, and underestimating it towards the end. However, with the current generation of radio telescopes, we are unlikely to be able to make a significant measurement of the power on such small scales in any case. 

Applying conservative foreground avoidance when averaging to obtain the spherically-averaged power discards the vast majority of the contaminated $(k_{\perp}, k_{\parallel})$ modes. However, far fewer scales are accessible in such a conservative regime, particularly at larger scales where cosmological information lies, so this is unlikely to be an optimal approach. 

For a horizon-limit foreground avoidance scheme, the characteristic flattening of the power spectrum seen during the HI transition period is still detectable, suggesting that for a realistic observation with SKA-Low at the AA* stage, we can expect to observe this transition signature.

\section{Conclusion}
We present semi-numerical simulations of the cosmological HI signal in the redshift range $3 \leq z \leq 7$, wherein the transition in the distribution of neutral hydrogen (from being continuous across the IGM, to being localised in dark matter haloes) takes place.

Our post-processing pipeline uses a watershed-based image segmentation algorithm to find haloes on a gridded density field produced by 21CMFAST, and a HI-halo relation to assign a total HI mass for each halo. HI density profiles are built according to a modified NFW profile, and convolved with the haloes falling into the appropriate mass bin. Finally, the HI mass field is converted to a brightness temperature field. Although the halo model of HI neglects variations in HI mass due to local halo environments, these variations will likely average out on the scales we observes during intensity mapping, and so we find the halo model to be suitable for this use case of modelling the anticipated signal as detected by an intensity mapping survey.

We present the spherically-averaged power spectra of the matter, halo, and HI brightness temperature field, alongside the HI bias and the expected thermal noise contribution for a given survey; as well as functions to calculate and plot other key objects such as the cosmological HI density $\Omega_{\text{HI}}$, the HMF, and the HI cylindrically-averaged power spectrum.

We find that our halo finder predicts halo abundance and position to a reasonable degree of accuracy that is suitable for its intended purpose of determining the HI mass distribution on cosmological scales. It is able to reproduce the halo distribution found by a FoF halo finder and the theoretical halo mass function across redshift.

The $\Omega_{\text{HI}}$ evolution of our simulation across redshift agrees with current observational constraints from DLA column density and C[II] measurements post-EoR.

In order to best observe the HI transition signal in an SKA-Low deep survey, we simulate the projected thermal noise for different stages of SKA-Low, observing times, and redshift bin widths for power spectrum measurment.
Assuming our fiducial model of astrophysics and cosmology, our results suggest that the optimal redshift bin width to use is $\Delta z=0.2$. This width strikes a balance between averaging down the thermal noise contribution from the instrument and maintaining a appropriate level of redshift precision for the data.

During the HI transition period, we find a flattening in the 21-cm power spectrum arising from the presence of small neutral IGM regions that persist to lower redshifts through the process of recombination. By $z\approx 5$, the contribution to the 21-cm power from halo-based HI begins to dominate over the IGM contribution, and the 21-cm power spectrum returns to tracing the matter power spectrum at large scales.

The SNR from our forecasts indicates that for the SKA-Low deep survey of collecting time $= 5000$ h and survey area $= 100$ deg$^2$ across $3 \leq z \leq 6$ using the AA* stage of SKA-Low, statistically significant detection of the HI signal should be possible at all redshifts, for large scales $k< 1$ $h$Mpc$^{-1}$ and frequency bin width $\Delta \nu = 10$ MHz. At smaller scales, noise becomes dominant once reionization has entirely concluded (i.e. the entire IGM is reionized, with no neutral bubbles remaining), but the exact wavenumber at which this occurs increases with decreasing redshift. For example, noise becomes dominant at $k\approx 1$ $h$ Mpc$^{-1}$ for $z=5$, but at $k\approx3$ $h$ Mpc$^{-1}$ for $z=3$.

We find that the HI transition signal is completely unaffected by the characteristics of the X-ray background and emission, as expected. However, both the timing and statistical characteristics of the transition are sensitive to the escape fraction of ionizing photons, and the star formation rate in galaxies. A lower value of these parameters leads to not only a longer HI transition period, but also a change in the characteristics of this minimum: for the evolution of $\Omega_{\text{HI}}(z)$ in particular, a delayed reionization beyond the fiducial estimate both delays and reduces the minimum at $z\approx 5.5$ seen in the fiducial model of reionization.

We expect the HI transition signal to still be detectable upon application of a horizon-limit foreground avoidance regime for calculating the spherically-averaged power spectra, although the largest scales become inaccessible.

In future work, we plan to investigate the effect of redshift-space distortions within the pipeline. This is important to ensure our simulation properly represents what will be observed by upcoming surveys, since they observe in redshift-space rather than real-space, since the distance is inferred from the observed frequency of the photons. We also wish to investigate how the cosmology affects the observed HI transition signal, and whether we may obtain meaningful constraints on the cosmology by observing this transition. 

\section*{Acknowledgements}
We would like to thank Andrei Mesinger and Jordan Flitter for useful comments.

JI is funded by the UK Research and Innovation Science and Technology Facilities Council PhD studentship. LW is a UK Research and Innovation Future Leaders Fellow [grant MR/V026437/1].

We would like to thank the authors of the Python libraries \textsc{NumPy} \citep{2020Natur.585..357H}, \textsc{SciPy} \citep{2020NatMe..17..261V}, \textsc{Matplotlib} \citep{2007CSE.....9...90H}, \textsc{scikit-image} \citep{2014PeerJ...2..453V}, and \textsc{Astropy} \citep{2013A&A...558A..33A}.

\section*{Data Availability}

The code is publicly available at \url{https://github.com/raxquax123/postEoR}. Issues found within the pipeline may be raised within this repository, or by contacting JI directly.



\bibliographystyle{mnras}
\bibliography{bib.bib} 

@ARTICLE{2011MNRAS.411..955M,
       author = {{Mesinger}, Andrei and {Furlanetto}, Steven and {Cen}, Renyue},
        title = "{21CMFAST: a fast, seminumerical simulation of the high-redshift 21-cm signal}",
      journal = {MNRAS},
         year = 2011,
        month = feb,
       volume = {411},
       number = {2},
        pages = {955-972},
          doi = {10.1111/j.1365-2966.2010.17731.x},
archivePrefix = {arXiv},
       eprint = {1003.3878},
 primaryClass = {astro-ph.CO},
       adsurl = {https://ui.adsabs.harvard.edu/abs/2011MNRAS.411..955M}
}

@ARTICLE{2014MNRAS.441.3271W,
       author = {{Wolz}, L. and {Abdalla}, F.~B. and {Blake}, C. and {Shaw}, J.~R. and {Chapman}, E. and {Rawlings}, S.},
        title = "{The effect of foreground subtraction on cosmological measurements from intensity mapping}",
      journal = {\mnras},
         year = 2014,
        month = jul,
       volume = {441},
       number = {4},
        pages = {3271-3283},
          doi = {10.1093/mnras/stu792},
archivePrefix = {arXiv},
       eprint = {1310.8144},
 primaryClass = {astro-ph.CO},
       adsurl = {https://ui.adsabs.harvard.edu/abs/2014MNRAS.441.3271W}
}

@ARTICLE{2000ApJ...530....1M,
       author = {{Miralda-Escud{\'e}}, Jordi and {Haehnelt}, Martin and {Rees}, Martin J.},
        title = "{Reionization of the Inhomogeneous Universe}",
      journal = {ApJ},
         year = 2000,
        month = feb,
       volume = {530},
       number = {1},
        pages = {1-16},
          doi = {10.1086/308330},
archivePrefix = {arXiv},
       eprint = {astro-ph/9812306},
 primaryClass = {astro-ph},
       adsurl = {https://ui.adsabs.harvard.edu/abs/2000ApJ...530....1M}
}

@ARTICLE{2012RPPh...75h6901P,
       author = {{Pritchard}, Jonathan R. and {Loeb}, Abraham},
        title = "{21 cm cosmology in the 21st century}",
      journal = {Rep. Prog. Phys.},
         year = 2012,
        month = aug,
       volume = {75},
       number = {8},
          eid = {086901},
        pages = {086901},
          doi = {10.1088/0034-4885/75/8/086901},
archivePrefix = {arXiv},
       eprint = {1109.6012},
 primaryClass = {astro-ph.CO},
       adsurl = {https://ui.adsabs.harvard.edu/abs/2012RPPh...75h6901P}
}

@ARTICLE{1991ApJ...379..440B,
       author = {{Bond}, J.~R. and {Cole}, S. and {Efstathiou}, G. and {Kaiser}, N.},
        title = "{Excursion Set Mass Functions for Hierarchical Gaussian Fluctuations}",
      journal = {ApJ},
         year = 1991,
        month = oct,
       volume = {379},
        pages = {440},
          doi = {10.1086/170520},
       adsurl = {https://ui.adsabs.harvard.edu/abs/1991ApJ...379..440B}
}

@ARTICLE{2004ApJ...613....1F,
       author = {{Furlanetto}, Steven R. and {Zaldarriaga}, Matias and {Hernquist}, Lars},
        title = "{The Growth of H II Regions During Reionization}",
      journal = {ApJ},
         year = 2004,
        month = sep,
       volume = {613},
       number = {1},
        pages = {1-15},
          doi = {10.1086/423025},
archivePrefix = {arXiv},
       eprint = {astro-ph/0403697},
 primaryClass = {astro-ph},
       adsurl = {https://ui.adsabs.harvard.edu/abs/2004ApJ...613....1F}
}

@ARTICLE{2004MNRAS.350.1195M,
       author = {{Meyer}, M.~J. and {Zwaan}, M.~A. and {Webster}, R.~L. and {Staveley-Smith}, L. and {Ryan-Weber}, E. and {Drinkwater}, M.~J. and {Barnes}, D.~G. and {Howlett}, M. and {Kilborn}, V.~A. and {Stevens}, J. and {Waugh}, M. and {Pierce}, M.~J. and {Bhathal}, R. and {de Blok}, W.~J.~G. and {Disney}, M.~J. and {Ekers}, R.~D. and {Freeman}, K.~C. and {Garcia}, D.~A. and {Gibson}, B.~K. and {Harnett}, J. and {Henning}, P.~A. and {Jerjen}, H. and {Kesteven}, M.~J. and {Knezek}, P.~M. and {Koribalski}, B.~S. and {Mader}, S. and {Marquarding}, M. and {Minchin}, R.~F. and {O'Brien}, J. and {Oosterloo}, T. and {Price}, R.~M. and {Putman}, M.~E. and {Ryder}, S.~D. and {Sadler}, E.~M. and {Stewart}, I.~M. and {Stootman}, F. and {Wright}, A.~E.},
        title = "{The HIPASS catalogue - I. Data presentation}",
      journal = {MNRAS},
         year = 2004,
        month = jun,
       volume = {350},
       number = {4},
        pages = {1195-1209},
          doi = {10.1111/j.1365-2966.2004.07710.x},
archivePrefix = {arXiv},
       eprint = {astro-ph/0406384},
 primaryClass = {astro-ph},
       adsurl = {https://ui.adsabs.harvard.edu/abs/2004MNRAS.350.1195M}
}

@ARTICLE{2018ApJ...866..135V,
       author = {{Villaescusa-Navarro}, Francisco and {Genel}, Shy and {Castorina}, Emanuele and {Obuljen}, Andrej and {Spergel}, David N. and {Hernquist}, Lars and {Nelson}, Dylan and {Carucci}, Isabella P. and {Pillepich}, Annalisa and {Marinacci}, Federico and {Diemer}, Benedikt and {Vogelsberger}, Mark and {Weinberger}, Rainer and {Pakmor}, R{\"u}diger},
        title = "{Ingredients for 21 cm Intensity Mapping}",
      journal = {ApJ},
         year = 2018,
        month = oct,
       volume = {866},
       number = {2},
          eid = {135},
        pages = {135},
          doi = {10.3847/1538-4357/aadba0},
archivePrefix = {arXiv},
       eprint = {1804.09180},
 primaryClass = {astro-ph.CO},
       adsurl = {https://ui.adsabs.harvard.edu/abs/2018ApJ...866..135V}
}

@ARTICLE{2020MNRAS.493.5434S,
       author = {{Spinelli}, Marta and {Zoldan}, Anna and {De Lucia}, Gabriella and {Xie}, Lizhi and {Viel}, Matteo},
        title = "{The atomic hydrogen content of the post-reionization Universe}",
      journal = {MNRAS},
         year = 2020,
        month = apr,
       volume = {493},
       number = {4},
        pages = {5434-5455},
          doi = {10.1093/mnras/staa604},
archivePrefix = {arXiv},
       eprint = {1909.02242},
 primaryClass = {astro-ph.CO},
       adsurl = {https://ui.adsabs.harvard.edu/abs/2020MNRAS.493.5434S}
}

@INPROCEEDINGS{2015aska.confE..19S,
       author = {{Santos}, M. and {Bull}, P. and {Alonso}, D. and {Camera}, S. and {Ferreira}, P. and {Bernardi}, G. and {Maartens}, R. and {Viel}, M. and {Villaescusa-Navarro}, F. and {Abdalla}, F.~B. and {Jarvis}, M. and {Metcalf}, R.~B. and {Pourtsidou}, A. and {Wolz}, L.},
        title = "{Cosmology from a SKA HI intensity mapping survey}",
    booktitle = {Advancing Astrophysics with the Square Kilometre Array (AASKA14)},
         year = 2015,
        month = apr,
          eid = {19},
        pages = {19},
          doi = {10.22323/1.215.0019},
archivePrefix = {arXiv},
       eprint = {1501.03989},
 primaryClass = {astro-ph.CO},
       adsurl = {https://ui.adsabs.harvard.edu/abs/2015aska.confE..19S}
}

@ARTICLE{2004MNRAS.355.1339B,
       author = {{Battye}, Richard A. and {Davies}, Rod D. and {Weller}, Jochen},
        title = "{Neutral hydrogen surveys for high-redshift galaxy clusters and protoclusters}",
      journal = {MNRAS},
         year = 2004,
        month = dec,
       volume = {355},
       number = {4},
        pages = {1339-1347},
          doi = {10.1111/j.1365-2966.2004.08416.x},
archivePrefix = {arXiv},
       eprint = {astro-ph/0401340},
 primaryClass = {astro-ph},
       adsurl = {https://ui.adsabs.harvard.edu/abs/2004MNRAS.355.1339B}
}

@ARTICLE{2009MNRAS.397.1926W,
       author = {{Wyithe}, J. Stuart B. and {Loeb}, Abraham},
        title = "{The 21-cm power spectrum after reionization}",
      journal = {MNRAS},
         year = 2009,
        month = aug,
       volume = {397},
       number = {4},
        pages = {1926-1934},
          doi = {10.1111/j.1365-2966.2009.15019.x},
archivePrefix = {arXiv},
       eprint = {0808.2323},
 primaryClass = {astro-ph},
       adsurl = {https://ui.adsabs.harvard.edu/abs/2009MNRAS.397.1926W}
}

@ARTICLE{2017MNRAS.464.4008P,
       author = {{Padmanabhan}, Hamsa and {Refregier}, Alexandre},
        title = "{Constraining a halo model for cosmological neutral hydrogen}",
      journal = {MNRAS},
         year = 2017,
        month = feb,
       volume = {464},
       number = {4},
        pages = {4008-4017},
          doi = {10.1093/mnras/stw2706},
archivePrefix = {arXiv},
       eprint = {1607.01021},
 primaryClass = {astro-ph.CO},
       adsurl = {https://ui.adsabs.harvard.edu/abs/2017MNRAS.464.4008P}
}

@ARTICLE{2017MNRAS.465..111K,
       author = {{Kim}, Han-Seek and {Wyithe}, J. Stuart. B. and {Baugh}, C.~M. and {Lagos}, C. d. P. and {Power}, C. and {Park}, Jaehong},
        title = "{The spatial distribution of neutral hydrogen as traced by low H I mass galaxies}",
      journal = {MNRAS},
         year = 2017,
        month = feb,
       volume = {465},
       number = {1},
        pages = {111-122},
          doi = {10.1093/mnras/stw2779},
archivePrefix = {arXiv},
       eprint = {1603.02383},
 primaryClass = {astro-ph.GA},
       adsurl = {https://ui.adsabs.harvard.edu/abs/2017MNRAS.465..111K}
}

@ARTICLE{2006PhR...433..181F,
       author = {{Furlanetto}, Steven R. and {Oh}, S. Peng and {Briggs}, Frank H.},
        title = "{Cosmology at low frequencies: The 21 cm transition and the high-redshift Universe}",
      journal = {Phys. Rep.},
         year = 2006,
        month = oct,
       volume = {433},
       number = {4-6},
        pages = {181-301},
          doi = {10.1016/j.physrep.2006.08.002},
archivePrefix = {arXiv},
       eprint = {astro-ph/0608032},
 primaryClass = {astro-ph},
       adsurl = {https://ui.adsabs.harvard.edu/abs/2006PhR...433..181F}
}

@ARTICLE{2020PASA...37....7S,
       author = {{SKA Cosmology Working Group} and {Bacon}, David J. and {Battye}, Richard A. and {Bull}, Philip and {Camera}, Stefano and {Ferreira}, Pedro G. and {Harrison}, Ian and {Parkinson}, David and {Pourtsidou}, Alkistis and {Santos}, M{\'a}rio G. and {Wolz}, Laura and {Abdalla}, Filipe and {Akrami}, Yashar and {Alonso}, David and {Andrianomena}, Sambatra and {Ballardini}, Mario and {Bernal}, Jos{\'e} Luis and {Bertacca}, Daniele and {Bengaly}, Carlos A.~P. and {Bonaldi}, Anna and {Bonvin}, Camille and {Brown}, Michael L. and {Chapman}, Emma and {Chen}, Song and {Chen}, Xuelei and {Cunnington}, Steven and {Davis}, Tamara M. and {Dickinson}, Clive and {Fonseca}, Jos{\'e} and {Grainge}, Keith and {Harper}, Stuart and {Jarvis}, Matt J. and {Maartens}, Roy and {Maddox}, Natasha and {Padmanabhan}, Hamsa and {Pritchard}, Jonathan R. and {Raccanelli}, Alvise and {Rivi}, Marzia and {Roychowdhury}, Sambit and {Sahl{\'e}n}, Martin and {Schwarz}, Dominik J. and {Siewert}, Thilo M. and {Viel}, Matteo and {Villaescusa-Navarro}, Francisco and {Xu}, Yidong and {Yamauchi}, Daisuke and {Zuntz}, Joe},
        title = "{Cosmology with Phase 1 of the Square Kilometre Array Red Book 2018: Technical specifications and performance forecasts}",
      journal = {PASA},
         year = 2020,
        month = mar,
       volume = {37},
          eid = {e007},
        pages = {e007},
          doi = {10.1017/pasa.2019.51},
archivePrefix = {arXiv},
       eprint = {1811.02743},
 primaryClass = {astro-ph.CO},
       adsurl = {https://ui.adsabs.harvard.edu/abs/2020PASA...37....7S}
}

@ARTICLE{2016MNRAS.462..804G,
       author = {{Geil}, Paul M. and {Mutch}, Simon J. and {Poole}, Gregory B. and {Angel}, Paul W. and {Duffy}, Alan R. and {Mesinger}, Andrei and {Wyithe}, J. Stuart B.},
        title = "{Dark-ages reionization and galaxy formation simulation V: morphology and statistical signatures of reionization}",
      journal = {MNRAS},
         year = 2016,
        month = oct,
       volume = {462},
       number = {1},
        pages = {804-817},
          doi = {10.1093/mnras/stw1718},
archivePrefix = {arXiv},
       eprint = {1512.00564},
 primaryClass = {astro-ph.CO},
       adsurl = {https://ui.adsabs.harvard.edu/abs/2016MNRAS.462..804G}
}

@ARTICLE{1996ApJ...462..563N,
       author = {{Navarro}, Julio F. and {Frenk}, Carlos S. and {White}, Simon D.~M.},
        title = "{The Structure of Cold Dark Matter Halos}",
      journal = {ApJ},
         year = 1996,
        month = may,
       volume = {462},
        pages = {563},
          doi = {10.1086/177173},
archivePrefix = {arXiv},
       eprint = {astro-ph/9508025},
 primaryClass = {astro-ph},
       adsurl = {https://ui.adsabs.harvard.edu/abs/1996ApJ...462..563N}
}

@ARTICLE{1997ApJ...490..493N,
       author = {{Navarro}, Julio F. and {Frenk}, Carlos S. and {White}, Simon D.~M.},
        title = "{A Universal Density Profile from Hierarchical Clustering}",
      journal = {ApJ},
         year = 1997,
        month = dec,
       volume = {490},
       number = {2},
        pages = {493-508},
          doi = {10.1086/304888},
archivePrefix = {arXiv},
       eprint = {astro-ph/9611107},
 primaryClass = {astro-ph},
       adsurl = {https://ui.adsabs.harvard.edu/abs/1997ApJ...490..493N}
}

@ARTICLE{2017MNRAS.470.3220W,
       author = {{Wolz}, L. and {Blake}, C. and {Wyithe}, J.~S.~B.},
        title = "{Determining the H I content of galaxies via intensity mapping cross-correlations}",
      journal = {MNRAS},
         year = 2017,
        month = sep,
       volume = {470},
       number = {3},
        pages = {3220-3226},
          doi = {10.1093/mnras/stx1388},
archivePrefix = {arXiv},
       eprint = {1703.08268},
 primaryClass = {astro-ph.GA},
       adsurl = {https://ui.adsabs.harvard.edu/abs/2017MNRAS.470.3220W}
}

@ARTICLE{2013MNRAS.433.1230W,
       author = {{Watson}, William A. and {Iliev}, Ilian T. and {D'Aloisio}, Anson and {Knebe}, Alexander and {Shapiro}, Paul R. and {Yepes}, Gustavo},
        title = "{The halo mass function through the cosmic ages}",
      journal = {MNRAS},
         year = 2013,
        month = aug,
       volume = {433},
       number = {2},
        pages = {1230-1245},
          doi = {10.1093/mnras/stt791},
archivePrefix = {arXiv},
       eprint = {1212.0095},
 primaryClass = {astro-ph.CO},
       adsurl = {https://ui.adsabs.harvard.edu/abs/2013MNRAS.433.1230W}
}

@ARTICLE{2013A&C.....3...23M,
       author = {{Murray}, S.~G. and {Power}, C. and {Robotham}, A.~S.~G.},
        title = "{HMFcalc: An online tool for calculating dark matter halo mass functions}",
      journal = {Astron. Comput.},
         year = 2013,
        month = nov,
       volume = {3},
          eid = {23},
        pages = {23},
          doi = {10.1016/j.ascom.2013.11.001},
archivePrefix = {arXiv},
       eprint = {1306.6721},
 primaryClass = {astro-ph.CO},
       adsurl = {https://ui.adsabs.harvard.edu/abs/2013A&C.....3...23M}
}

@ARTICLE{2020A&A...641A...6P,
       author = {{Planck Collaboration} and {Aghanim}, N. and {Akrami}, Y. and {Ashdown}, M. and {Aumont}, J. and {Baccigalupi}, C. and {Ballardini}, M. and {Banday}, A.~J. and {Barreiro}, R.~B. and {Bartolo}, N. and {Basak}, S. and {Battye}, R. and {Benabed}, K. and {Bernard}, J. -P. and {Bersanelli}, M. and {Bielewicz}, P. and {Bock}, J.~J. and {Bond}, J.~R. and {Borrill}, J. and {Bouchet}, F.~R. and {Boulanger}, F. and {Bucher}, M. and {Burigana}, C. and {Butler}, R.~C. and {Calabrese}, E. and {Cardoso}, J. -F. and {Carron}, J. and {Challinor}, A. and {Chiang}, H.~C. and {Chluba}, J. and {Colombo}, L.~P.~L. and {Combet}, C. and {Contreras}, D. and {Crill}, B.~P. and {Cuttaia}, F. and {de Bernardis}, P. and {de Zotti}, G. and {Delabrouille}, J. and {Delouis}, J. -M. and {Di Valentino}, E. and {Diego}, J.~M. and {Dor{\'e}}, O. and {Douspis}, M. and {Ducout}, A. and {Dupac}, X. and {Dusini}, S. and {Efstathiou}, G. and {Elsner}, F. and {En{\ss}lin}, T.~A. and {Eriksen}, H.~K. and {Fantaye}, Y. and {Farhang}, M. and {Fergusson}, J. and {Fernandez-Cobos}, R. and {Finelli}, F. and {Forastieri}, F. and {Frailis}, M. and {Fraisse}, A.~A. and {Franceschi}, E. and {Frolov}, A. and {Galeotta}, S. and {Galli}, S. and {Ganga}, K. and {G{\'e}nova-Santos}, R.~T. and {Gerbino}, M. and {Ghosh}, T. and {Gonz{\'a}lez-Nuevo}, J. and {G{\'o}rski}, K.~M. and {Gratton}, S. and {Gruppuso}, A. and {Gudmundsson}, J.~E. and {Hamann}, J. and {Handley}, W. and {Hansen}, F.~K. and {Herranz}, D. and {Hildebrandt}, S.~R. and {Hivon}, E. and {Huang}, Z. and {Jaffe}, A.~H. and {Jones}, W.~C. and {Karakci}, A. and {Keih{\"a}nen}, E. and {Keskitalo}, R. and {Kiiveri}, K. and {Kim}, J. and {Kisner}, T.~S. and {Knox}, L. and {Krachmalnicoff}, N. and {Kunz}, M. and {Kurki-Suonio}, H. and {Lagache}, G. and {Lamarre}, J. -M. and {Lasenby}, A. and {Lattanzi}, M. and {Lawrence}, C.~R. and {Le Jeune}, M. and {Lemos}, P. and {Lesgourgues}, J. and {Levrier}, F. and {Lewis}, A. and {Liguori}, M. and {Lilje}, P.~B. and {Lilley}, M. and {Lindholm}, V. and {L{\'o}pez-Caniego}, M. and {Lubin}, P.~M. and {Ma}, Y. -Z. and {Mac{\'\i}as-P{\'e}rez}, J.~F. and {Maggio}, G. and {Maino}, D. and {Mandolesi}, N. and {Mangilli}, A. and {Marcos-Caballero}, A. and {Maris}, M. and {Martin}, P.~G. and {Martinelli}, M. and {Mart{\'\i}nez-Gonz{\'a}lez}, E. and {Matarrese}, S. and {Mauri}, N. and {McEwen}, J.~D. and {Meinhold}, P.~R. and {Melchiorri}, A. and {Mennella}, A. and {Migliaccio}, M. and {Millea}, M. and {Mitra}, S. and {Miville-Desch{\^e}nes}, M. -A. and {Molinari}, D. and {Montier}, L. and {Morgante}, G. and {Moss}, A. and {Natoli}, P. and {N{\o}rgaard-Nielsen}, H.~U. and {Pagano}, L. and {Paoletti}, D. and {Partridge}, B. and {Patanchon}, G. and {Peiris}, H.~V. and {Perrotta}, F. and {Pettorino}, V. and {Piacentini}, F. and {Polastri}, L. and {Polenta}, G. and {Puget}, J. -L. and {Rachen}, J.~P. and {Reinecke}, M. and {Remazeilles}, M. and {Renzi}, A. and {Rocha}, G. and {Rosset}, C. and {Roudier}, G. and {Rubi{\~n}o-Mart{\'\i}n}, J.~A. and {Ruiz-Granados}, B. and {Salvati}, L. and {Sandri}, M. and {Savelainen}, M. and {Scott}, D. and {Shellard}, E.~P.~S. and {Sirignano}, C. and {Sirri}, G. and {Spencer}, L.~D. and {Sunyaev}, R. and {Suur-Uski}, A. -S. and {Tauber}, J.~A. and {Tavagnacco}, D. and {Tenti}, M. and {Toffolatti}, L. and {Tomasi}, M. and {Trombetti}, T. and {Valenziano}, L. and {Valiviita}, J. and {Van Tent}, B. and {Vibert}, L. and {Vielva}, P. and {Villa}, F. and {Vittorio}, N. and {Wandelt}, B.~D. and {Wehus}, I.~K. and {White}, M. and {White}, S.~D.~M. and {Zacchei}, A. and {Zonca}, A.},
        title = "{Planck 2018 results. VI. Cosmological parameters}",
      journal = {A&A},
         year = 2020,
        month = sep,
       volume = {641},
          eid = {A6},
        pages = {A6},
          doi = {10.1051/0004-6361/201833910},
archivePrefix = {arXiv},
       eprint = {1807.06209},
 primaryClass = {astro-ph.CO},
       adsurl = {https://ui.adsabs.harvard.edu/abs/2020A&A...641A...6P}
}

@ARTICLE{2018AJ....156..160H,
       author = {{Hand}, Nick and {Feng}, Yu and {Beutler}, Florian and {Li}, Yin and {Modi}, Chirag and {Seljak}, Uro{\v{s}} and {Slepian}, Zachary},
        title = "{nbodykit: An Open-source, Massively Parallel Toolkit for Large-scale Structure}",
      journal = {AJ},
         year = 2018,
        month = oct,
       volume = {156},
       number = {4},
          eid = {160},
        pages = {160},
          doi = {10.3847/1538-3881/aadae0},
archivePrefix = {arXiv},
       eprint = {1712.05834},
 primaryClass = {astro-ph.IM},
       adsurl = {https://ui.adsabs.harvard.edu/abs/2018AJ....156..160H}
}

@ARTICLE{2020JOSS....5.2582M,
       author = {{Murray}, Steven and {Greig}, Bradley and {Mesinger}, Andrei and {Mu{\~n}oz}, Julian and {Qin}, Yuxiang and {Park}, Jaehong and {Watkinson}, Catherine},
        title = "{21cmFAST v3: A Python-integrated C code for generating 3D realizations of the cosmic 21cm signal.}",
      journal = {JOSS},
         year = 2020,
        month = oct,
       volume = {5},
       number = {54},
        pages = {2582},
          doi = {10.21105/joss.02582},
}

@ARTICLE{2014JCAP...09..050V,
       author = {{Villaescusa-Navarro}, Francisco and {Viel}, Matteo and {Datta}, Kanan K. and {Choudhury}, T. Roy},
        title = "{Modeling the neutral hydrogen distribution in the post-reionization Universe: intensity mapping}",
      journal = {JCAP},
         year = 2014,
        month = sep,
       volume = {2014},
       number = {9},
        pages = {050-050},
          doi = {10.1088/1475-7516/2014/09/050},
archivePrefix = {arXiv},
       eprint = {1405.6713},
 primaryClass = {astro-ph.CO},
       adsurl = {https://ui.adsabs.harvard.edu/abs/2014JCAP...09..050V}
}

@ARTICLE{2019MNRAS.484.1007W,
       author = {{Wolz}, L. and {Murray}, S.~G. and {Blake}, C. and {Wyithe}, J.~S.},
        title = "{Intensity mapping cross-correlations II: HI halo models including shot noise}",
      journal = {MNRAS},
         year = 2019,
        month = mar,
       volume = {484},
       number = {1},
        pages = {1007-1020},
          doi = {10.1093/mnras/sty3142},
archivePrefix = {arXiv},
       eprint = {1803.02477},
 primaryClass = {astro-ph.CO},
       adsurl = {https://ui.adsabs.harvard.edu/abs/2019MNRAS.484.1007W}
}

@ARTICLE{2019arXiv191212699B,
       author = {{Braun}, Robert and {Bonaldi}, Anna and {Bourke}, Tyler and {Keane}, Evan and {Wagg}, Jeff},
        title = "{Anticipated Performance of the Square Kilometre Array -- Phase 1 (SKA1)}",
      journal = {arXiv e-prints},
         year = 2019,
        month = dec,
          eid = {arXiv:1912.12699},
        pages = {arXiv:1912.12699},
          doi = {10.48550/arXiv.1912.12699},
archivePrefix = {arXiv},
       eprint = {1912.12699},
 primaryClass = {astro-ph.IM},
       adsurl = {https://ui.adsabs.harvard.edu/abs/2019arXiv191212699B}
}

@ARTICLE{2015ApJ...803...21B,
       author = {{Bull}, Philip and {Ferreira}, Pedro G. and {Patel}, Prina and {Santos}, M{\'a}rio G.},
        title = "{Late-time Cosmology with 21 cm Intensity Mapping Experiments}",
      journal = {ApJ},
         year = 2015,
        month = apr,
       volume = {803},
       number = {1},
          eid = {21},
        pages = {21},
          doi = {10.1088/0004-637X/803/1/21},
archivePrefix = {arXiv},
       eprint = {1405.1452},
 primaryClass = {astro-ph.CO},
       adsurl = {https://ui.adsabs.harvard.edu/abs/2015ApJ...803...21B}
}

@article{Lewis_2000,
   title={Efficient Computation of Cosmic Microwave Background Anisotropies in Closed Friedmann‐Robertson‐Walker Models},
   volume={538},
   ISSN={1538-4357},
   url={http://dx.doi.org/10.1086/309179},
   DOI={10.1086/309179},
   number={2},
   journal={ApJ},
   publisher={American Astronomical Society},
   author={Lewis, Antony and Challinor, Anthony and Lasenby, Anthony},
   year={2000},
   month=aug, pages={473–476} }

@ARTICLE{2009ApJS..182..608K,
       author = {{Knollmann}, Steffen R. and {Knebe}, Alexander},
        title = "{AHF: Amiga's Halo Finder}",
      journal = {ApJS},
         year = 2009,
        month = jun,
       volume = {182},
       number = {2},
        pages = {608-624},
          doi = {10.1088/0067-0049/182/2/608},
archivePrefix = {arXiv},
       eprint = {0904.3662},
 primaryClass = {astro-ph.CO},
       adsurl = {https://ui.adsabs.harvard.edu/abs/2009ApJS..182..608K}
}

@ARTICLE{2011MNRAS.415.2293K,
       author = {{Knebe}, Alexander and {Knollmann}, Steffen R. and {Muldrew}, Stuart I. and {Pearce}, Frazer R. and {Aragon-Calvo}, Miguel Angel and {Ascasibar}, Yago and {Behroozi}, Peter S. and {Ceverino}, Daniel and {Colombi}, Stephane and {Diemand}, Juerg and {Dolag}, Klaus and {Falck}, Bridget L. and {Fasel}, Patricia and {Gardner}, Jeff and {Gottl{\"o}ber}, Stefan and {Hsu}, Chung-Hsing and {Iannuzzi}, Francesca and {Klypin}, Anatoly and {Luki{\'c}}, Zarija and {Maciejewski}, Michal and {McBride}, Cameron and {Neyrinck}, Mark C. and {Planelles}, Susana and {Potter}, Doug and {Quilis}, Vicent and {Rasera}, Yann and {Read}, Justin I. and {Ricker}, Paul M. and {Roy}, Fabrice and {Springel}, Volker and {Stadel}, Joachim and {Stinson}, Greg and {Sutter}, P.~M. and {Turchaninov}, Victor and {Tweed}, Dylan and {Yepes}, Gustavo and {Zemp}, Marcel},
        title = "{Haloes gone MAD: The Halo-Finder Comparison Project}",
      journal = {MNRAS},
         year = 2011,
        month = aug,
       volume = {415},
       number = {3},
        pages = {2293-2318},
          doi = {10.1111/j.1365-2966.2011.18858.x},
archivePrefix = {arXiv},
       eprint = {1104.0949},
 primaryClass = {astro-ph.CO},
       adsurl = {https://ui.adsabs.harvard.edu/abs/2011MNRAS.415.2293K}
}

@ARTICLE{2018MNRAS.475..624N,
       author = {{Nelson}, Dylan and {Pillepich}, Annalisa and {Springel}, Volker and {Weinberger}, Rainer and {Hernquist}, Lars and {Pakmor}, R{\"u}diger and {Genel}, Shy and {Torrey}, Paul and {Vogelsberger}, Mark and {Kauffmann}, Guinevere and {Marinacci}, Federico and {Naiman}, Jill},
        title = "{First results from the IllustrisTNG simulations: the galaxy colour bimodality}",
      journal = {MNRAS},
         year = 2018,
        month = mar,
       volume = {475},
       number = {1},
        pages = {624-647},
          doi = {10.1093/mnras/stx3040},
archivePrefix = {arXiv},
       eprint = {1707.03395},
 primaryClass = {astro-ph.GA},
       adsurl = {https://ui.adsabs.harvard.edu/abs/2018MNRAS.475..624N}
}

@ARTICLE{1965ApJ...142.1633G,
       author = {{Gunn}, James E. and {Peterson}, Bruce A.},
        title = "{On the Density of Neutral Hydrogen in Intergalactic Space.}",
      journal = {ApJ},
         year = 1965,
        month = nov,
       volume = {142},
        pages = {1633-1636},
          doi = {10.1086/148444},
       adsurl = {https://ui.adsabs.harvard.edu/abs/1965ApJ...142.1633G}
}

@ARTICLE{2015PASA...32...45B,
       author = {{Becker}, George D. and {Bolton}, James S. and {Lidz}, Adam},
        title = "{Reionisation and High-Redshift Galaxies: The View from Quasar Absorption Lines}",
      journal = {PASA},
         year = 2015,
        month = dec,
       volume = {32},
          eid = {e045},
        pages = {e045},
          doi = {10.1017/pasa.2015.45},
archivePrefix = {arXiv},
       eprint = {1510.03368},
 primaryClass = {astro-ph.CO},
       adsurl = {https://ui.adsabs.harvard.edu/abs/2015PASA...32...45B}
}

@ARTICLE{2021ApJ...923...87C,
       author = {{Christenson}, Holly M. and {Becker}, George D. and {Furlanetto}, Steven R. and {Davies}, Frederick B. and {Malkan}, Matthew A. and {Zhu}, Yongda and {Boera}, Elisa and {Trapp}, Adam},
        title = "{Constraints on the End of Reionization from the Density Fields Surrounding Two Highly Opaque Quasar Sightlines}",
      journal = {ApJ},
         year = 2021,
        month = dec,
       volume = {923},
       number = {1},
          eid = {87},
        pages = {87},
          doi = {10.3847/1538-4357/ac2a34},
archivePrefix = {arXiv},
       eprint = {2109.13170},
 primaryClass = {astro-ph.CO},
       adsurl = {https://ui.adsabs.harvard.edu/abs/2021ApJ...923...87C}
}

@ARTICLE{2023MNRAS.518.6262C,
       author = {{Cunnington}, Steven and {Li}, Yichao and {Santos}, Mario G. and {Wang}, Jingying and {Carucci}, Isabella P. and {Irfan}, Melis O. and {Pourtsidou}, Alkistis and {Spinelli}, Marta and {Wolz}, Laura and {Soares}, Paula S. and {Blake}, Chris and {Bull}, Philip and {Engelbrecht}, Brandon and {Fonseca}, Jos{\'e} and {Grainge}, Keith and {Ma}, Yin-Zhe},
        title = "{H I intensity mapping with MeerKAT: power spectrum detection in cross-correlation with WiggleZ galaxies}",
      journal = {MNRAS},
         year = 2023,
        month = feb,
       volume = {518},
       number = {4},
        pages = {6262-6272},
          doi = {10.1093/mnras/stac3060},
archivePrefix = {arXiv},
       eprint = {2206.01579},
 primaryClass = {astro-ph.CO},
       adsurl = {https://ui.adsabs.harvard.edu/abs/2023MNRAS.518.6262C}
}

@ARTICLE{2013MNRAS.434L..46S,
       author = {{Switzer}, E.~R. and {Masui}, K.~W. and {Bandura}, K. and {Calin}, L. -M. and {Chang}, T. -C. and {Chen}, X. -L. and {Li}, Y. -C. and {Liao}, Y. -W. and {Natarajan}, A. and {Pen}, U. -L. and {Peterson}, J.~B. and {Shaw}, J.~R. and {Voytek}, T.~C.},
        title = "{Determination of z \raisebox{-0.5ex}\textasciitilde 0.8 neutral hydrogen fluctuations using the 21cm  intensity mapping autocorrelation.}",
      journal = {MNRAS},
         year = 2013,
        month = jul,
       volume = {434},
        pages = {L46-L50},
          doi = {10.1093/mnrasl/slt074},
archivePrefix = {arXiv},
       eprint = {1304.3712},
 primaryClass = {astro-ph.CO},
       adsurl = {https://ui.adsabs.harvard.edu/abs/2013MNRAS.434L..46S}
}

@ARTICLE{2013ApJ...763L..20M,
       author = {{Masui}, K.~W. and {Switzer}, E.~R. and {Banavar}, N. and {Bandura}, K. and {Blake}, C. and {Calin}, L. -M. and {Chang}, T. -C. and {Chen}, X. and {Li}, Y. -C. and {Liao}, Y. -W. and {Natarajan}, A. and {Pen}, U. -L. and {Peterson}, J.~B. and {Shaw}, J.~R. and {Voytek}, T.~C.},
        title = "{Measurement of 21 cm Brightness Fluctuations at z \raisebox{-0.5ex}\textasciitilde 0.8 in Cross-correlation}",
      journal = {ApJL},
         year = 2013,
        month = jan,
       volume = {763},
       number = {1},
          eid = {L20},
        pages = {L20},
          doi = {10.1088/2041-8205/763/1/L20},
archivePrefix = {arXiv},
       eprint = {1208.0331},
 primaryClass = {astro-ph.CO},
       adsurl = {https://ui.adsabs.harvard.edu/abs/2013ApJ...763L..20M}
}

@ARTICLE{2022MNRAS.510.3495W,
       author = {{Wolz}, Laura and {Pourtsidou}, Alkistis and {Masui}, Kiyoshi W. and {Chang}, Tzu-Ching and {Bautista}, Julian E. and {M{\"u}ller}, Eva-Maria and {Avila}, Santiago and {Bacon}, David and {Percival}, Will J. and {Cunnington}, Steven and {Anderson}, Chris and {Chen}, Xuelei and {Kneib}, Jean-Paul and {Li}, Yi-Chao and {Liao}, Yu-Wei and {Pen}, Ue-Li and {Peterson}, Jeffrey B. and {Rossi}, Graziano and {Schneider}, Donald P. and {Yadav}, Jaswant and {Zhao}, Gong-Bo},
        title = "{H I constraints from the cross-correlation of eBOSS galaxies and Green Bank Telescope intensity maps}",
      journal = {MNRAS},
         year = 2022,
        month = mar,
       volume = {510},
       number = {3},
        pages = {3495-3511},
          doi = {10.1093/mnras/stab3621},
archivePrefix = {arXiv},
       eprint = {2102.04946},
 primaryClass = {astro-ph.CO},
       adsurl = {https://ui.adsabs.harvard.edu/abs/2022MNRAS.510.3495W}
}

@ARTICLE{2018MNRAS.476.3382A,
       author = {{Anderson}, C.~J. and {Luciw}, N.~J. and {Li}, Y. -C. and {Kuo}, C.~Y. and {Yadav}, J. and {Masui}, K.~W. and {Chang}, T. -C. and {Chen}, X. and {Oppermann}, N. and {Liao}, Y. -W. and {Pen}, U. -L. and {Price}, D.~C. and {Staveley-Smith}, L. and {Switzer}, E.~R. and {Timbie}, P.~T. and {Wolz}, L.},
        title = "{Low-amplitude clustering in low-redshift 21-cm intensity maps cross-correlated with 2dF galaxy densities}",
      journal = {MNRAS},
         year = 2018,
        month = may,
       volume = {476},
       number = {3},
        pages = {3382-3392},
          doi = {10.1093/mnras/sty346},
archivePrefix = {arXiv},
       eprint = {1710.00424},
 primaryClass = {astro-ph.CO},
       adsurl = {https://ui.adsabs.harvard.edu/abs/2018MNRAS.476.3382A}
}

@ARTICLE{2025MNRAS.537..321G,
       author = {{Geda}, Robel and {Teyssier}, Romain},
        title = "{Constructing merger trees of density peaks using phase-space watershed segmentation algorithm}",
      journal = {MNRAS},
         year = 2025,
        month = feb,
       volume = {537},
       number = {1},
        pages = {321-331},
          doi = {10.1093/mnras/staf016},
archivePrefix = {arXiv},
       eprint = {2501.08399},
 primaryClass = {astro-ph.GA},
       adsurl = {https://ui.adsabs.harvard.edu/abs/2025MNRAS.537..321G}
}

@ARTICLE{2015ComAC...2....5B,
       author = {{Bleuler}, Andreas and {Teyssier}, Romain and {Carassou}, S{\'e}bastien and {Martizzi}, Davide},
        title = "{PHEW: a parallel segmentation algorithm for three-dimensional AMR datasets. Application to structure detection in self-gravitating flows}",
      journal = {CompAC},
         year = 2015,
        month = jun,
       volume = {2},
          eid = {5},
        pages = {5},
          doi = {10.1186/s40668-015-0009-7},
archivePrefix = {arXiv},
       eprint = {1412.0510},
 primaryClass = {astro-ph.IM},
       adsurl = {https://ui.adsabs.harvard.edu/abs/2015ComAC...2....5B}
}

@ARTICLE{2008MNRAS.386.2101N,
       author = {{Neyrinck}, Mark C.},
        title = "{ZOBOV: a parameter-free void-finding algorithm}",
      journal = {MNRAS},
         year = 2008,
        month = jun,
       volume = {386},
       number = {4},
        pages = {2101-2109},
          doi = {10.1111/j.1365-2966.2008.13180.x},
archivePrefix = {arXiv},
       eprint = {0712.3049},
 primaryClass = {astro-ph},
       adsurl = {https://ui.adsabs.harvard.edu/abs/2008MNRAS.386.2101N}
}

@ARTICLE{2007MNRAS.380..551P,
       author = {{Platen}, Erwin and {van de Weygaert}, Rien and {Jones}, Bernard J.~T.},
        title = "{A cosmic watershed: the WVF void detection technique}",
      journal = {MNRAS},
         year = 2007,
        month = sep,
       volume = {380},
       number = {2},
        pages = {551-570},
          doi = {10.1111/j.1365-2966.2007.12125.x},
archivePrefix = {arXiv},
       eprint = {0706.2788},
 primaryClass = {astro-ph},
       adsurl = {https://ui.adsabs.harvard.edu/abs/2007MNRAS.380..551P}
}

@ARTICLE{2017MNRAS.469.2323P,
       author = {{Padmanabhan}, Hamsa and {Refregier}, Alexandre and {Amara}, Adam},
        title = "{A halo model for cosmological neutral hydrogen : abundances and clustering}",
      journal = {MNRAS},
         year = 2017,
        month = aug,
       volume = {469},
       number = {2},
        pages = {2323-2334},
          doi = {10.1093/mnras/stx979},
archivePrefix = {arXiv},
       eprint = {1611.06235},
 primaryClass = {astro-ph.CO},
       adsurl = {https://ui.adsabs.harvard.edu/abs/2017MNRAS.469.2323P}
}

@ARTICLE{2016MNRAS.461.1760H,
       author = {{Hirschmann}, Michaela and {De Lucia}, Gabriella and {Fontanot}, Fabio},
        title = "{Galaxy assembly, stellar feedback and metal enrichment: the view from the GAEA model}",
      journal = {MNRAS},
         year = 2016,
        month = sep,
       volume = {461},
       number = {2},
        pages = {1760-1785},
          doi = {10.1093/mnras/stw1318},
archivePrefix = {arXiv},
       eprint = {1512.04531},
 primaryClass = {astro-ph.GA},
       adsurl = {https://ui.adsabs.harvard.edu/abs/2016MNRAS.461.1760H}
}

@ARTICLE{2017MNRAS.464.3812F,
       author = {{Fontanot}, Fabio and {De Lucia}, Gabriella and {Hirschmann}, Michaela and {Bruzual}, Gustavo and {Charlot}, St{\'e}phane and {Zibetti}, Stefano},
        title = "{Variations of the stellar initial mass function in semi-analytical models: implications for the mass assembly and the chemical enrichment of galaxies in the GAEA model}",
      journal = {MNRAS},
         year = 2017,
        month = feb,
       volume = {464},
       number = {4},
        pages = {3812-3824},
          doi = {10.1093/mnras/stw2612},
archivePrefix = {arXiv},
       eprint = {1606.01908},
 primaryClass = {astro-ph.GA},
       adsurl = {https://ui.adsabs.harvard.edu/abs/2017MNRAS.464.3812F}
}

@ARTICLE{2017MNRAS.466L..88D,
       author = {{De Lucia}, Gabriella and {Fontanot}, Fabio and {Hirschmann}, Michaela},
        title = "{AGN feedback and the origin of the {\ensuremath{\alpha}} enhancement in early-type galaxies - insights from the GAEA model}",
      journal = {MNRAS},
         year = 2017,
        month = mar,
       volume = {466},
       number = {1},
        pages = {L88-L92},
          doi = {10.1093/mnrasl/slw242},
archivePrefix = {arXiv},
       eprint = {1611.04597},
 primaryClass = {astro-ph.GA},
       adsurl = {https://ui.adsabs.harvard.edu/abs/2017MNRAS.466L..88D}
}

@ARTICLE{2005Natur.435..629S,
       author = {{Springel}, Volker and {White}, Simon D.~M. and {Jenkins}, Adrian and {Frenk}, Carlos S. and {Yoshida}, Naoki and {Gao}, Liang and {Navarro}, Julio and {Thacker}, Robert and {Croton}, Darren and {Helly}, John and {Peacock}, John A. and {Cole}, Shaun and {Thomas}, Peter and {Couchman}, Hugh and {Evrard}, August and {Colberg}, J{\"o}rg and {Pearce}, Frazer},
        title = "{Simulations of the formation, evolution and clustering of galaxies and quasars}",
      journal = {Nature},
         year = 2005,
        month = jun,
       volume = {435},
       number = {7042},
        pages = {629-636},
          doi = {10.1038/nature03597},
archivePrefix = {arXiv},
       eprint = {astro-ph/0504097},
 primaryClass = {astro-ph},
       adsurl = {https://ui.adsabs.harvard.edu/abs/2005Natur.435..629S}
}

@ARTICLE{2009MNRAS.398.1150B,
       author = {{Boylan-Kolchin}, Michael and {Springel}, Volker and {White}, Simon D.~M. and {Jenkins}, Adrian and {Lemson}, Gerard},
        title = "{Resolving cosmic structure formation with the Millennium-II Simulation}",
      journal = {MNRAS},
         year = 2009,
        month = sep,
       volume = {398},
       number = {3},
        pages = {1150-1164},
          doi = {10.1111/j.1365-2966.2009.15191.x},
archivePrefix = {arXiv},
       eprint = {0903.3041},
 primaryClass = {astro-ph.CO},
       adsurl = {https://ui.adsabs.harvard.edu/abs/2009MNRAS.398.1150B}
}

@ARTICLE{2018arXiv181009572C,
       author = {{Cosmic Visions 21 cm Collaboration} and {Ansari}, R{\'e}za and {Arena}, Evan J. and {Bandura}, Kevin and {Bull}, Philip and {Castorina}, Emanuele and {Chang}, Tzu-Ching and {Chen}, Shi-Fan and {Connor}, Liam and {Foreman}, Simon and {Frisch}, Josef and {Green}, Daniel and {Johnson}, Matthew C. and {Karagiannis}, Dionysios and {Liu}, Adrian and {Masui}, Kiyoshi W. and {Meerburg}, P. Daniel and {M{\"u}nchmeyer}, Moritz and {Newburgh}, Laura B. and {Obuljen}, Andrej and {O'Connor}, Paul and {Padmanabhan}, Hamsa and {Shaw}, J. Richard and {Sheehy}, Christopher and {Slosar}, An{\v{z}}e and {Smith}, Kendrick and {Stankus}, Paul and {Stebbins}, Albert and {Timbie}, Peter and {Villaescusa-Navarro}, Francisco and {Wallisch}, Benjamin and {White}, Martin},
        title = "{Inflation and Early Dark Energy with a Stage II Hydrogen Intensity Mapping Experiment}",
      journal = {arXiv e-prints},
         year = 2018,
        month = oct,
          eid = {arXiv:1810.09572},
        pages = {arXiv:1810.09572},
          doi = {10.48550/arXiv.1810.09572},
archivePrefix = {arXiv},
       eprint = {1810.09572},
 primaryClass = {astro-ph.CO},
       adsurl = {https://ui.adsabs.harvard.edu/abs/2018arXiv181009572C}
}

@ARTICLE{2020PASP..132f2001L,
       author = {{Liu}, Adrian and {Shaw}, J. Richard},
        title = "{Data Analysis for Precision 21 cm Cosmology}",
      journal = {\pasp},
         year = 2020,
        month = jun,
       volume = {132},
       number = {1012},
          eid = {062001},
        pages = {062001},
          doi = {10.1088/1538-3873/ab5bfd},
archivePrefix = {arXiv},
       eprint = {1907.08211},
 primaryClass = {astro-ph.IM},
       adsurl = {https://ui.adsabs.harvard.edu/abs/2020PASP..132f2001L}
}

@ARTICLE{2001JApA...22...21B,
       author = {{Bharadwaj}, Somnath and {Nath}, Biman B. and {Sethi}, Shiv K.},
        title = "{Using HI to probe large scale structures at z{\ensuremath{\sim}}3}",
      journal = {Journal of Astrophysics and Astronomy},
         year = 2001,
        month = mar,
       volume = {22},
       number = {1},
        pages = {21-34},
          doi = {10.1007/BF02933588},
archivePrefix = {arXiv},
       eprint = {astro-ph/0003200},
 primaryClass = {astro-ph},
       adsurl = {https://ui.adsabs.harvard.edu/abs/2001JApA...22...21B}
}

@ARTICLE{2013MNRAS.435.1618K,
       author = {{Knebe}, Alexander and {Pearce}, Frazer R. and {Lux}, Hanni and {Ascasibar}, Yago and {Behroozi}, Peter and {Casado}, Javier and {Moran}, Christine Corbett and {Diemand}, Juerg and {Dolag}, Klaus and {Dominguez-Tenreiro}, Rosa and {Elahi}, Pascal and {Falck}, Bridget and {Gottl{\"o}ber}, Stefan and {Han}, Jiaxin and {Klypin}, Anatoly and {Luki{\'c}}, Zarija and {Maciejewski}, Michal and {McBride}, Cameron K. and {Merch{\'a}n}, Manuel E. and {Muldrew}, Stuart I. and {Neyrinck}, Mark and {Onions}, Julian and {Planelles}, Susana and {Potter}, Doug and {Quilis}, Vicent and {Rasera}, Yann and {Ricker}, Paul M. and {Roy}, Fabrice and {Ruiz}, Andr{\'e}s N. and {Sgr{\'o}}, Mario A. and {Springel}, Volker and {Stadel}, Joachim and {Sutter}, P.~M. and {Tweed}, Dylan and {Zemp}, Marcel},
        title = "{Structure finding in cosmological simulations: the state of affairs}",
      journal = {\mnras},
         year = 2013,
        month = oct,
       volume = {435},
       number = {2},
        pages = {1618-1658},
          doi = {10.1093/mnras/stt1403},
archivePrefix = {arXiv},
       eprint = {1304.0585},
 primaryClass = {astro-ph.CO},
       adsurl = {https://ui.adsabs.harvard.edu/abs/2013MNRAS.435.1618K}
}

@ARTICLE{2025arXiv250417254D,
       author = {{Davies}, James E. and {Mesinger}, Andrei and {Murray}, Steven},
        title = "{Efficient simulation of discrete galaxy populations and associated radiation fields during the first billion years}",
      journal = {arXiv e-prints},
         year = 2025,
        month = apr,
          eid = {arXiv:2504.17254},
        pages = {arXiv:2504.17254},
          doi = {10.48550/arXiv.2504.17254},
archivePrefix = {arXiv},
       eprint = {2504.17254},
 primaryClass = {astro-ph.CO},
       adsurl = {https://ui.adsabs.harvard.edu/abs/2025arXiv250417254D}
}

@ARTICLE{2020MNRAS.491.1736K,
       author = {{Keating}, Laura C. and {Weinberger}, Lewis H. and {Kulkarni}, Girish and {Haehnelt}, Martin G. and {Chardin}, Jonathan and {Aubert}, Dominique},
        title = "{Long troughs in the Lyman-{\ensuremath{\alpha}} forest below redshift 6 due to islands of neutral hydrogen}",
      journal = {\mnras},
         year = 2020,
        month = jan,
       volume = {491},
       number = {2},
        pages = {1736-1745},
          doi = {10.1093/mnras/stz3083},
archivePrefix = {arXiv},
       eprint = {1905.12640},
 primaryClass = {astro-ph.CO},
       adsurl = {https://ui.adsabs.harvard.edu/abs/2020MNRAS.491.1736K}
}

@ARTICLE{2025PASA...42..107C,
       author = {{Cain}, Christopher and {D'Aloisio}, Anson and {Mu{\~n}oz}, Julian},
        title = "{New constraints on the galactic ionising efficiency and escape fraction at 2.5 < z < 6 based on quasar absorption spectra}",
      journal = {\pasa},
         year = 2025,
        month = jul,
       volume = {42},
          eid = {e107},
        pages = {e107},
          doi = {10.1017/pasa.2025.10071},
archivePrefix = {arXiv},
       eprint = {2503.08778},
 primaryClass = {astro-ph.GA},
       adsurl = {https://ui.adsabs.harvard.edu/abs/2025PASA...42..107C}
}

@ARTICLE{2016MNRAS.459.3025P,
       author = {{Poole}, Gregory B. and {Angel}, Paul W. and {Mutch}, Simon J. and {Power}, Chris and {Duffy}, Alan R. and {Geil}, Paul M. and {Mesinger}, Andrei and {Wyithe}, Stuart B.},
        title = "{Dark-ages Reionization and Galaxy formation simulation - I. The dynamical lives of high-redshift galaxies}",
      journal = {\mnras},
         year = 2016,
        month = jul,
       volume = {459},
       number = {3},
        pages = {3025-3039},
          doi = {10.1093/mnras/stw674},
archivePrefix = {arXiv},
       eprint = {1512.00559},
 primaryClass = {astro-ph.GA},
       adsurl = {https://ui.adsabs.harvard.edu/abs/2016MNRAS.459.3025P}
}

@ARTICLE{2016MNRAS.462..250M,
       author = {{Mutch}, Simon J. and {Geil}, Paul M. and {Poole}, Gregory B. and {Angel}, Paul W. and {Duffy}, Alan R. and {Mesinger}, Andrei and {Wyithe}, J. Stuart B.},
        title = "{Dark-ages reionization and galaxy formation simulation - III. Modelling galaxy formation and the epoch of reionization}",
      journal = {\mnras},
         year = 2016,
        month = oct,
       volume = {462},
       number = {1},
        pages = {250-276},
          doi = {10.1093/mnras/stw1506},
archivePrefix = {arXiv},
       eprint = {1512.00562},
 primaryClass = {astro-ph.GA},
       adsurl = {https://ui.adsabs.harvard.edu/abs/2016MNRAS.462..250M}
}

@INPROCEEDINGS{2016SPIE.9906E..5XN,
       author = {{Newburgh}, L.~B. and {Bandura}, K. and {Bucher}, M.~A. and {Chang}, T. -C. and {Chiang}, H.~C. and {Cliche}, J.~F. and {Dav{\'e}}, R. and {Dobbs}, M. and {Clarkson}, C. and {Ganga}, K.~M. and {Gogo}, T. and {Gumba}, A. and {Gupta}, N. and {Hilton}, M. and {Johnstone}, B. and {Karastergiou}, A. and {Kunz}, M. and {Lokhorst}, D. and {Maartens}, R. and {Macpherson}, S. and {Mdlalose}, M. and {Moodley}, K. and {Ngwenya}, L. and {Parra}, J.~M. and {Peterson}, J. and {Recnik}, O. and {Saliwanchik}, B. and {Santos}, M.~G. and {Sievers}, J.~L. and {Smirnov}, O. and {Stronkhorst}, P. and {Taylor}, R. and {Vanderlinde}, K. and {Van Vuuren}, G. and {Weltman}, A. and {Witzemann}, A.},
        title = "{HIRAX: a probe of dark energy and radio transients}",
    booktitle = {Ground-based and Airborne Telescopes VI},
         year = 2016,
       editor = {{Hall}, Helen J. and {Gilmozzi}, Roberto and {Marshall}, Heather K.},
       series = {Society of Photo-Optical Instrumentation Engineers (SPIE) Conference Series},
       volume = {9906},
        month = aug,
          eid = {99065X},
        pages = {99065X},
          doi = {10.1117/12.2234286},
archivePrefix = {arXiv},
       eprint = {1607.02059},
 primaryClass = {astro-ph.IM},
       adsurl = {https://ui.adsabs.harvard.edu/abs/2016SPIE.9906E..5XN}
}

@ARTICLE{2022ApJS..261...29C,
       author = {{CHIME Collaboration} and {Amiri}, Mandana and {Bandura}, Kevin and {Boskovic}, Anja and {Chen}, Tianyue and {Cliche}, Jean-Fran{\c{c}}ois and {Deng}, Meiling and {Denman}, Nolan and {Dobbs}, Matt and {Fandino}, Mateus and {Foreman}, Simon and {Halpern}, Mark and {Hanna}, David and {Hill}, Alex S. and {Hinshaw}, Gary and {H{\"o}fer}, Carolin and {Kania}, Joseph and {Klages}, Peter and {Landecker}, T.~L. and {MacEachern}, Joshua and {Masui}, Kiyoshi and {Mena-Parra}, Juan and {Milutinovic}, Nikola and {Mirhosseini}, Arash and {Newburgh}, Laura and {Nitsche}, Rick and {Ordog}, Anna and {Pen}, Ue-Li and {Pinsonneault-Marotte}, Tristan and {Polzin}, Ava and {Reda}, Alex and {Renard}, Andre and {Shaw}, J. Richard and {Siegel}, Seth R. and {Singh}, Saurabh and {Smegal}, Rick and {Tretyakov}, Ian and {van Gassen}, Kwinten and {Vanderlinde}, Keith and {Wang}, Haochen and {Wiebe}, Donald V. and {Willis}, James S. and {Wulf}, Dallas},
        title = "{An Overview of CHIME, the Canadian Hydrogen Intensity Mapping Experiment}",
      journal = {\apjs},
         year = 2022,
        month = aug,
       volume = {261},
       number = {2},
          eid = {29},
        pages = {29},
          doi = {10.3847/1538-4365/ac6fd9},
archivePrefix = {arXiv},
       eprint = {2201.07869},
 primaryClass = {astro-ph.IM},
       adsurl = {https://ui.adsabs.harvard.edu/abs/2022ApJS..261...29C}
}

@INPROCEEDINGS{2014SPIE.9145E..22B,
       author = {{Bandura}, Kevin and {Addison}, Graeme E. and {Amiri}, Mandana and {Bond}, J. Richard and {Campbell-Wilson}, Duncan and {Connor}, Liam and {Cliche}, Jean-Fran{\c{c}}ois and {Davis}, Greg and {Deng}, Meiling and {Denman}, Nolan and {Dobbs}, Matt and {Fandino}, Mateus and {Gibbs}, Kenneth and {Gilbert}, Adam and {Halpern}, Mark and {Hanna}, David and {Hincks}, Adam D. and {Hinshaw}, Gary and {H{\"o}fer}, Carolin and {Klages}, Peter and {Landecker}, Tom L. and {Masui}, Kiyoshi and {Mena Parra}, Juan and {Newburgh}, Laura B. and {Pen}, Ue-li and {Peterson}, Jeffrey B. and {Recnik}, Andre and {Shaw}, J. Richard and {Sigurdson}, Kris and {Sitwell}, Mike and {Smecher}, Graeme and {Smegal}, Rick and {Vanderlinde}, Keith and {Wiebe}, Don},
        title = "{Canadian Hydrogen Intensity Mapping Experiment (CHIME) pathfinder}",
    booktitle = {Ground-based and Airborne Telescopes V},
         year = 2014,
       editor = {{Stepp}, Larry M. and {Gilmozzi}, Roberto and {Hall}, Helen J.},
       series = {Society of Photo-Optical Instrumentation Engineers (SPIE) Conference Series},
       volume = {9145},
        month = jul,
          eid = {914522},
        pages = {914522},
          doi = {10.1117/12.2054950},
archivePrefix = {arXiv},
       eprint = {1406.2288},
 primaryClass = {astro-ph.IM},
       adsurl = {https://ui.adsabs.harvard.edu/abs/2014SPIE.9145E..22B}
}

@article{wyithe2008baryonic,
  title={Baryonic acoustic oscillations in 21-cm emission: a probe of dark energy out to high redshifts},
  author={Wyithe, J Stuart B and Loeb, Abraham and Geil, Paul M},
  journal={Monthly Notices of the Royal Astronomical Society},
  volume={383},
  number={3},
  pages={1195--1209},
  year={2008},
  publisher={Blackwell Publishing Ltd Oxford, UK}
}

@article{chang2008baryon,
  title={Baryon acoustic oscillation intensity mapping of dark energy},
  author={Chang, Tzu-Ching and Pen, Ue-Li and Peterson, Jeffrey B and McDonald, Patrick},
  journal={Physical Review Letters},
  volume={100},
  number={9},
  pages={091303},
  year={2008},
  publisher={APS}
}

@ARTICLE{2014PhRvD..90b3018L,
       author = {{Liu}, Adrian and {Parsons}, Aaron R. and {Trott}, Cathryn M.},
        title = "{Epoch of reionization window. I. Mathematical formalism}",
      journal = {\prd},
         year = 2014,
        month = jul,
       volume = {90},
       number = {2},
          eid = {023018},
        pages = {023018},
          doi = {10.1103/PhysRevD.90.023018},
archivePrefix = {arXiv},
       eprint = {1404.2596},
 primaryClass = {astro-ph.CO},
       adsurl = {https://ui.adsabs.harvard.edu/abs/2014PhRvD..90b3018L}
}

@ARTICLE{2025MNRAS.541..476M,
       author = {{Mazumder}, Aishrila and {Wolz}, Laura and {Chen}, Zhaoting and {Paul}, Sourabh and {Santos}, Mario G. and {Jarvis}, Matt and {Townsend}, Junaid and {Sekhar}, Srikrishna and {Taylor}, Russ},
        title = "{HI intensity mapping with the MIGHTEE Survey: first results of the HI power spectrum}",
      journal = {\mnras},
         year = 2025,
        month = jul,
       volume = {541},
       number = {1},
        pages = {476-493},
          doi = {10.1093/mnras/staf975},
archivePrefix = {arXiv},
       eprint = {2501.17564},
 primaryClass = {astro-ph.CO},
       adsurl = {https://ui.adsabs.harvard.edu/abs/2025MNRAS.541..476M}
}

@ARTICLE{2025A&A...693A.276M,
       author = {{Munshi}, S. and {Mertens}, F.~G. and {Koopmans}, L.~V.~E. and {Offringa}, A.~R. and {Ceccotti}, E. and {Brackenhoff}, S.~A. and {Chege}, J.~K. and {Gehlot}, B.~K. and {Ghosh}, S. and {H{\"o}fer}, C. and {Mevius}, M.},
        title = "{Beyond the horizon: Quantifying the full sky foreground wedge in the cylindrical power spectrum}",
      journal = {\aap},
         year = 2025,
        month = jan,
       volume = {693},
          eid = {A276},
        pages = {A276},
          doi = {10.1051/0004-6361/202451181},
archivePrefix = {arXiv},
       eprint = {2407.10686},
 primaryClass = {astro-ph.CO},
       adsurl = {https://ui.adsabs.harvard.edu/abs/2025A&A...693A.276M}
}

@ARTICLE{2014MNRAS.440.1662S,
       author = {{Sobacchi}, Emanuele and {Mesinger}, Andrei},
        title = "{Inhomogeneous recombinations during cosmic reionization}",
      journal = {\mnras},
         year = 2014,
        month = may,
       volume = {440},
       number = {2},
        pages = {1662-1673},
          doi = {10.1093/mnras/stu377},
archivePrefix = {arXiv},
       eprint = {1402.2298},
 primaryClass = {astro-ph.CO},
       adsurl = {https://ui.adsabs.harvard.edu/abs/2014MNRAS.440.1662S}
}

@ARTICLE{2024MNRAS.533.2364G,
       author = {{Giri}, Sambit K. and {Bianco}, Michele and {Schaeffer}, Timoth{\'e}e and {Iliev}, Ilian T. and {Mellema}, Garrelt and {Schneider}, Aurel},
        title = "{The 21-cm signal during the end stages of reionization}",
      journal = {\mnras},
         year = 2024,
        month = sep,
       volume = {533},
       number = {2},
        pages = {2364-2378},
          doi = {10.1093/mnras/stae1999},
archivePrefix = {arXiv},
       eprint = {2403.04838},
 primaryClass = {astro-ph.CO},
       adsurl = {https://ui.adsabs.harvard.edu/abs/2024MNRAS.533.2364G}
}

@ARTICLE{2017ComAC...4....2P,
       author = {{Potter}, Douglas and {Stadel}, Joachim and {Teyssier}, Romain},
        title = "{PKDGRAV3: beyond trillion particle cosmological simulations for the next era of galaxy surveys}",
      journal = {Computational Astrophysics and Cosmology},
         year = 2017,
        month = may,
       volume = {4},
       number = {1},
          eid = {2},
        pages = {2},
          doi = {10.1186/s40668-017-0021-1},
archivePrefix = {arXiv},
       eprint = {1609.08621},
 primaryClass = {astro-ph.IM},
       adsurl = {https://ui.adsabs.harvard.edu/abs/2017ComAC...4....2P}
}

@ARTICLE{2006NewA...11..374M,
       author = {{Mellema}, Garrelt and {Iliev}, Ilian T. and {Alvarez}, Marcelo A. and {Shapiro}, Paul R.},
        title = "{C $^{2}$-ray: A new method for photon-conserving transport of ionizing radiation}",
      journal = {\na},
         year = 2006,
        month = mar,
       volume = {11},
       number = {5},
        pages = {374-395},
          doi = {10.1016/j.newast.2005.09.004},
archivePrefix = {arXiv},
       eprint = {astro-ph/0508416},
 primaryClass = {astro-ph},
       adsurl = {https://ui.adsabs.harvard.edu/abs/2006NewA...11..374M}
}

@ARTICLE{2024A&C....4800861H,
       author = {{Hirling}, P. and {Bianco}, M. and {Giri}, S.~K. and {Iliev}, I.~T. and {Mellema}, G. and {Kneib}, J.-P.},
        title = "{pyC2Ray: A flexible and GPU-accelerated radiative transfer framework for simulating the cosmic epoch of reionization}",
      journal = {Astronomy and Computing},
         year = 2024,
        month = jul,
       volume = {48},
          eid = {100861},
        pages = {100861},
          doi = {10.1016/j.ascom.2024.100861},
archivePrefix = {arXiv},
       eprint = {2311.01492},
 primaryClass = {astro-ph.CO},
       adsurl = {https://ui.adsabs.harvard.edu/abs/2024A&C....4800861H}
}

@ARTICLE{1996MNRAS.282..263E,
       author = {{Eke}, Vincent R. and {Cole}, Shaun and {Frenk}, Carlos S.},
        title = "{Cluster evolution as a diagnostic for Omega}",
      journal = {\mnras},
         year = 1996,
        month = sep,
       volume = {282},
        pages = {263-280},
          doi = {10.1093/mnras/282.1.263},
archivePrefix = {arXiv},
       eprint = {astro-ph/9601088},
 primaryClass = {astro-ph},
       adsurl = {https://ui.adsabs.harvard.edu/abs/1996MNRAS.282..263E}
}

@ARTICLE{2010ApJ...724..341H,
       author = {{Heger}, Alexander and {Woosley}, S.~E.},
        title = "{Nucleosynthesis and Evolution of Massive Metal-free Stars}",
      journal = {\apj},
         year = 2010,
        month = nov,
       volume = {724},
       number = {1},
        pages = {341-373},
          doi = {10.1088/0004-637X/724/1/341},
archivePrefix = {arXiv},
       eprint = {0803.3161},
 primaryClass = {astro-ph},
       adsurl = {https://ui.adsabs.harvard.edu/abs/2010ApJ...724..341H}
}

@ARTICLE{2008MNRAS.390.1496G,
       author = {{Geil}, Paul M. and {Wyithe}, J. Stuart B. and {Petrovic}, Nada and {Oh}, S. Peng},
        title = "{The effect of Galactic foreground subtraction on redshifted 21-cm observations of quasar HII regions}",
      journal = {\mnras},
         year = 2008,
        month = nov,
       volume = {390},
       number = {4},
        pages = {1496-1504},
          doi = {10.1111/j.1365-2966.2008.13798.x},
archivePrefix = {arXiv},
       eprint = {0805.0038},
 primaryClass = {astro-ph},
       adsurl = {https://ui.adsabs.harvard.edu/abs/2008MNRAS.390.1496G}
}

@ARTICLE{2013PASA...30...31B,
       author = {{Bowman}, Judd D. and {Cairns}, Iver and {Kaplan}, David L. and {Murphy}, Tara and {Oberoi}, Divya and {Staveley-Smith}, Lister and {Arcus}, Wayne and {Barnes}, David G. and {Bernardi}, Gianni and {Briggs}, Frank H. and {Brown}, Shea and {Bunton}, John D. and {Burgasser}, Adam J. and {Cappallo}, Roger J. and {Chatterjee}, Shami and {Corey}, Brian E. and {Coster}, Anthea and {Deshpande}, Avinash and {deSouza}, Ludi and {Emrich}, David and {Erickson}, Philip and {Goeke}, Robert F. and {Gaensler}, B.~M. and {Greenhill}, Lincoln J. and {Harvey-Smith}, Lisa and {Hazelton}, Bryna J. and {Herne}, David and {Hewitt}, Jacqueline N. and {Johnston-Hollitt}, Melanie and {Kasper}, Justin C. and {Kincaid}, Barton B. and {Koenig}, Ronald and {Kratzenberg}, Eric and {Lonsdale}, Colin J. and {Lynch}, Mervyn J. and {Matthews}, Lynn D. and {McWhirter}, S. Russell and {Mitchell}, Daniel A. and {Morales}, Miguel F. and {Morgan}, Edward H. and {Ord}, Stephen M. and {Pathikulangara}, Joseph and {Prabu}, Thiagaraj and {Remillard}, Ronald A. and {Robishaw}, Timothy and {Rogers}, Alan E.~E. and {Roshi}, Anish A. and {Salah}, Joseph E. and {Sault}, Robert J. and {Shankar}, N. Udaya and {Srivani}, K.~S. and {Stevens}, Jamie B. and {Subrahmanyan}, Ravi and {Tingay}, Steven J. and {Wayth}, Randall B. and {Waterson}, Mark and {Webster}, Rachel L. and {Whitney}, Alan R. and {Williams}, Andrew J. and {Williams}, Christopher L. and {Wyithe}, J. Stuart B.},
        title = "{Science with the Murchison Widefield Array}",
      journal = {\pasa},
         year = 2013,
        month = apr,
       volume = {30},
          eid = {e031},
        pages = {e031},
          doi = {10.1017/pas.2013.009},
archivePrefix = {arXiv},
       eprint = {1212.5151},
 primaryClass = {astro-ph.IM},
       adsurl = {https://ui.adsabs.harvard.edu/abs/2013PASA...30...31B}
}

@ARTICLE{2001MNRAS.320..153B,
       author = {{Benson}, A.~J. and {Nusser}, Adi and {Sugiyama}, Naoshi and {Lacey}, C.~G.},
        title = "{Non-uniform reionization by galaxies and its effect on the cosmic microwave background}",
      journal = {\mnras},
         year = 2001,
        month = jan,
       volume = {320},
       number = {1},
        pages = {153-176},
          doi = {10.1046/j.1365-8711.2001.03957.x},
archivePrefix = {arXiv},
       eprint = {astro-ph/0002457},
 primaryClass = {astro-ph},
       adsurl = {https://ui.adsabs.harvard.edu/abs/2001MNRAS.320..153B}
}

@ARTICLE{2000ApJ...534..507C,
       author = {{Chiu}, Weihsueh A. and {Ostriker}, Jeremiah P.},
        title = "{A Semianalytic Model for Cosmological Reheating and Reionization Due to the Gravitational Collapse of Structure}",
      journal = {\apj},
         year = 2000,
        month = may,
       volume = {534},
       number = {2},
        pages = {507-532},
          doi = {10.1086/308780},
archivePrefix = {arXiv},
       eprint = {astro-ph/9907220},
 primaryClass = {astro-ph},
       adsurl = {https://ui.adsabs.harvard.edu/abs/2000ApJ...534..507C}
}

@ARTICLE{2004ApJ...610....9G,
       author = {{Gnedin}, Nickolay Y.},
        title = "{Reionization, Sloan, and WMAP: Is the Picture Consistent?}",
      journal = {\apj},
         year = 2004,
        month = jul,
       volume = {610},
       number = {1},
        pages = {9-13},
          doi = {10.1086/421450},
archivePrefix = {arXiv},
       eprint = {astro-ph/0403699},
 primaryClass = {astro-ph},
       adsurl = {https://ui.adsabs.harvard.edu/abs/2004ApJ...610....9G}
}

@ARTICLE{2023arXiv230111943P,
       author = {{Paul}, Sourabh and {Santos}, Mario G. and {Chen}, Zhaoting and {Wolz}, Laura},
        title = "{A first detection of neutral hydrogen intensity mapping on Mpc scales at $z\approx 0.32$ and $z\approx 0.44$}",
      journal = {arXiv e-prints},
         year = 2023,
        month = jan,
          eid = {arXiv:2301.11943},
        pages = {arXiv:2301.11943},
          doi = {10.48550/arXiv.2301.11943},
archivePrefix = {arXiv},
       eprint = {2301.11943},
 primaryClass = {astro-ph.CO},
       adsurl = {https://ui.adsabs.harvard.edu/abs/2023arXiv230111943P}
}

@ARTICLE{2021MNRAS.502.5259C,
       author = {{Chen}, Zhaoting and {Wolz}, Laura and {Spinelli}, Marta and {Murray}, Steven G.},
        title = "{Extracting H I astrophysics from interferometric intensity mapping}",
      journal = {\mnras},
         year = 2021,
        month = apr,
       volume = {502},
       number = {4},
        pages = {5259-5276},
          doi = {10.1093/mnras/stab386},
archivePrefix = {arXiv},
       eprint = {2010.07985},
 primaryClass = {astro-ph.CO},
       adsurl = {https://ui.adsabs.harvard.edu/abs/2021MNRAS.502.5259C}
}

@ARTICLE{2025JCAP...04..003H,
       author = {{Hitz}, Pascal and {Berner}, Pascale and {Crichton}, Devin and {Hennig}, John and {Refregier}, Alexandre},
        title = "{Fast simulation of cosmological neutral hydrogen based on the halo model}",
      journal = {\jcap},
         year = 2025,
        month = apr,
       volume = {2025},
       number = {4},
          eid = {003},
        pages = {003},
          doi = {10.1088/1475-7516/2025/04/003},
archivePrefix = {arXiv},
       eprint = {2410.01694},
 primaryClass = {astro-ph.CO},
       adsurl = {https://ui.adsabs.harvard.edu/abs/2025JCAP...04..003H}
}

@ARTICLE{2019MNRAS.485.4060P,
       author = {{Padmanabhan}, Hamsa and {Refregier}, Alexandre and {Amara}, Adam},
        title = "{Impact of astrophysics on cosmology forecasts for 21 cm surveys}",
      journal = {\mnras},
         year = 2019,
        month = may,
       volume = {485},
       number = {3},
        pages = {4060-4070},
          doi = {10.1093/mnras/stz683},
archivePrefix = {arXiv},
       eprint = {1804.10627},
 primaryClass = {astro-ph.CO},
       adsurl = {https://ui.adsabs.harvard.edu/abs/2019MNRAS.485.4060P}
}

@ARTICLE{2014PeerJ...2..453V,
       author = {{van der Walt}, Stefan and {Sch{\"o}nberger}, Johannes L. and {Nunez-Iglesias}, Juan and {Boulogne}, Fran{\c{c}}ois and {Warner}, Joshua D. and {Yager}, Neil and {Gouillart}, Emmanuelle and {Yu}, Tony and {scikit-image Contributors}},
        title = "{scikit-image: Image processing in Python}",
      journal = {PeerJ},
         year = 2014,
        month = jan,
       volume = {2},
          eid = {e453},
        pages = {e453},
          doi = {10.7717/peerj.453},
archivePrefix = {arXiv},
       eprint = {1407.6245},
 primaryClass = {cs.MS},
       adsurl = {https://ui.adsabs.harvard.edu/abs/2014PeerJ...2..453V}
}

@ARTICLE{2007CSE.....9...90H,
       author = {{Hunter}, John D.},
        title = "{Matplotlib: A 2D Graphics Environment}",
      journal = {Computing in Science and Engineering},
         year = 2007,
        month = jan,
       volume = {9},
       number = {3},
        pages = {90-95},
          doi = {10.1109/MCSE.2007.55},
       adsurl = {https://ui.adsabs.harvard.edu/abs/2007CSE.....9...90H}
}

@ARTICLE{2020NatMe..17..261V,
       author = {{Virtanen}, Pauli and {Gommers}, Ralf and {Oliphant}, Travis E. and {Haberland}, Matt and {Reddy}, Tyler and {Cournapeau}, David and {Burovski}, Evgeni and {Peterson}, Pearu and {Weckesser}, Warren and {Bright}, Jonathan and {van der Walt}, St{\'e}fan J. and {Brett}, Matthew and {Wilson}, Joshua and {Millman}, K. Jarrod and {Mayorov}, Nikolay and {Nelson}, Andrew R.~J. and {Jones}, Eric and {Kern}, Robert and {Larson}, Eric and {Carey}, C.~J. and {Polat}, {\.I}lhan and {Feng}, Yu and {Moore}, Eric W. and {VanderPlas}, Jake and {Laxalde}, Denis and {Perktold}, Josef and {Cimrman}, Robert and {Henriksen}, Ian and {Quintero}, E.~A. and {Harris}, Charles R. and {Archibald}, Anne M. and {Ribeiro}, Ant{\^o}nio H. and {Pedregosa}, Fabian and {van Mulbregt}, Paul and {SciPy 1.  0 Contributors}},
        title = "{SciPy 1.0: fundamental algorithms for scientific computing in Python}",
      journal = {Nature Medicine},
         year = 2020,
        month = feb,
       volume = {17},
        pages = {261-272},
          doi = {10.1038/s41592-019-0686-2},
archivePrefix = {arXiv},
       eprint = {1907.10121},
 primaryClass = {cs.MS},
       adsurl = {https://ui.adsabs.harvard.edu/abs/2020NatMe..17..261V}
}

@ARTICLE{2020Natur.585..357H,
       author = {{Harris}, Charles R. and {Millman}, K. Jarrod and {van der Walt}, St{\'e}fan J. and {Gommers}, Ralf and {Virtanen}, Pauli and {Cournapeau}, David and {Wieser}, Eric and {Taylor}, Julian and {Berg}, Sebastian and {Smith}, Nathaniel J. and {Kern}, Robert and {Picus}, Matti and {Hoyer}, Stephan and {van Kerkwijk}, Marten H. and {Brett}, Matthew and {Haldane}, Allan and {del R{\'\i}o}, Jaime Fern{\'a}ndez and {Wiebe}, Mark and {Peterson}, Pearu and {G{\'e}rard-Marchant}, Pierre and {Sheppard}, Kevin and {Reddy}, Tyler and {Weckesser}, Warren and {Abbasi}, Hameer and {Gohlke}, Christoph and {Oliphant}, Travis E.},
        title = "{Array programming with NumPy}",
      journal = {\nat},
         year = 2020,
        month = sep,
       volume = {585},
       number = {7825},
        pages = {357-362},
          doi = {10.1038/s41586-020-2649-2},
archivePrefix = {arXiv},
       eprint = {2006.10256},
 primaryClass = {cs.MS},
       adsurl = {https://ui.adsabs.harvard.edu/abs/2020Natur.585..357H}
}

@ARTICLE{2013A&A...558A..33A,
       author = {{Astropy Collaboration} and {Robitaille}, Thomas P. and {Tollerud}, Erik J. and {Greenfield}, Perry and {Droettboom}, Michael and {Bray}, Erik and {Aldcroft}, Tom and {Davis}, Matt and {Ginsburg}, Adam and {Price-Whelan}, Adrian M. and {Kerzendorf}, Wolfgang E. and {Conley}, Alexander and {Crighton}, Neil and {Barbary}, Kyle and {Muna}, Demitri and {Ferguson}, Henry and {Grollier}, Fr{\'e}d{\'e}ric and {Parikh}, Madhura M. and {Nair}, Prasanth H. and {Unther}, Hans M. and {Deil}, Christoph and {Woillez}, Julien and {Conseil}, Simon and {Kramer}, Roban and {Turner}, James E.~H. and {Singer}, Leo and {Fox}, Ryan and {Weaver}, Benjamin A. and {Zabalza}, Victor and {Edwards}, Zachary I. and {Azalee Bostroem}, K. and {Burke}, D.~J. and {Casey}, Andrew R. and {Crawford}, Steven M. and {Dencheva}, Nadia and {Ely}, Justin and {Jenness}, Tim and {Labrie}, Kathleen and {Lim}, Pey Lian and {Pierfederici}, Francesco and {Pontzen}, Andrew and {Ptak}, Andy and {Refsdal}, Brian and {Servillat}, Mathieu and {Streicher}, Ole},
        title = "{Astropy: A community Python package for astronomy}",
      journal = {\aap},
         year = 2013,
        month = oct,
       volume = {558},
          eid = {A33},
        pages = {A33},
          doi = {10.1051/0004-6361/201322068},
archivePrefix = {arXiv},
       eprint = {1307.6212},
 primaryClass = {astro-ph.IM},
       adsurl = {https://ui.adsabs.harvard.edu/abs/2013A&A...558A..33A}
}

@ARTICLE{2010ApJ...724..526D,
       author = {{Datta}, A. and {Bowman}, J.~D. and {Carilli}, C.~L.},
        title = "{Bright Source Subtraction Requirements for Redshifted 21 cm Measurements}",
      journal = {\apj},
         year = 2010,
        month = nov,
       volume = {724},
       number = {1},
        pages = {526-538},
          doi = {10.1088/0004-637X/724/1/526},
archivePrefix = {arXiv},
       eprint = {1005.4071},
 primaryClass = {astro-ph.CO},
       adsurl = {https://ui.adsabs.harvard.edu/abs/2010ApJ...724..526D}
}

@ARTICLE{2009arXiv0908.2854W,
       author = {{Wyithe}, Stuart and {Brown}, Michael J.~I. and {Zwaan}, Martin A. and {Meyer}, Martin J.},
        title = "{The Halo Occupation Distribution of HI Galaxies}",
      journal = {arXiv e-prints},
         year = 2009,
        month = aug,
          eid = {arXiv:0908.2854},
        pages = {arXiv:0908.2854},
          doi = {10.48550/arXiv.0908.2854},
archivePrefix = {arXiv},
       eprint = {0908.2854},
 primaryClass = {astro-ph.CO},
       adsurl = {https://ui.adsabs.harvard.edu/abs/2009arXiv0908.2854W}
}

@ARTICLE{2002ApJ...564..576D,
       author = {{Di Matteo}, Tiziana and {Perna}, Rosalba and {Abel}, Tom and {Rees}, Martin J.},
        title = "{Radio Foregrounds for the 21 Centimeter Tomography of the Neutral Intergalactic Medium at High Redshifts}",
      journal = {\apj},
         year = 2002,
        month = jan,
       volume = {564},
       number = {2},
        pages = {576-580},
          doi = {10.1086/324293},
archivePrefix = {arXiv},
       eprint = {astro-ph/0109241},
 primaryClass = {astro-ph},
       adsurl = {https://ui.adsabs.harvard.edu/abs/2002ApJ...564..576D}
}

@ARTICLE{2012ApJ...757..101T,
       author = {{Trott}, Cathryn M. and {Wayth}, Randall B. and {Tingay}, Steven J.},
        title = "{The Impact of Point-source Subtraction Residuals on 21 cm Epoch of Reionization Estimation}",
      journal = {\apj},
         year = 2012,
        month = sep,
       volume = {757},
       number = {1},
          eid = {101},
        pages = {101},
          doi = {10.1088/0004-637X/757/1/101},
archivePrefix = {arXiv},
       eprint = {1208.0646},
 primaryClass = {astro-ph.CO},
       adsurl = {https://ui.adsabs.harvard.edu/abs/2012ApJ...757..101T}
}

@ARTICLE{2012ApJ...745..176V,
       author = {{Vedantham}, Harish and {Udaya Shankar}, N. and {Subrahmanyan}, Ravi},
        title = "{Imaging the Epoch of Reionization: Limitations from Foreground Confusion and Imaging Algorithms}",
      journal = {\apj},
         year = 2012,
        month = feb,
       volume = {745},
       number = {2},
          eid = {176},
        pages = {176},
          doi = {10.1088/0004-637X/745/2/176},
archivePrefix = {arXiv},
       eprint = {1106.1297},
 primaryClass = {astro-ph.IM},
       adsurl = {https://ui.adsabs.harvard.edu/abs/2012ApJ...745..176V}
}

@ARTICLE{2014arXiv1408.4695C,
       author = {{Chapman}, Emma and {Zaroubi}, Saleem and {Abdalla}, Filipe and {Dulwich}, Fred and {Jeli{\'c}}, Vibor and {Mort}, Benjamin},
        title = "{The Effect of Foreground Mitigation Strategy on EoR Window Recovery}",
      journal = {arXiv e-prints},
         year = 2014,
        month = aug,
          eid = {arXiv:1408.4695},
        pages = {arXiv:1408.4695},
          doi = {10.48550/arXiv.1408.4695},
archivePrefix = {arXiv},
       eprint = {1408.4695},
 primaryClass = {astro-ph.CO},
       adsurl = {https://ui.adsabs.harvard.edu/abs/2014arXiv1408.4695C}
}

@ARTICLE{2007PhRvD..76h3005L,
       author = {{Lewis}, Antony and {Challinor}, Anthony},
        title = "{21cm angular-power spectrum from the dark ages}",
      journal = {\prd},
         year = 2007,
        month = oct,
       volume = {76},
       number = {8},
          eid = {083005},
        pages = {083005},
          doi = {10.1103/PhysRevD.76.083005},
archivePrefix = {arXiv},
       eprint = {astro-ph/0702600},
 primaryClass = {astro-ph},
       adsurl = {https://ui.adsabs.harvard.edu/abs/2007PhRvD..76h3005L}
}




\appendix


\bsp	
\label{lastpage}
\end{document}